\documentclass[11pt]{article}
\usepackage{jheppub}
\usepackage{amsmath,amssymb,amsfonts,graphicx,slashed,color,amsthm,mathtools,upgreek, enumerate, tensor, subfig}
\usepackage{arydshln}

\usepackage{hyperref}
\usepackage[utf8]{inputenc}
\usepackage[titletoc]{appendix}
\numberwithin{equation}{section} 
\allowdisplaybreaks

\allowdisplaybreaks

\newcommand{\beq}{\begin{equation}}
\newcommand{\eeq}{\end{equation}}
\newcommand{\bea}{\begin{equation}\begin{aligned}}
\newcommand{\eea}{\end{aligned}\end{equation}}
\newcommand{\bra}[1]{\langle #1 |}
\newcommand{\ket}[1]{| #1 \rangle}
\newcommand{\inner}[2]{\langle #1 | #2 \rangle}
\newcommand{\avg}[1]{\langle #1 \rangle}

\DeclareMathOperator{\tr}{tr}

\subheader{\small CERN-TH-2019-207}

\title{
Holographic Probes of Inner Horizons
}

\author[a, b]{Vijay Balasubramanian}
\author[a]{\!, Arjun Kar}
\author[a,b,c]{\!, G\'{a}bor S\'{a}rosi}

\affiliation[\,a]{David Rittenhouse Laboratory, University of Pennsylvania,\\
209 S.33rd Street, Philadelphia, PA 19104, U.S.A.}
\affiliation[\,b]{Theoretische Natuurkunde, Vrije Universiteit Brussel (VUB), and \\ International Solvay Institutes, Pleinlaan 2, B-1050 Brussels, Belgium.}
\affiliation[\,c]{CERN, Theoretical Physics Department, 1211 Geneva 23, Switzerland}

\emailAdd{vijay@physics.upenn.edu}
\emailAdd{arjunkar@sas.upenn.edu}
\emailAdd{gabor.sarosi@cern.ch}

\abstract{
We study the inner horizons of rotating and charged black holes in anti-de Sitter space. These black holes have a classical analytic extension through the inner horizon to additional asymptotic regions. If this extension survives in the quantum theory, it requires particular analytic properties in a dual CFT, which give a prescription for calculating correlation functions for operators placed on any asymptotic boundary of the maximally extended spacetime.  We show that for charged black holes in three or greater dimensions, and rotating black holes in four or greater dimensions, these analytic properties are inconsistent in the dual CFT, implying the absence of an analytic extension for quantum fields past the inner horizon.
Thus, we find that strong cosmic censorship holds for all AdS black holes except rotating BTZ.  To further study the latter case, we insert classical perturbations near the boundary at late times, producing shockwaves traveling along the inner horizon. We holographically compute CFT correlators in this background that probe a high energy scattering process near the inner horizon and argue that the shockwave does not destabilize the inner horizon violently enough to prevent signaling between different asymptotic regions of the Penrose diagram.   This provides evidence that the rotating BTZ black hole does violate the strong cosmic censorship conjecture.
}


\begin{document}

\maketitle

\parskip=10pt

\section{Introduction}

\begin{figure}
    \centering
    \includegraphics[scale=.75]{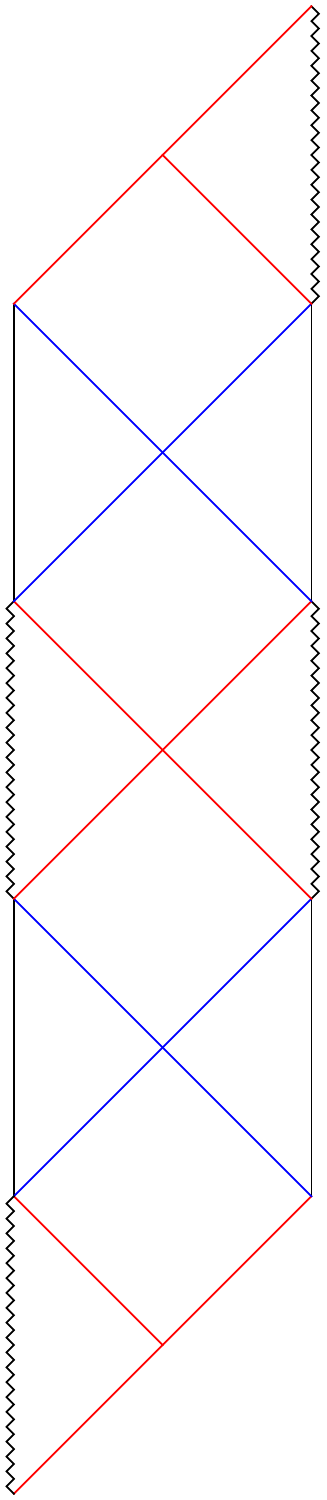}
    \caption{The Penrose diagram of a non-extremal rotating or charged black hole in anti-de Sitter space.
    The diagram repeats infinitely in the vertical direction.
    Blue lines are outer horizons, red lines are inner horizons, and jagged lines are the timelike singularity.}
    \label{fig:unlabeled-penrose}
\end{figure}

The inner horizon of a rotating or charged black hole presents several conceptual problems.
Perhaps the most disturbing of these is already apparent at the level of the Penrose diagram (Fig.~\ref{fig:unlabeled-penrose}).\footnote{
We will refer to diagrams like Fig.~\ref{fig:unlabeled-penrose} as Penrose diagrams throughout this paper, though they are more properly called projection diagrams since rotating spacetimes break spherical symmetry explicitly.
This causes confusions if one interprets these pictures as proper Penrose diagrams, which are supposed to obey certain conditions related to causal structure; we will return to this point in Sec.~\ref{sec:embedding}.
}
The maximal analytic extension of the spacetime continues infinitely in the past and future directions, and includes an infinite number of asymptotic regions.
This is in sharp contrast to the static case, where the Penrose diagram is finite and has only two asymptotic regions.
In principle, an observer could leave one asymptotic region, travel through the outer and inner horizons, view the timelike singularity, exit the black hole, and then arrive at another asymptotic region to report his or her observations.
As this thought experiment suggests, the inner horizon represents a breakdown of predictability in general relativity.
The evolution of fields is not unique beyond the inner horizon, as there is a choice of boundary conditions to be made at the singularity.
Even the metric suffers from this non-uniqueness, if we only require a finite degree of differentiability.
The strong cosmic censorship conjecture \cite{Penrose1978} was created to save the determinism of general relativity, and effectively states that performing the described experiment is impossible.

Much has been said about the strong cosmic censorship conjecture in the context of rotating and charged black holes in de Sitter, flat, and anti-de Sitter (AdS) backgrounds from the standpoint of classical gravity.
In many examples, the conjecture is enforced by an instability to perturbations \cite{Simpson:1973ua,McNamara:1978,Chandrasekhar:1982,Poisson:1990eh,Dafermos:2003wr}; the slightest perturbation (even a single particle) causes the inner horizon to collapse into a true spacelike curvature singularity. 
Classically, these analyses usually attempt to analytically extend perturbed initial data across the inner horizon, and discover that derivatives of the metric diverge in a more severe way than is usually allowed \cite{Ori:1991zz,Dafermos:2017dbw,Luk:2017jxq}.  Semiclassical analyses employ quantum field theory on curved backgrounds to understand the stress tensor sourced by a point particle as it approaches the inner horizon \cite{Hiscock:1977qe,Birrell:1978th,Balasubramanian:2004zu,Dias:2019ery,Sela:2018xko,Steif:1993zv}.
These approaches have yielded interesting insights, and several recent works have suggested that strong cosmic censorship can in fact be violated \cite{Dias:2019ery,Papadodimas:2019msp}.\footnote{
For just a small collection of older work on strong cosmic censorship, see \cite{Ori:1991zz,Mellor:1989ac,Brady:1996za,Bhattacharjee:2016zof}.
For the recent resurgence of interest, see \cite{Dias:2018etb,Dias:2018ynt,Luna:2018jfk,Luk:2017jxq,Hod:2018lmi,Mo:2018nnu,Dias:2018ufh,Hod:2018dpx,Cardoso:2017soq} and particularly the introductions of \cite{Dias:2019ery,Papadodimas:2019msp} and references therein.
}
Nevertheless, it is safe to say that the validity of strong cosmic censorship is still an open question, and may not be fully resolved until a complete theory of quantum gravity is understood nonperturbatively.

In asymptotically AdS backgrounds, such a nonperturbative ultraviolet completion of classical gravity is famously given by a conformal field theory (CFT) \cite{Maldacena:1997re,Witten:1998qj}.    In this case the infinite number of asymptotic regions in Fig.~\ref{fig:unlabeled-penrose} na\"{i}vely suggests that the maximally extended spacetime should be dual to an infinite product of field theories, one on each asymptotic boundary.  But this picture must be wrong for two reasons.  First, the various asymptotic regions are timelike separated, and so are not independent.  Secondly, we know
that eternal black holes in AdS space are described by thermofield double states \cite{Maldacena:2001kr} in a product of just two Hilbert spaces even in the charged and rotating cases.  From this perspective also the additional asymptotic regions must not be independent of the first two, again suggesting some version of cosmic censorship (see \cite{Engelhardt:2015gla} for a related discussion). The physics of the interiors of AdS black holes and horizon stability have been perturbatively studied using the AdS/CFT correspondence \cite{Hemming:2002kd,Kraus:2002iv,Levi:2003cx,Fidkowski:2003nf,Kaplan:2004qe,Balasubramanian:2004zu,Brecher:2004gn}.
In this context, some of the most intriguing results have been extracted by studying the analytic behavior of CFT correlation functions.
The authors of \cite{Fidkowski:2003nf,Festuccia:2005pi} observed that the bulk geometry predicts the existence of a spurious lightcone singularity in the CFT two-point function, but that such a singularity is only present on the second sheet of the analytically continued correlator.
Later, analytic properties of wave equations in rotating black hole spacetimes were used to explain how information about the inner horizon appears in CFT correlators \cite{Castro:2013kea}.

We develop these ideas further to study the implications of extending the black hole metric beyond the inner horizon on CFT correlation functions.  The potential ambiguities in this extension of spacetime (see above) are fixed if we require analyticity; thus the maximal analytic extension of is unique. As discussed above, is clear that if this extended spacetime means anything from the CFT point of view, a prescription to compute correlators of operators between any asymptotic boundary from correlators of a single CFT should exist.\footnote{There should be a prescription to compute the correlations in a two-sided thermofield double state as we discussed above, but, as is well known, such correlators can be obtained by analytic continuation from a one-sided correlator.}  The analytic structure of the bulk allows us to deduce how to do this via analytic continuation of correlators on a single boundary\footnote{See \cite{Kraus:2002iv} for a similar prescription for scattering through the singularity in static BTZ}.  We use this prescription to conduct two tests of inner horizon stability in AdS black holes. First, we examine conditions for the procedure to be well defined, by requiring Lorentzian bulk correlators to be single valued when we move AdS operators around Lorentzian loops that do not cross light cones of other insertions (see \cite{Castro:2013kea} for a closely related discussion). For loops enclosing the outer bifurcate horizon, we find that single valuedness of boundary correlators is equivalent to the well-known KMS condition, satisfied by any thermal correlation function. We find a similar, but inequivalent, periodicity condition coming from loops enclosing the inner bifurcate horizon.\footnote{We will argue that these periodicity conditions are sensitive to the structure of the vacuum state of the bulk fields, analogously to those probed in the smoothness test of \cite{Papadodimas:2019msp}.}
We  find that the  KMS condition coming from the inner horizon is in clash with unitarity and causality in the boundary CFT for charged, and in higher than three bulk dimensions, also rotating black holes. However, for rotating BTZ black holes, the two conditions are satisfied simultaneously and correspond to having a separate KMS periodicity condition for leftmovers and rightmovers in the dual CFT.  We will examine the implications of these results and interpret them as saying that in quantum gravity the black hole spacetime cannot be extended beyond the inner horizon for charged black holes in any dimension, and for rotating black holes in more than three dimensions.\footnote{We focus on the non-extremal case;  extremal black holes have  very different Penrose diagrams.}

Clearly, the rotating BTZ black hole \cite{Banados:1992wn,Banados:1992gq,Carlip:1995qv} is special, since we find no obstructions to define correlation functions between boundaries separated by the inner horizon via analytic continuation. We therefore move on to study such correlation functions with the aim to look for signatures of an instability under perturbations. The idea is to study a two point function between two sides of the inner horizon in a state that is slightly perturbed away from the rotating thermal ensemble. We realize this perturbation via the insertion of another local operator, and therefore the problem reduces to studying a certain analytic continuation of a CFT four point function. We will determine this four point function in two ways. First, in the high temperature limit, in which case one can map the four point function to the plane via the exponential map. We show that the relevant analytic continuation probes the second sheet of the vacuum correlator, which is the same that is relevant for out of time ordered correlators (OTOC) and the butterfly effect \cite{Roberts:2014ifa}.\footnote{OTOCs have been computed in many different backgrounds and theories\cite{Caputa:2016tgt,Caputa:2017rkm,Stikonas:2018ane,Balasubramanian:2019stt,Craps:2019rbj,Jahnke:2019gxr,Ahn:2019rnq}, and studying OTOCs has led to bounds on chaos and the butterfly effect in quantum field theory\cite{Maldacena:2015waa,Mezei:2019dfv}.} One can then give an expression for holographic theories using large $c$ Virasoro vacuum block approximation\cite{Fitzpatrick:2014vua,Roberts:2014ifa}. Second, we perform a bulk high energy scattering experiment near the inner horizon using the elastic eikonal approximation along the lines of \cite{Shenker:2013pqa,Roberts:2014isa,Shenker:2014cwa}. In this setup, the perturbing operator produces an exact shockwave solution at the inner horizon.\footnote{See \cite{Stikonas:2018ane,Jahnke:2019gxr} for OTOC calculations involving the outer horizon of rotating BTZ.
}

With such a four point function in hand, we can ask what kind of effects of instability we are looking for. In generic situations, four point functions in the CFT factorize to leading order in $G_N$. Having an instability implies that we cannot neglect backreaction and therefore we are looking for a kinematical enhancement of $G_N$ corrections (i.e., secular effects). This is precisely something that we expect from OTO-like correlation functions, which manifest such effects associated with Lyapunov growth in the context of dynamical chaos. Three possible scenarios are:
\begin{enumerate}
    \item \label{pt:mild}They are limited to certain special configurations of points between the two sides of the Cauchy horizon: in this case we would say cosmic censorship is violated because, for most pairs of points on opposite sides of the horizon, the correlator remains unchanged, suggesting that bulk probes are able to pass through largely undisturbed.
    \item They affect all configurations of points between the two sides of the Cauchy horizon, but the resummed correlator is non-trivial: in this case, some form of cosmic censorship occurs,  but a careful study of the endpoint of the instability will be required to decide what happens in the end.
    \item The resummed effect of the perturbation causes correlators between two sides of the inner horizon factorize $\langle W_{\rm up}W_{\rm down}\rangle \approx \langle W_{\rm up}\rangle \langle W_{\rm down}\rangle$. We would interpret this as saying that the spacetime breaks into two disconnected pieces along the inner horizon.
\end{enumerate}
We will find scenario \ref{pt:mild} for the four point function in the rotating BTZ black hole. As further evidence for  stability, we will explore the structure of null geodesics in the presence of the shockwave at the inner horizon and find that the shockwave makes it easier to send signals between boundaries separated by the Cauchy horizon. Thus, the rotating BTZ black hole appears to violate cosmic censorship in agreement with other recent work \cite{Dias:2019ery,Papadodimas:2019msp}. 

The paper is organized as follows.
In Sec.~\ref{sec:kms}, we present a general CFT argument against a smooth, unique spacetime beyond the inner horizon for any AdS black hole (excluding rotating BTZ) using monodromy ideas similar to \cite{Castro:2013kea}.
We also present a proposal for analytic continuation of CFT correlators which places operators in different boundaries of the Penrose diagram for a rotating or charged thermal ensemble.
In Sec.~\ref{sec:embedding}, we review the embedding space construction of the rotating BTZ black hole and introduce various useful coordinate patches.
In Sec.~\ref{sec:geodesics}, we analyze boundary-anchored null geodesics in rotating BTZ with a shockwave on the inner horizon.\footnote{Similar shockwaves were studied for flat space 4d black holes in \cite{Marolf:2011dj}.
}
In Sec.~\ref{sec:otoc}, we explain the relevance of OTOC methods for inner horizon stability, and calculate these quantities in the boosted black brane and rotating BTZ backgrounds.  We conclude with a discussion in Sec.~\ref{sec:discussion}.

\section{Monodromy, No-Go Theorems, and Multiboundary Correlators}\label{sec:kms}

In this section, we exploit the analytic structure of the black hole geometry to infer contraints on the analytic structure of  correlation functions in the holographically dual CFT. Specifically, we will move probe operators from a boundary of the geometry, through the bulk, and back again to the boundary, and analyze the structure of two-point functions obtained in this way at leading order in  $G_N$.\footnote{Higher point correlators factorize at this order.}
In this way, we \ derive two monodromy relations for the two-point function  on an asymptotic boundary by demanding single-valuedness of the correlator when we move an operator around a bifurcate horizon.
For charged black holes, and rotating black holes in dimensions greater than 3, this constraint (which probes how local operators sense the quantum gravity vacuum for these CFT ensembles) will already be enough to see signatures of the breakdown of smoothness at the inner horizon: we present two no-go theorems based on these constraints, one for charged and another for rotating black holes, showing that the bulk spacetime does not have a consistent analytic extension beyond the inner horizon.

These no-go theorems leave open the possibility that, uniquely, the rotating BTZ black hole in three dimensions has a maximal analytic extension of the standard form which is well-behaved in quantum gravity. The classical extension has an infinite series of timelike separated boundaries past the inner horizon (Fig.~\ref{fig:unlabeled-penrose}).  We propose a definition for the correlation function of operators placed on multiple boundaries of this extended geometry.  Our prescription can be thought of as a fractional combination of the  monodromy relations around the bifurcate horizons, and extends the standard method of calculating the correlation function of operators on the spacelike separated boundaries of the non-rotating  black hole.  We apply our prescription to the rotating BTZ geometry in Sec.~\ref{sec:otoc}.

\subsection{Outer KMS conditions}

Consider the eternal AdS-Schwarzschild wormhole. This geometry is understood as the dual to the thermofield double state (TFD) on two copies of the boundary CFT \cite{Maldacena:2001kr}. The TFD state is created by slicing up the path integral that calculates the thermal partition function.  In this interpretation, the partition function is the norm of the TFD state. The Euclidean geometry has a thermal circle that is cut at two locations, giving  two spatial slices. This leads to two copies of the CFT. In this Euclidean set up, it is easy to see that two sided correlators (i.e. correlators of operators placed in the two CFT copies) are obtained from one sided correlators in the thermal theory by sending operators halfway around the thermal circle. That is
\beq
\label{eq:TFDprescription}
\langle V_R(t,x)V_L(0,0)\rangle_{\beta} = \langle V_L(t-i \beta/2,x)V_L(0,0)\rangle_{\beta},
\eeq
where $\beta$ is the inverse temperature, and we require the operators to be spacelike separated so that the correlator is on the Euclidean sheet, that is $|t|<|x|$.
On the other hand, if we send an operator all the way around the thermal circle, we recover a condition on the one-sided correlator
\beq
\label{eq:outerKMS1}
\langle V(x,t-i\beta)V(0,0)\rangle_\beta =\langle V(x,t)V(0,0)\rangle_\beta , \quad |x|>|t|.
\eeq
This is the well known KMS condition that is true for any thermal correlation function and follows simply from the cyclicity of the thermal trace and the fact that the logarithm of the thermal density matrix is the Hamiltonian.\footnote{When the ensemble is charged, neutral operators satisfy the same condition because they commute with the charge.} In terms of Euclidean correlation functions, it is simply the statement that the Euclidean manifold has a compact thermal circle of length $\beta$.

\begin{figure}
\centering
\includegraphics[scale=.75]{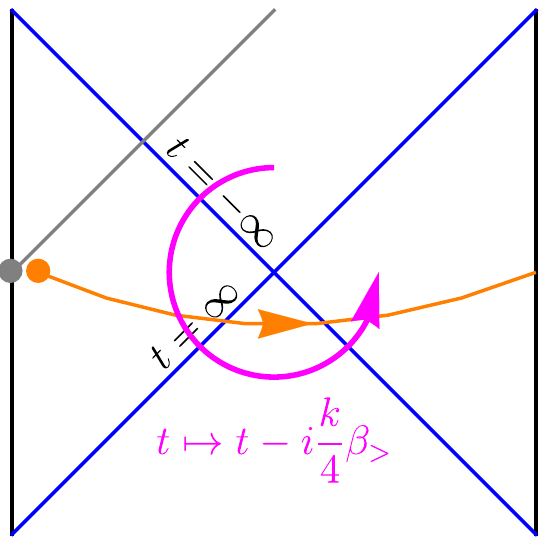}
\caption{Moving an operator between spacelike separated boundaries in the analytically extended spacetime. First, we start on the Euclidean sheet of the correlator, such that the operator we want to move (orange dot) is spacelike separated from the rest of the insertions (gray dot). Then we move the operator from one side to the other in the analytically extended Lorentzian spacetime, and we can do this without crossing light cones. In terms of Killing time coordinate $t$, this amounts to sending $t\mapsto \pm i \beta_{>}/2$ in the position of the operator (wile keeping a co-rotating coordinate fixed, if this is a rotating black hole).
}
\label{fig:movingoperators0}
\end{figure}

We propose an alternative understanding of this prescription that comes directly from the Lorentzian holographic geometry. The AdS-Schwarzschild metric has the general form
\beq
\label{eq:generalBH}
ds^2 = -F(r)dt^2+\frac{1}{F(r)}dr^2 + h_{ij}(r,x)dx^i dx^j,
\eeq
where $F(r)$ is the ``blackening factor''. Here we allow for a general $F$ so that we can also treat charged black holes.\footnote{See \cite{Chamblin:1999tk} for the explicit form of $F$ in case of charged black holes in AdS. We will discuss rotating black holes later, in Sec.~\ref{sec:rotatingkms}.} The outer horizon is the largest simple zero of $F(r)$ at $r_+$. The standard way of obtaining the maximal analytic extension is to introduce a tortoise coordinate via $dr_*=dr/F(r)$ as a step towards changing to Kruskal coordinates\footnote{See \cite{chrusciel2015geometry} for a thorough review.}
\beq
\label{eq:generalkruskal}
U=-e^{-\kappa_+ (t-r_*)}, \quad \quad V=e^{\kappa_+(t+r_*)}, \quad \quad \kappa_+=F'(r_+)/2.
\eeq
Here $\kappa_+=2\pi/\beta$ is the surface gravity, and $\beta \equiv \beta_{>}$ is the inverse temperature of the outer horizon, which is also the physical inverse temperature of the dual state.
We will later refer to the outer and inner horizon temperatures as $\beta_>$ and $\beta_<$, respectively, though for now we only need the outer temperature which we refer to as $\beta$.
In these coordinates, the horizon is at $UV=0$ and we obtain four wedges which are labeled by the four different signs that $U$ and $V$ can take. Approaching this horizon from the right outer tortoise coordinates we have $t\rightarrow \infty$ and $r_*\rightarrow -\infty$, so $U=0$ and $V$ finite. It is apparent from \eqref{eq:generalkruskal} that we can understand the change of sign in $U$ and therefore crossing to the interior in these coordinates in terms of an imaginary shift at this infinity
\beq
\label{eq:exampleshift}
t \rightarrow t-i \beta/4, \quad r_* \rightarrow r_*+i \beta/4,
\eeq
since this flips the sign of $U$ but keeps the sign of $V$. Crossing to the left exterior happens at $t=-\infty$ and $r_*=-\infty$ and here we need to perform the shift 
\beq
t \rightarrow t-i \beta/4, \quad r_* \rightarrow r_*-i \beta/4,
\eeq
so that we flip the sign of $V$ at crossing $V=0$ but keep the sign of the finite $U$. The shifts in $r_*$ cancel and we end up with the shift $t \rightarrow t-i \beta/2$ between right and left AdS-Schwarzschild coordinates. We propose that we can understand the prescription \eqref{eq:TFDprescription} in the Lorentzian spacetime, by thinking about boundary correlators as the boundary limit of bulk correlators, and then picking an operator that is space-like separated from the others and sending it to the other side of the wormhole via the Lorentzian bulk, while keeping track of the trajectory of its complexified coordinates. We do not cross any lightcones in the process and the result of this procedure is clearly \eqref{eq:TFDprescription}. This is illustrated in Fig. \ref{fig:movingoperators0}.

To be consistent and important  condition must be satisfied:  as long as we do not cross light cones of other insertions, the result should not depend on the bulk path through which we carry the operator. In other words, correlation functions should be single valued in the bulk, up to operator ordering. Since the Killing time coordinate in a four wedge gluing, as in Fig. \ref{fig:movingoperators0}, picks up imaginary shifts in going around the wedges, non-trivial constraints arise from moving operators around such bifurcate horizons. 

In fact, we have already met an example of such a constraint.
Consider the boundary two point function of a neutral operator in a (possibly charged) black hole background, where the operators are spacelike separated, and carry an operator around the outer bifurcate horizon (left panel of Fig. \ref{fig:lorentziankms}). This gives rise to the periodicity condition\footnote{Here we are being a bit schematic, since the boundary manifold might not be flat space. By $|x|>|t|$ we mean that the real parts of the separation is spacelike. By choosing $|x|$ much smaller than the curvature radius of the boundary manifold, the flat space formulas do apply.}
\beq
\langle V(x,t-i\beta)V(0,0)\rangle_\beta =\langle V(x,t)V(0,0)\rangle_\beta , \quad |x|>|t|.
\eeq
which is precisely the KMS condition \eqref{eq:outerKMS1} in the CFT.
This requirement is equivalent to the usual way of determining the temperature by requiring absence of a conical singularity in the Euclidean bulk geometry, and therefore single valued correlation functions in the Euclidean bulk. However, the Euclidean geometry does not explicitly represent the interior of the black hole, so it is more difficult to relate properties of the inner horizon to Euclidean boundary correlators. This is why we find the above Lorentzian single valuedness condition convenient: it can be directly applied to the inner bifurcate horizon.

\subsection{Inner KMS conditions and consequences of its violation}

\begin{figure}
\centering
\includegraphics[width=0.3\textwidth]{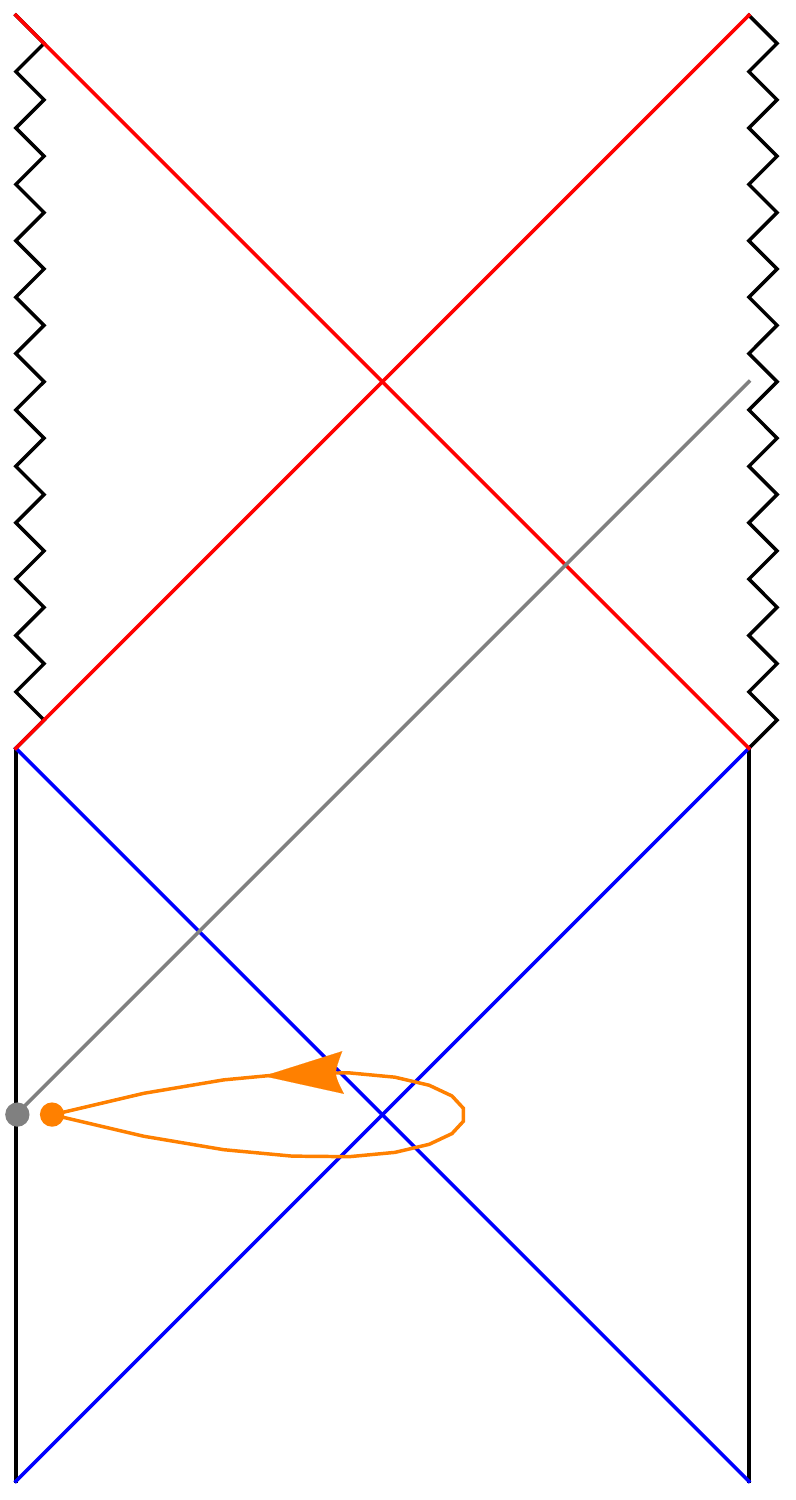}\hspace{1cm}
\includegraphics[width=0.3\textwidth]{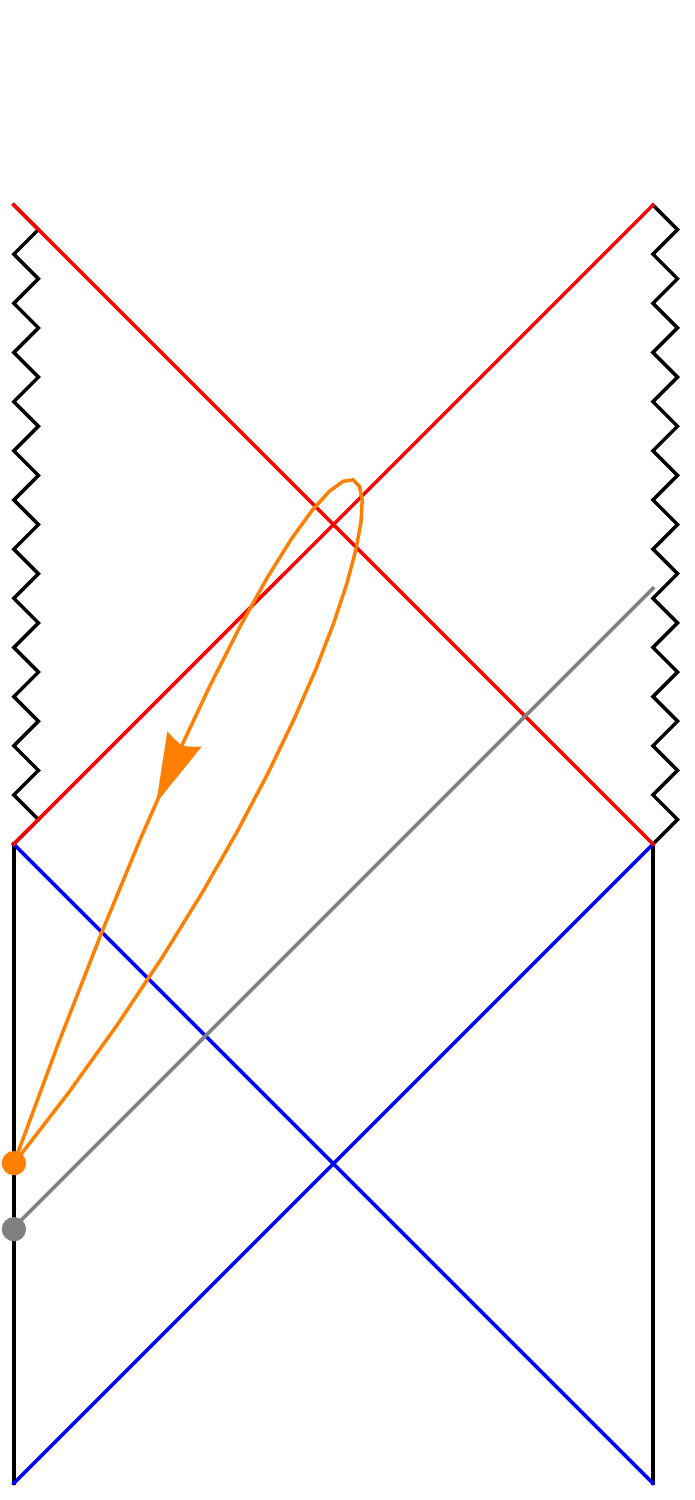}
\caption{Left: carrying an operator around the outer bifurcate horizon gives rise to the KMS condition. Right: carrying an operator around the inner horizon gives rise to a new periodicity condition.
}
\label{fig:lorentziankms}
\end{figure}

Let us consider a boundary two point function\footnote{To leading order in $G_N$, the condition follows for higher point functions from large $N$ factorization. Here we restrict attention to this situation.} where the operators are time-like separated, and we move the futuremost operator around a loop encircling the inner bifurcate horizon, where the loop is such that it stays in the future of the other insertion. Such a loop is illustrated on the right of Fig. \ref{fig:lorentziankms}. 
We can extract the monodromy directly from the Kruskal coordinates.
However, the Kruskal patch \eqref{eq:generalkruskal} does not cover the inner horizon.
There is an analogous Kruskal transformation which covers four wedges of the Penrose diagram around the inner horizon, given for charged black holes by
\beq
\label{eq:generalinnerkruskal}
U=-e^{-\kappa_- (t-r_*)}, \quad \quad V=e^{\kappa_-(t+r_*)}, \quad \quad \kappa_-= -F'(r_-)/2,
\eeq
where the inner horizon temperature $\beta_{<}$ is defined via $\kappa_- \equiv 2\pi/\beta_{<}$.
Notice that both the outer \eqref{eq:generalkruskal} and inner \eqref{eq:generalinnerkruskal} Kruskal transformations have an invariance under shifting the exponent by $2\pi i$, which can be achieved by
\begin{equation}
    t \to t + \frac{2\pi i}{\kappa_+}, \quad \quad   t \to t + \frac{2\pi i}{\kappa_-} ,
\end{equation}
in the outer and inner horizon cases, respectively.
The inner shift then gives the condition
\beq
\label{eq:innerKMS1}
\langle V(x,t-i\beta_{<})V(0,0)\rangle_{\beta_>\beta_<} =\langle V(x,t)V(0,0)\rangle_{\beta_>\beta_<} , \quad |x|<|t|.
\eeq
We will call this the ``inner KMS condition". 
Also, we now denote expectation values in rotating or charged ensembles with two temperatures.
These two temperatures can be either the outer and inner horizon temperatures $\beta_>$ and $\beta_<$, or the chiral CFT temperatures $\beta_+$ and $\beta_-$.
Of course, these quantities are related via
\begin{equation}
\begin{split}
    \beta_> & = \frac{1}{2}(\beta_+ + \beta_-) , \\
    \beta_< & = \frac{1}{2}(\beta_+ - \beta_-) .
\end{split}
\end{equation}

To infer the inner KMS condition \eqref{eq:innerKMS1}, we are moving the bulk operator past the Cauchy horizon; thus we are testing the conditions imposed by the existence of a CFT dual on the
evolution of bulk fields past this horizon. So let us spell out what it means for this evolution if we find that \eqref{eq:innerKMS1} fails in the CFT. The infinite $N$ two point function in the bulk is determined by solving a linear wave equation in the analytically extended spacetime. This solution is unique and unaffected by the singularity in the diamond that is between the outer and inner horizons. This diamond is part of the inner Kruskal patch \eqref{eq:generalinnerkruskal}, and we can  ask if we can analytically continue this solution past the inner horizon. If an analytic continuation exists to some regions beyond the initial diamond $\text{Re }U,\text{Re }V>0$, it is of course unique because of the identity theorem, though possibly multi-valued. There are then three possibilities for this  extension:
\begin{enumerate}
    \item \label{pt:nice}It is analytic in a small neighborhood of $U=V=0$ in which case it must be single valued and obey the inner condition \eqref{eq:innerKMS1}.
    \item \label{pt:notsonice}It has a branch point at $U=V=0$ in which case there is a discrete label on the possible continuations for real $U,V<0$ past the inner horizon. This evolution is then not single valued and we would find a violation of \eqref{eq:innerKMS1} in the boundary CFT.
    \item  \label{pt:notnice}There is an essential singularity on the lines $\text{Re }U=0$, $\text{Re }V=0$ and no continuation exists past the diamond.
\end{enumerate}

Now let us examine the dependence of the solutions to the wave equation on the boundary conditions on the timelike singularities. These boundary conditions can only change the solution past the Cauchy horizon. Therefore, also taking into account the linearity of the equation, the generic solution of the wave equation must have the form
\beq
\phi = \phi_{\rm analytic} + \phi_{\rm non-analytic},
\eeq
where $\phi_{\rm analytic}$ is the analytic extension (if it exists) discussed in the previous paragraph, and $\phi_{\rm non-analytic}$ is a solution that is identically zero in the diamond between the inner and outer horizons (and therefore nowhere analytic in inner Kruskal coordinates on the inner horizon). All the dependence on the timelike singularities is in $\phi_{\rm non-analytic}$, and it is by definition single valued around the bifurcate horizon, since it starts from zero and goes to zero. It follows that the monodromy of the solution around $U=V=0$ is determined only by $\phi_{\rm analytic}$. 

We conclude that if we find that \eqref{eq:innerKMS1} is violated in the CFT, we can rule out possibility \ref{pt:nice}, i.e. either the bulk is not unique around the inner bifurcate horizon, or the bulk two point function has an essential singularity everywhere on the inner horizon. Both of these scenarios are in some sense worse than just having say some bombs thrown in from the timelike singularity along the inner horizon: neither of them allows for a  sensible definition of spacetime past the inner horizon. In scenario \ref{pt:notnice}, we would in fact see a complete decoupling between the two sides of the horizon, since in that case the two independent pieces of the solution are localized on different sides.

This test is  similar to the one proposed in \cite{Papadodimas:2019msp}, as it probes smoothness of the bulk quantum vacuum on a fixed background.  The argument in \cite{Papadodimas:2019msp} is that a finite energy state of effective field theory in the bulk requires the bulk two point function to have a short distance singularity that is identical with what one finds in the flat space vacuum state, and having this preferred short distance singularity near both the inner and outer horizons at the same time is a constraint on the bulk evolution of the modes that is difficult to satisfy. 
We will comment further on the relationship with the smoothness condition of \cite{Papadodimas:2019msp} in Sec.~\ref{sec:discussion}. 

\subsection{Charged operators}

Once again, for both \eqref{eq:outerKMS1} and \eqref{eq:innerKMS1} we restrict to charged black holes and we assume the operator $V$ to be neutral. For a charged operator, we use the fact that the correlation function in the boundary is written in the form $\text{Tr}[e^{-\beta_> H+\mu Q} V(t)V^\dagger(0)]$ and assume that $[Q,V]=q V$. From this, we obtain the charged KMS condition
\beq
\label{eq:chargedKMS}
\langle V(x,t-i\beta_>)V^\dagger (0,0)\rangle_{\beta_>\beta_<} =e^{-\mu q} \langle V(x,t)V(0,0)^\dagger\rangle_{\beta_>\beta_<}, \quad |x|>t.
\eeq
The extra factor of $e^{-\mu q}$ can be eliminated by making the state dependent redefinition 
\beq
\label{eq:dressedV}
V(t)\rightarrow e^{-i\frac{\mu q}{\beta_>}t}V(t).
\eeq
One can understand this from the Lorentzian bulk, by representing the bulk two point function in the first quantized formalism\footnote{Strickly speaking, one would have to look at the monodromies of the bulk Green's function, following \cite{Castro:2013kea}, as before, we bypass this with a heuristic argument.}
\beq
\label{eq:chargedfirstq}
\langle \Phi_V(x)\Phi_V(y)^\dagger \rangle \sim \int_{X_i=x, X_f=y} \mathcal{D}X e^{-i m_V \int \sqrt{\dot{X}^2}-i q\int A_\mu {\dot X}^\mu}.
\eeq
Here, $\Phi_V$ is the bulk field dual to $V$ and $m_V$ its mass, $q$ its charge, and $A$ the vector potential of the black hole background \cite{Chamblin:1999tk}
\beq
A=\left( -\frac{C}{r^{d-2}}+\Phi \right) dt,
\eeq
where $C$ is a constant proportional to the charge of the black hole \cite{Chamblin:1999tk} and $\Phi$ is a free parameter. We expect \eqref{eq:chargedfirstq} again to give a single valued function in the bulk, apart from a possible monodromy around bifurcate horizons. This monodromy is fixed by the value of the constant $\Phi$, which we can change by a gauge transformation $A \mapsto A+ d\chi$, $\chi=\chi_0 t$, but it is important that $\chi$ is not single valued in the Euclidean bulk, since Euclidean time is periodic. Different choice of $\chi_0$ correspond to different dressings of the operator as in \eqref{eq:dressedV}. However, once we have fixed this dressing, it is not possible to remove it for the monodromy around the inner bifurcate horizon. 
So in a gauge where we remove the exponential factor for the outer KMS condition, we expect to obtain conditions of the form
\bea
\label{eq:chargedoperatordoublekms}
\langle V(x,t-i\beta_>)V^\dagger (0,0)\rangle_{\beta_>\beta_<} = \langle V(x,t)V(0,0)^\dagger\rangle_{\beta_>\beta_<}, \quad |x|>t, \quad \text{(outer)}\\
\langle V(x,t-i\beta_{<})V^\dagger (0,0)\rangle_{\beta_>\beta_<} =e^{-\tilde \mu q} \langle V(x,t)V(0,0)^\dagger\rangle_{\beta_>\beta_<}, \quad |x|<t, \quad \text{(inner)},
\eea
where $\tilde \mu$ is a function of $\mu$ and possibly $\beta$. We refrain from determining its explicit form, because we will not need it in the following. 


\subsection{KMS conditions in rotating black holes}
\label{sec:rotatingkms}

We now turn to the rotating AdS black hole.
When we have rotation, the metric is not of the form \eqref{eq:generalBH}, but an analogous construction of Kruskal coordinates exists. The important new feature in this case is that we need to keep co-rotating coordinates fixed as we analytically continue between different wedges in the Penrose diagram. These are related to boundary spatial coordinates by shifts in time, therefore we end up with a prescription that is similar to \eqref{eq:outerKMS1}, except that it also involves complex shifts in spatial coordinates. We will work this out explicitly in the case of the rotating BTZ black hole below.
Subsequently, we will show how it works for all dimensions $D \geq 4$.

Our strategy for dealing with the rotating black hole begins with Schwarzschild coordinates which cover a single wedge on the Penrose diagram. We will uncover the appropriate Kruskal transformation to analytically extend around the bifurcate point.\footnote{
The naming of coordinate systems for the rotating black hole is confusing.
The analog of Schwarzschild coordinates is usually referred to as Boyer-Lindquist coordinates, despite the fact that this system is only a small modification of Kerr's own \cite{Kerr:1963ud}.
The major novel contribution of Boyer and Lindquist \cite{Boyer:1966qh} was actually, among other things, to construct the analog of Kruskal coordinates for the Kerr black hole.
We will keep the Schwarzschild and Kruskal nomenclature even for the rotating black hole.
}
The form of this transformation will yield the appropriate analytic continuation of boundary coordinates, just as it did for the charged black hole.
In the 3-dimensional case, the Kruskal transformation for the (co-rotating) asymptotic region is well known and is given by \cite{Carlip:1995qv}
\begin{equation}
    U = -e^{-\kappa_+ (t - r_*)}, \hspace{.5cm} V = e^{\kappa_+ (t+r_*)} ,
\end{equation}
where $r_*$ is a tortoise coordinate and we can express the surface gravities for rotating BTZ in terms of the inner and outer horizon radii via
\begin{equation}
    \kappa_\pm \equiv \frac{r_+^2-r_-^2}{r_\pm} .
\end{equation}
Just as in the charged case, there is an invariance of these coordinates in the complex $t$ plane corresponding to a shift by $2\pi i$ of the exponential argument, and this shift in the bulk sends us around the bifurcate horizon once.
The necessary shift is simply
\begin{equation}
\label{eq:BTZouter1}
    t \to t + \frac{2\pi i}{\kappa_+} ,
\end{equation}
and since we are keeping the co-rotating angle $\varphi - \frac{r_-}{r_+} t$ fixed, this induces a shift in the boundary angle
\begin{equation}
\label{eq:BTZouter2}
    \varphi \to \varphi + \frac{r_-}{r_+}\frac{2\pi i}{\kappa_+} = \varphi + \frac{2\pi i}{\kappa_-} .
\end{equation}
For the interior region (next to the singularity), the Kruskal transformation is instead
\begin{equation}
    U = -e^{-\kappa_- (t - r_*)}, \hspace{.5cm} V = e^{\kappa_- (t+r_*)} ,
\end{equation}
and so correspondingly we have for the inner shift (note the inner co-rotating angle $\varphi - \frac{r_+}{r_-}t$ differs from the outer one)
\begin{equation}
\label{eq:BTZinner}
    t \to t + \frac{2\pi i}{\kappa_-} , \hspace{.5cm} \varphi \to \varphi + \frac{2\pi i}{\kappa_+} .
\end{equation}
The monodromy prescriptions for the correlator are therefore
\begin{equation}
\label{eq:rotatingKMS}
    \avg{V (t,\varphi) V(0,0)}_{\beta_{>}\beta_{<}} = \avg{V \left(t+\frac{2\pi i}{\kappa_\pm},\varphi+\frac{2\pi i}{\kappa_\mp}\right) V(0,0)}_{\beta_>\beta_{<}} .
\end{equation}
These are the two KMS conditions for rotating BTZ.
We will also be interested in the static limit of these formulas, and the monodromies at $r_- \to 0$ are
\begin{equation}
\begin{split}
    t & \to t + \frac{2\pi i}{r_+} , \hspace{.5cm} \varphi \to \varphi \hspace{.5cm}\text{ (outer) },\\
    t & \to t , \hspace{.5cm} \varphi \to \varphi + \frac{2\pi i}{r_+} \hspace{.5cm}\text{ (inner) }.
\end{split}
\end{equation}

For the $D \geq 4$-dimensional generalizations, we employ the results and formalism of \cite{Gibbons:2004uw}, and the metric of an AdS black hole with rotation parameters is given in Appendix E of \cite{Gibbons:2004uw}.
We reproduce this metric here:
\begin{equation}
    \begin{split}
        ds^2 = & -W(r^2+1) d\tau^2 + \frac{2M}{SF} \left( W d\tau - \sum_{k=1}^N \frac{a_k \mu_k^2 d\varphi_k}{1-a_k^2} \right)^2 + \sum_{k=1}^N \frac{r^2+a_k^2}{1-a_k^2} \mu_k^2 d\varphi_k^2 \\
        & + \frac{SF dr^2}{S-2M} + \sum_{k=1}^{N+\epsilon} \frac{r^2+a_k^2}{1-a_k^2} d\mu_k^2 - \frac{1}{W(r^2+1)} \left( \sum_{k=1}^{N+\epsilon} \frac{r^2+a_k^2}{1-a_k^2} \mu_k d\mu_k \right)^2 ,
    \end{split}
\end{equation}
where we have defined $N$ implicitly by $D = 2N + \epsilon + 1$, and 
\begin{equation}
\label{eq:adskerr}
    \begin{alignedat}{2}
        W & \equiv \sum_{k=1}^{N+\epsilon} \frac{\mu_k^2}{1-a_k^2} , \hspace{.5cm} & 
        F & \equiv \frac{1}{r^2+1} \sum_{k=1}^{N+\epsilon} \frac{r^2 \mu_k^2}{r^2+a_k^2} , \\
        S & \equiv r^{\epsilon-2} (r^2+1) \prod_{k=1}^N (r^2+a_k^2) , \hspace{.5cm} & 
        \epsilon & \equiv D - 1 \text{ mod } 2.
    \end{alignedat}
\end{equation}
The $\mu_k$ are latitudinal coordinates obeying the constraint $\sum_{k=1}^{N+\epsilon} \mu_k^2=1$, while the $a_k$, $k=1,...,N$ are  rotation parameters of the solution, with $a_{N+1}=0$ in the even dimensional case. Before proceeding to the monodromy, we discuss a subtle point.
As is well known, in dimension 5 and above, there are rotating extended black objects with non-spherical horizon topologies.
The simplest of these are black rings \cite{Emparan:2001wn}.
In AdS/CFT, if there are multiple Euclidean geometries which 1) are saddle points of the Euclidean Einstein-Hilbert action and 2) obey the boundary conditions set by the CFT thermal ensemble, then the bulk dual geometry is the one with the lowest free energy.
The question of which geometry is the true thermodynamic saddle for a given temperature or chemical potential is, to our knowledge, still unresolved in general for $D \geq 5$.
This is in part because the full set of solutions has not even been classified, and constraints from quantum gravity appear to play a role \cite{McInnes:2019}.
However, much has been said about black rings and black holes \cite{Emparan:2007wm,Altamirano:2014tva}, and these results suggest that the rotating AdS black hole dominates the thermodynamics at large temperature and equal rotation rates.\footnote{The usual Hawking-Page transition also occurs in higher dimensions, and for equal rotation rates the Kerr-AdS solution is dominant over thermal AdS \cite{Carter:2005uw}.}

Thus, our higher dimensional statements should be understood with the caveat that we assume that the rotating black hole with an inner horizon dominates the thermodynamics on a line in moduli space that passes through the static black hole.
We take this line to be the equal rotation rate line, where all rotation parameters in higher dimensions are equal, and we therefore set
\begin{equation}
    a_k \equiv a .
\end{equation}
There is evidence that our assumption is true, at least in $D = 5,6$ \cite{Hawking:1998kw,Altamirano:2014tva}.
The reason we did not address this point in our discussion of charged black holes is because, in that case, the full spherical symmetry constrains the possible solutions \cite{Schleich:2009uj}. 
Therefore, we are more confident Reissner-Nordstr\"{o}m-AdS is the true dual geometry for large enough temperature since there are no rings or other extended objects with which it must compete.\footnote{In \cite{Schleich:2009uj}
 Birkhoff's theorem is proved for 4d AdS black holes, but not for charged AdS black holes in general dimension.  In flat space there are results for black holes in any dimension \cite{Bronnikov:1994ja} and charged black holes in 4 dimensions \cite{Hoffmann1932}.
}

Rotating black holes can be understood in the framework of Kerr-Schild theory \cite{Kerr2009RepublicationOA}, where the full geometry $g$ is understood as a perturbation of a background $\bar{g}$ (in our case, AdS) by a particular null vector field $k$, so the metric is
\begin{equation}
    g_{\mu\nu} = \bar{g}_{\mu\nu} + \frac{2M}{U} k_\mu k_\nu ,
\end{equation}
where $M$ is the mass of the black hole and $U$ is some function of the coordinates.
By construction, the vector $k$ has the interesting property that $k^2 = 0$ in both the background metric $\bar{g}$ \textit{and} the full metric $g$.
What this means is that if we interpret the $k_\mu k_\nu$ term as a perturbation and ensure it solves the linearized Einstein equations around the $\bar{g}$ background, the perturbed metric $g$ actually becomes a solution to the full nonlinear Einstein equations.
This is in part because the exact inverse metric is $g^{\mu\nu} = \bar{g}^{\mu\nu} - \frac{2M}{U} k^\mu k^\nu$.

The strategy to determine Kruskal coordinates, following \cite{Boyer:1966qh}, is to un-twist the null vector $k$ at the horizon in question and then to effect a transformation which makes the integral curves of $k$ lie on constant coordinate hypersurfaces.
The special null vector in our case is
\begin{equation}
    k^\mu \partial_\mu = \partial_r - \frac{S}{S-2M} \left( \frac{1}{r^2+1} \partial_\tau + \sum_{i=1}^N \frac{a}{r^2+a^2} \partial_{\varphi_i} \right).
\end{equation}
The equation which determines horizon locations plays a crucial role.
It is
\begin{equation}
    S - 2M = 0 .
    \label{eq:horizon-equation}
\end{equation}
There are two positive real roots $r_\pm > 0$ which we define implicitly as\footnote{An oddity in $D \geq 5$ is that, if one of the rotation parameters vanishes, there is only one positive real root in \eqref{eq:horizon-equation}.
The second root is a direct consequence of the diverging piece $r^{\epsilon-2}$ in $S$, and if this is cancelled by a bare $r^2$, the inner horizon disappears.
We have nothing to say about black holes that lack inner horizons, so we will not discuss this further beyond noting that there are brief comments related to this issue at the end of \cite{Hawking:1998kw}.}
\begin{equation}
    S(r_\pm(M,a)) = 2M.
    \label{eq:horizon-roots}
\end{equation}
To untwist $k$ at $r_\pm$, we make the angular shift\footnote{The coefficient of the shift is, as usual, equal to the angular potential.  It does not match the angular potential quoted in \cite{Gibbons:2004uw} because we are using a coordinate system which differs at infinity from that which \cite{Gibbons:2004uw} used to extract the potential; see \cite{Carter:2005uw} for the angular potential in our coordinates.} 
\begin{equation}
    \phi_i^\pm \equiv \varphi_i - \frac{a(r_\pm^2+1)}{r_\pm^2+a^2} \tau .
\end{equation}
This transformation leaves the $\tau$ and $r$ directions unchanged, and the integral curves of $k$ therefore obey 
\begin{equation}
    d\tau = \frac{S}{2M-S} \frac{dr}{r^2+1} ,
\end{equation}
and we wish to integrate both sides.
Let us write
\begin{equation}
    r^{2-\epsilon} (2M-S) (r^2+1) = (r-r_+)(r-r_-) H(r) ,
    \label{eq:H-defn}
\end{equation}
where $H(r)$ is a polynomial in $r$ that does not vanish at $r_+$ or $r_-$.
The $r$ integral can be computed via (for the Kruskal patch around $r=r_+$)
\begin{align*}
    \tau + \text{const.} & = \int dr \frac{r^{2-\epsilon} S}{r^{2-\epsilon}(2M-S)(r^2+1)} \\
    & = \int dr \left[ -1 + \frac{2 M r_+^{2-\epsilon}}{(r_+-r_-)(r-r_+)H(r_+)} + \frac{2M r_-^{2-\epsilon}}{(r_--r_+)(r-r_-)H(r_-)} + \dots \right] \\
    & = -r + \frac{2M r_+^{2-\epsilon}}{(r_+-r_-)H(r_+)} \left[ \log(r-r_+) - \frac{r_-^{2-\epsilon}H(r_+)}{r_+^{2-\epsilon} H(r_-)} \log (r-r_-) + \dots \right].
\end{align*}
On the second line we have employed a partial fraction decomposition, which we are free to do since both the numerator and denominator are polynomials in $r$.
For the terms associated with $r_+$ and $r_-$ in this decomposition, which we have written explicitly, the numerator is simplified with \eqref{eq:horizon-roots}, and it will turn out that none of the other terms matter for our analysis so we have neglected to write them.
In order to obtain regular coordinates around a given horizon, we must ensure that the logarithm associated with that horizon in the square brackets above comes with unit coefficient, as this will exponentiate to a linear function which is  regular everywhere.
In the above, we have written the outer horizon logarithm with a unit coefficient, but we can easily obtain the analogous expression for the inner horizon by rearranging some constants.
Multiplying through by the coefficient outside the square brackets and exponentiating both sides, we find
\begin{equation}
    \left( \frac{r-r_+}{2M} \right) \left( \frac{r-r_-}{2M} \right)^{-\nu} \left( \dots \right) \exp \left[ \frac{-r-\tau}{\sigma_+} \right] = \text{const.} ,
    \label{eq:outer-U}
\end{equation}
where we have defined
\begin{equation}
    \nu \equiv \frac{r_-^{2-\epsilon} H(r_+)}{r_+^{2-\epsilon}H(r_-)} , \hspace{.5cm} \sigma_\pm \equiv \frac{2Mr_\pm^{2-\epsilon}}{(r_+-r_-)H(r_\pm)} ,
\end{equation}
and the corresponding expression for the inner Kruskal patch is
\begin{equation}
    \left( \frac{r_+-r}{2M} \right)^{-1/\nu} \left( \frac{r_--r}{2M} \right) \left( \dots\right) \exp \left[ \frac{r+\tau}{\sigma_-} \right] = \text{const.} 
    \label{eq:inner-U}
\end{equation}
Within the dots in the above expressions we have suppressed various extra terms associated with the integrals
\begin{equation}
    \int dr \frac{r_k^{2-\epsilon}S(r_k)}{(r_k-r_+)(r_k-r_-)H_k(r_k)(r-r_k)} ,
\end{equation}
where by $H_k(r_k)$ we mean to replace $r = r_k$ in $H(r)$ except for the factor $(r-r_k)$, which we drop.  Since $r_k$ is not necessarily real, we cannot conclude $S(r_k) = 2M$.
These terms do contribute functions of $r$ contained in the $(\dots )$ above, but since $H(r)$ has no real positive roots, it is strictly positive for all $r > 0$ and thus these functions must all be well-behaved in this region, i.e. they have no divergences and are analytic in $r$.
Therefore, the outer Kruskal transformation obtained by setting \eqref{eq:outer-U} equal to $U^2$ and the $\tau \to -\tau$ version of \eqref{eq:outer-U} equal to $V^2$ gives a metric with analytic components in all four patches around $r = r_+$, and similarly for the inner Kruskal transformation constants \eqref{eq:inner-U} and the patches around $r = r_-$.

In terms of boundary coordinates, the outer and inner monodromies around the bifurcate horizons are therefore given by 
\begin{equation}
    \tau \to \tau + 4\pi i \sigma_\pm  , \hspace{.5cm} \varphi_k \to \varphi_k + \frac{4 \pi i \sigma_\pm  a (r_\pm^2+1)}{r_\pm^2+a^2} .
\end{equation}
The corresponding correlator monodromy relations are
\begin{equation}
\label{eq:rotatinginnerkms}
    \avg{V (t,\varphi_k) V(0,0)}_{\beta_{>}\beta_{<}} = \avg{V \left(t+4\pi i \sigma_\pm,\varphi_k +\frac{4\pi i \sigma_\pm a (r_\pm^2+1)}{r_\pm^2+a^2}\right) V(0,0)}_{\beta_{>}\beta_{<}}
\end{equation}
In the static limit $a \to 0$, which implicitly sends $r_- \to 0$.
The outer horizon monodromy is simply
\begin{equation}
    \tau \to \tau + \frac{8\pi i M r_+^{1-\epsilon}}{H(r_+)} , \hspace{.5cm} \varphi_k \to \varphi_k .
\end{equation}
For the inner horizon monodromy, the calculation is more subtle, since we must deal with $H(r_-)$.

To begin, we want to understand the vanishing of $r_-$ with $a$.
We make a perturbative ansatz
\begin{equation}
    r_-(a) = \sum_{n=1}^\infty r_n a^n .
\end{equation}
We see there is a sort of mismatch between the two sides of the horizon equation \eqref{eq:horizon-roots}, where the left hand side has a very high power of $a$ no matter what we choose for $r_n$, and the right hand side has a very small power.
To match the polynomials, we must produce at minimum a term of order $2N$ on the right hand side, since there will always be a term like $a^{2N}$ on the left.
The lowest nonzero term must therefore be $r_{2N/(2-\epsilon)}$.
By matching constants, we have
\begin{equation}
    r_{2N/(2-\epsilon)} = \frac{1}{2M} .
\end{equation}
So, the root $r_-$ vanishes like
\begin{equation}
    r_-(a) \sim \frac{1}{2M} a^{\frac{2N}{2-\epsilon}}.
\end{equation}
We now turn to the function $H(r)$, defined in \eqref{eq:H-defn}.
We want to understand the behavior of $H(r_-)$ as $a \to 0$.
Let us write
\begin{equation}
    H(r) = \sum_{n=0}^{2N+2} h_n r^n .
\end{equation}
The degree of vanishing of this function is a subtle question because the coefficients $h_n$ can scale with $a$ also.
Let us begin with the lowest order term.
The smallest power appearing on the left hand side of \eqref{eq:H-defn} comes from $r^{2-\epsilon} S$, which actually contributes a constant (in $r$) piece $a^{2N}$.
Since the constant terms on the left and right must match, we conclude that the constant piece of the polynomial $H$ (which is simply its value at $r=0$) obeys
\begin{equation}
    a^{2N} = r_+ r_- h_0 .
\end{equation}
We have seen that $r_-$ vanishes with a particular power, and we conclude that $h_0$ must vanish like
\begin{equation}
    h_0 = \frac{a^{2N}}{r_+ r_-} \sim \frac{2M}{r_+} a^{2N \frac{1-\epsilon}{2-\epsilon}} .
\end{equation}
It is actually sufficient to stop our analysis of $H(r_-)$ at this order to understand the qualitative behavior of the remnant monodromy.
This is because even if higher order terms contribute faster or more slowly vanishing terms, the monodromy behavior will be unchanged.
To see this, first note that $H(r_-)$ cannot diverge in the limit $a \to 0$ because $H(r)$ is a perfectly well defined polynomial in that limit.

We now consider two cases: 1) $H(r_-)$ contains terms which vanish faster than the $h_0$ term we have found and 2) $H(r_-)$ vanishes more slowly than the $h_0$ term we have found (possibly not vanishing at all, as with $\epsilon = 1$).
In the first case, the vanishing of $\sigma_-$ is unaffected since $H(r_-)$ appears in the denominator so only its most slowly vanishing term will contribute.
In the second case, the vanishing of $\sigma_-$ will be strengthened compared to our result, but this only forces the remnant monodromy to vanish faster.
We find that the inner monodromy shifts in the static limit are vanishing in both cases with $a \to 0$ at least as fast as:
\begin{equation}
\label{eq:rotatingsmalla}
    \begin{alignedat}{3}
        \sigma_- & \sim a^{2N}, & \hspace{.5cm} \frac{\sigma_- a}{r_-^2+a^2} & \sim a^{2N-1}, \hspace{.5cm} & (\epsilon = 1), \\
        \sigma_- & \sim a^N, & \hspace{.5cm} \frac{\sigma_- a}{r_-^2+a^2} & \sim a^{N-1} , \hspace{.5cm} & (\epsilon = 0) .
    \end{alignedat}
\end{equation}
Since we have $N > 0$ for even $D$ and $N > 1$ for odd $D$, the inner monodromy remnant becomes degenerate in higher dimensions.
Notice that for $D=4$ ($N=1$, $\epsilon=1$) and $D=5$ ($N=2$, $\epsilon=0$) we have that the monodromies vanish linearly with $a$, which is as slowly as possible.
Thus the contribution of $h_0$ is actually marginal in these cases, and we can  guarantee that the monodromies vanish precisely like this and no faster.

We will now present, using the various KMS-type conditions derived above for CFT two-point functions, a boundary argument for the instability of the inner horizon.
We first address charged black holes in $D \geq 3$ and then address rotating black holes in $D \geq 4$.
We then explain why our arguments fail for rotating BTZ.

\subsection{No-go for charged black holes}

We will restrict to neutral fields probing the charged black hole. This will be sufficient for our purposes since in the boundary we always have a stress tensor that is neutral, and we will find that the two KMS conditions are not compatible with each other already for neutral operators. Let us first assume that we find a point in the boundary that is space-like separated from a point on the inner bifurcate horizon. If this is the case, we would be required to have \eqref{eq:outerKMS1} and \eqref{eq:innerKMS1} in the same regions, that is, as a function of Euclidean time, the two point function would need to have two real periodicities. This is not possible unless $\beta_{<}$ divides $\beta_{>}$, a condition that is not needed in the microscopic definition $\langle V(x,t)V(0,0)^\dagger\rangle\sim \text{Tr}[e^{-\beta H+\mu Q} V(t)V^\dagger(0)]$. 

\begin{figure}
\centering
\includegraphics[width=0.4\textwidth]{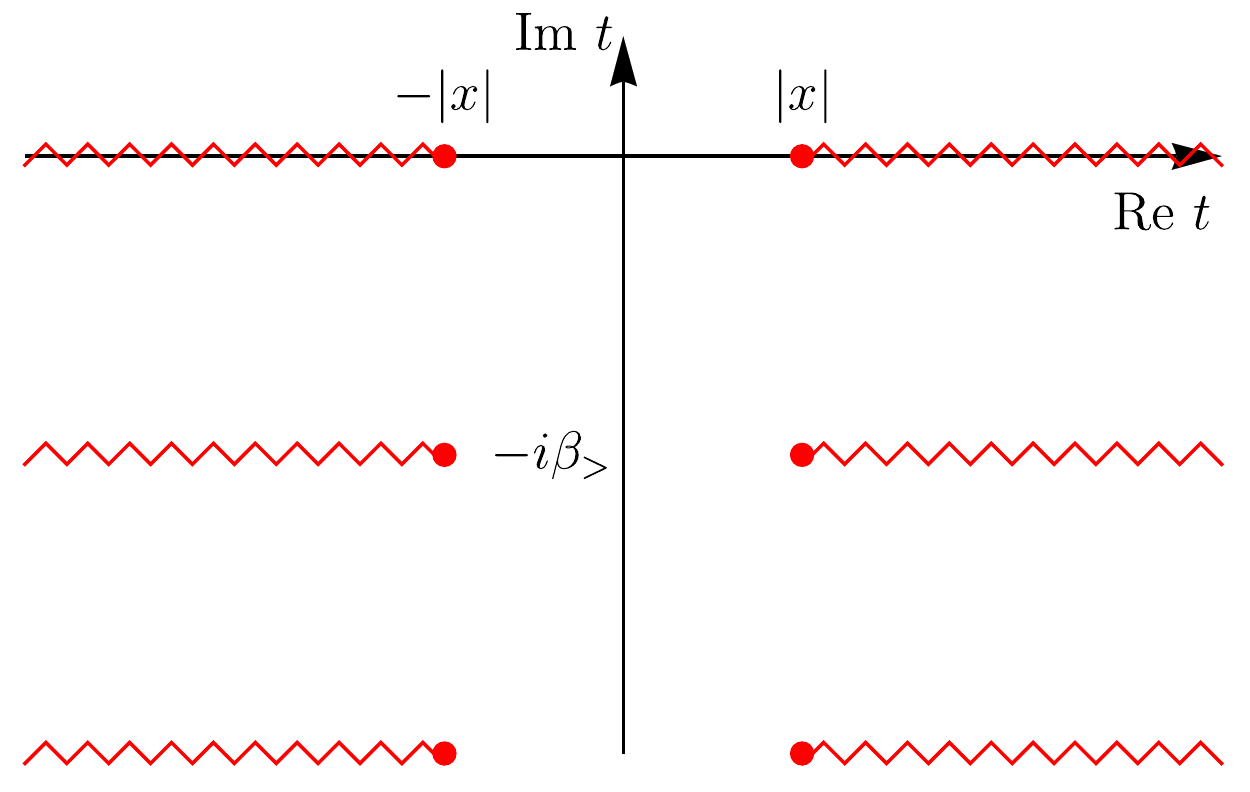}\hspace{1cm}
\includegraphics[width=0.4\textwidth]{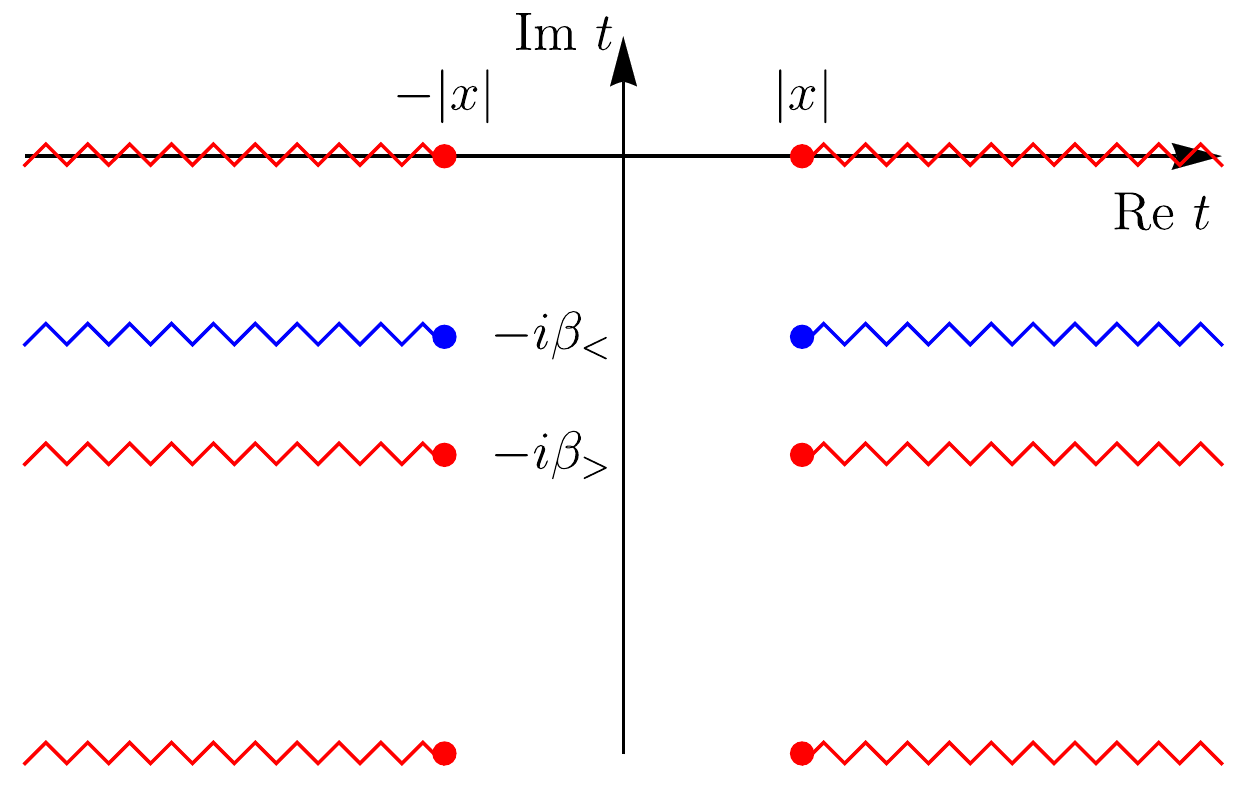}
\caption{Left: Analytic structure of the two point function imposed by unitarity and causality. The function is analytic in the strip, has light cone singularities at the red points, which are also branch points, and periodically extended using the KMS relation. Right: the inner KMS condition would give rise to a copy of the light cone singularity inside the region where the function must be analytic.
}
\label{fig:branchcuts}
\end{figure}

Alternatively assume that there is no pair of points on the boundary and the inner horizon that are space-like separated. In this case \eqref{eq:outerKMS1} and \eqref{eq:innerKMS1} are enforced on different regions of complexified time. However, a standard result\footnote{This is derived by examining where the trace converges, and analytically extending the correlator into a periodic function using the KMS relation and the edge of the wedge theorem. It also applies to the case when a chemical potential is turned on, provided the insertions cannot connect states with arbitrarily different charge, which is the case for neutral operators.} in a causal, unitary relativistic field theory is that the correlator is analytic in $t$ except for branch point singularities at $t=\pm |x|+i k \beta$, $k\in \mathbb{Z}$ (see left of Fig. \ref{fig:branchcuts}). Then, the inner KMS condition \eqref{eq:outerKMS1} implies that
\beq
g(t)=\langle V(x,t-i\beta_{<})V(0,0)^\dagger\rangle_{\beta_>\beta_<}-\langle V(x,t)V(0,0)^\dagger\rangle_{\beta_{>}\beta_{<}},
\eeq
is an analytic function (apart from the aformentioned branch points and their images) that is identically zero when $|\text{Re }t|>|x|$. By the identity theorem, $g(t)=0$ everywhere, except possibly its branch points. This shows that we also have \eqref{eq:outerKMS1} for $t<|x|$, and the argument from the previous paragraph about the double periodicity applies here too. Therefore, the evolution of a neutral field past the inner Cauchy horizon (as dictated by the CFT) cannot be single valued.

Now let us briefly discuss charged operators, in which case the conditions look like \eqref{eq:chargedoperatordoublekms}. We can again use the identity theorem to extend the inner KMS condition to the region $\text{Re }t<|x|$. We then get that imposing both of these conditions gives a quantization condition on the charge $q$ of the operator in terms of the inner and outer horizon temperatures.\footnote{This comes from requiring both $\langle V(x,t)V(0,0)^\dagger\rangle$ and $e^{-i\frac{\tilde \mu q}{\beta_{\rm inner}}t}\langle V(x,t)V(0,0)^\dagger\rangle$ to have the appropriate discrete Matsubara frequencies in Euclidean time.} This seems unnatural in a generic situation, but could arise when the charge comes from a KK reduction; in that case we think of the relations \eqref{eq:chargedoperatordoublekms} as coming from Fourier transforming non-degenerate double periodicity conditions, such as \eqref{eq:rotatingKMS}.

\subsection{No-go for rotating black holes in $D\geq 4$}

Let us first focus on the case $D=d+1=4$, where we have a single rotation parameter. To ease the formulas, we can take a large $M$ limit with $a$ fixed, so that we have a kind of $1/M$ expansion. In this case the two roots of $S=2M$ (see \eqref{eq:adskerr},\eqref{eq:horizon-roots}) are
\bea
r_+&=(2M)^{\frac{1}{3}}-\frac{1+a^2}{3}(2M)^{-\frac{1}{3}}-\frac{a^2}{6M}+ \cdots, \\
r_-&=\frac{a^2}{2M}+ \cdots,
\eea
the inverse surface gravities are
\bea
\label{eq:rotatingtemporal}
\frac{2\pi}{\kappa_+}&=\frac{2\pi}{3}(4/M)^{\frac{1}{3}} + \frac{2 a^2 \pi}{3M} + \cdots, \\
\frac{2\pi}{\kappa_-}&=\frac{2\pi a^2}{M} + \cdots,
\eea
while the ratios relevant for the angular part of the identification are
\bea
\label{eq:rotatingspatial}
\Omega_+\frac{2\pi}{\kappa_+}&= \frac{2\pi a(1-a^2)}{3M} + \cdots, \\
\Omega_-\frac{2\pi}{\kappa_-}&= \frac{2\pi a(1-a^2)}{M} + \cdots.
\eea
The largest in magnitude of these numbers is $\frac{2\pi}{\kappa_+}$, i.e. the physical inverse temperature in this large $M$ limit. We see that the angular identification degenerates for $a\rightarrow 0$. However, the identification associated to the inner horizon is not decompactifying as $a\rightarrow 0$, it is the opposite, the identification spacing becomes very small. This forces the correlation function to be constant in that particular direction, which seems pathological. More precisely, since $1/\kappa_-$ vanishes as $a^2$, while $\Omega_-/\kappa_-$ vanishes as $a$, in the static limit, the inner KMS condition turns into $a \partial_\varphi G_{\rm static} + O(a^2)=0$, i.e. the static correlator should be independent of the polar angle $\varphi$. This is impossible since due to spherical symmetry in the static case, the correlator should depend only on the geodesic distance on the spatial sphere.

It is clear how to generalize this argument for higher dimensions: we have shown in \eqref{eq:rotatingsmalla} that the inner horizon monodromies have a parametrically small period when the rotation parameters are small, and since the angular identification vanishes slower than the temporal one, this small identification points in the angular directions. 
 Therefore the inner KMS condition \eqref{eq:rotatinginnerkms} enforces the correlator to be independent of the combination  $\sum_i \varphi_i$ of the azimuthal angles in the static limit which breaks spherical symmetry and hence is a contradiction. 

Let us give a simple argument in $D=4$ for a direct contradiction with boundary unitarity. We write the boundary correlation function as
\beq
\label{eq:rotatingcorrelator}
\text{Tr}[e^{-\beta (H+\Omega J)} V(t,\varphi)V(0,0)] = \text{Tr}[e^{-\beta_- H_--\beta_+ H_+} V(t,\varphi)V(0,0)] 
\eeq
(we imagine $J$ acts by translating the angle $\varphi$ and we suppress dependence on other angular coordinates), where $H_{\pm}=\frac{H\pm J}{2}$ are positive operators due to the spinning unitarity bound $\Delta-|j|\geq d-2$ that we are applying now in radial quantization. The $\beta_{\pm}$ here are defined by the equation, $\beta_{\pm}=\beta(1 \pm \Omega)$. We can deduce that under the analytic continuation $t\mapsto t-i\tau$, $\varphi \mapsto \varphi - i \psi$, the correlator is analytic when
\beq
\label{eq:rotatinganaliticregion}
0 \leq \tau \pm \psi \leq \beta_{\pm}.
\eeq
Note that there is an obvious OPE singularity at $\tau=\psi=0$ and $|t|=|\varphi|=0$. The outer horizon identification enforces an image of this singularity at a particular other value of $\tau$ and $\psi$ determined by the respective first lines of \eqref{eq:rotatingtemporal} and \eqref{eq:rotatingspatial}, and therefore these should correspond to the $\beta_{\pm}$ in \eqref{eq:rotatinganaliticregion} that determine the region of analyticity (and the real KMS periodicity). Therefore we must have
\beq
\beta_{\pm} = \frac{2\pi}{\kappa_+}(1 \pm \Omega_+) =\frac{2\pi}{3}(4/M)^{\frac{1}{3}} + \frac{2 a^2 \pi}{3M} \pm \frac{2\pi a(1-a^2)}{3M} + \cdots
\eeq
In the $1/M$ expansion, $\beta_{\pm}=\beta + O(a/M)$. On the other hand, the inner horizon identifications (second lines of \eqref{eq:rotatingtemporal} and \eqref{eq:rotatingspatial}) will translate to
\beq
\beta^{<}_{\pm} = \frac{2\pi}{\kappa_-}(1 \pm \Omega_-) =\frac{2\pi a^2}{M} \pm \frac{2\pi a(1-a^2)}{M} \cdots
\eeq
that is, $\beta_{\pm}^{<}=O(a/M)$. The inner horizon periodicities are therefore parametrically smaller in $M$ than the outer horizon ones: $\beta_{\pm}^{<} \ll \beta_{\pm}$. When $\sqrt{5}-1<2a<\sqrt{5}+1$, $\beta^{<}_+$ and $\beta^{<}_-$ have the same sign, so they induce a copy of the OPE singularity at $\tau=\psi=0$ and $|t|=|\varphi|=0$ that is \textit{inside} the region of analyticity \eqref{eq:rotatinganaliticregion} that is enforced by boundary unitarity. Note that we could have run this argument in the charged case too; see the right of Fig. \ref{fig:branchcuts} for an illustration.

A possible problem with this argument is of course large $N$. One might argue that these forbidden singularities could develop in the infinite $N$ limit in otherwise healthy boundary correlators.\footnote{An example of this is the bulk point singularity \cite{Maldacena:2015iua}.} This scenario still seems pathological: notice that as $a\rightarrow 0$, the correlator must develop a line of singularities coming from the copies of the light cone singularity.\footnote{We imagine first taking $N\rightarrow \infty$ and $a\rightarrow 0$. A way to escape this argument is to have a situation when these to limits do not commute, but this also seems a bit pathological.} This line of singularities is not compatible with the large $N$ thermal correlator without rotation.

We have discussed the $D=4$ case explicitly in this subsection, and outlined two separate arguments which appear to lead to contradictions in the boundary theory.
For the second argument, dealing with boundary unitarity, it would be interesting to check the $D > 4$ cases explicitly, to find the precise range for the rotation parameter $a$ which is excluded.
Our arguments above excluded certain ranges of rotation parameters; so it is in principle, possible for black holes outside this range (like rapidly rotating objects) to be consistent with all of our conditions.

\subsection{Yes-go for the rotating BTZ black hole}

We have seen that the inner and outer KMS conditions are inconsistent with each other in the boundary for neutral operators in a charged black hole and rotating black holes in $d+1>3$. However, now we show that they are consistent for rotating BTZ black holes. The conditions are given in \eqref{eq:BTZouter1}, \eqref{eq:BTZouter2} for the outer and \eqref{eq:BTZinner} for the inner horizon. In terms of light cone coordinates $x^{\pm}=\varphi \pm t$ and chiral temperatures
\begin{equation}
\frac{2\pi}{\kappa_+}\equiv \beta_{>}=\frac{1}{2}(\beta_++\beta_-),\hspace{.5cm} \frac{2\pi}{\kappa_-} \equiv \beta_{<}= \frac{1}{2}(\beta_+-\beta_-),
\end{equation}
where we are assuming $\beta_+>\beta_-$, these identifications are equivalent with invariance of the two point function under
\bea
(x^- \mapsto x^- - 2i\beta_-,  x^+ \mapsto x^+) 
&\quad  \text{and} & \quad 
(x^- \mapsto x^- ,    x^+ \mapsto x^+- 2i\beta_+),
\eea
i.e. one has independent KMS conditions for left and right movers. In particular, both conditions have a non-degenerate static limit, as opposed to the higher dimensional situation in \eqref{eq:rotatingsmalla}. The boundary correlation function in rotating BTZ is given by the method of images correlation function
\beq
\label{eq:BTZmethodofimages}
\langle V(x^-,x^+)V(0,0)\rangle_{\beta_+\beta_-} \sim \sum_n \frac{1}{\left(\sinh \frac{\pi(x^-+2\pi n)}{\beta_-} \sinh \frac{\pi(x^++2\pi n)}{\beta_+} \right)^{\Delta_V}},
\eeq
which indeed satisfies these two independent KMS conditions. Note that in the limit of $\beta_{\pm} \ll x^{\pm}$, these correlation functions are universal in any 2d CFT, but the finite size correlator is theory dependent. One can check in 2d CFTs that are not holographic but the torus correlator is known, (such as minimal models) that the inner horizon KMS condition is not satisfied. It is then possible for $1/c$ corrections to the method of images propagator to break this extra symmetry.  In the bulk $1/c$ corrections would correspond to backreaction effects of probes.

\subsection{A prescription for multiboundary correlation functions}
\label{sec:multiboundary}

\begin{figure}
\centering
\includegraphics[scale=.75]{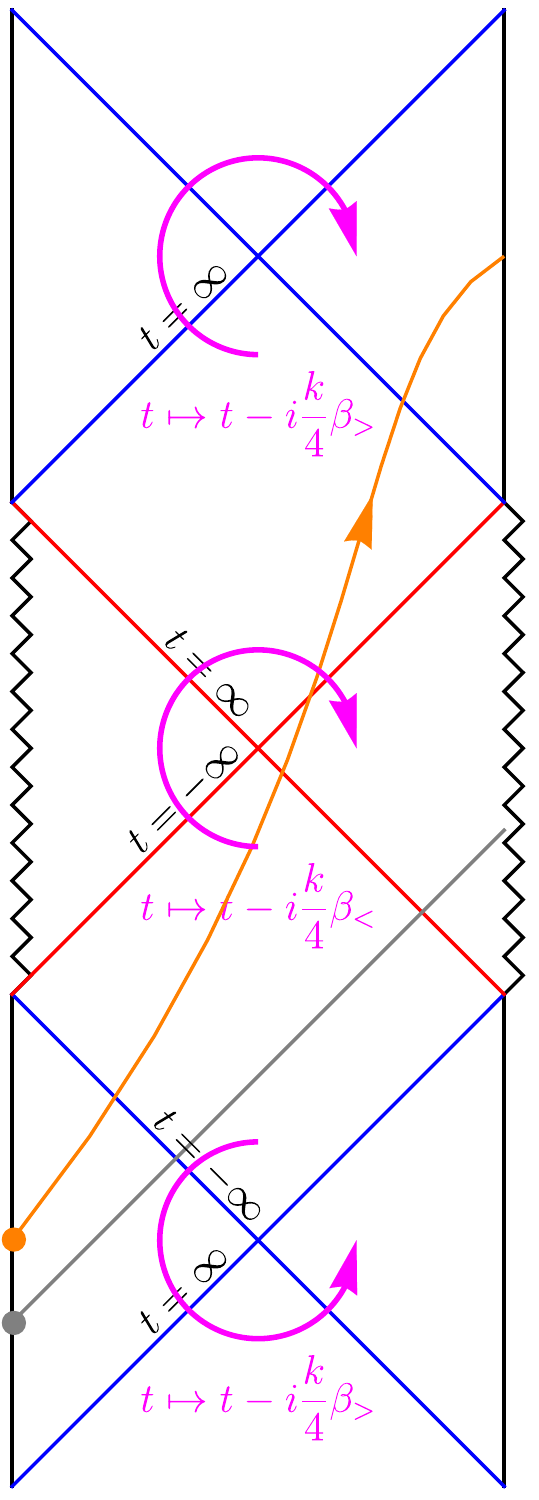}
\caption{Moving an operator between time-like separated boundary points in the analytically extended spacetime. First, we send the operator we want to move (orange dot) to the future of all the other insertions (gray dot). Now we can move in the bulk along the orange curve without crossing light cones. In terms of the Killing time, this amounts to sending $t\mapsto + i \beta_{>}/2 + i \beta_{<}/2$. When we deal with a rotating black hole, we need to be careful that we need to fix different co-rotating coordinates at the outer and inner horizons, which leads to a continuation that affects other coordinates besides $t$.
}
\label{fig:movingoperators}
\end{figure}

Next we explain our proposal for computing multiboundary correlation functions.
We have already reviewed, in our discussion of the KMS condition, how a half-KMS shift places an operator on the opposite asymptotic region of the eternal AdS wormhole.
We now wish to generalize this procedure to time-like separated boundaries that arise  in charged or rotating black holes. Note that when we say ``time-like separated boundaries", we mean that there are parts of the boundaries that are time-like separated.  In fact there are can also be points that are null separated between these boundaries.  This may seem counter-intuitive from the structure of the Penrose diagrams, as in Fig. \ref{fig:movingoperators}, but lightrays with angular momentum can make it through from a lower boundary to an upper boundary. In fact, in a charged black hole in AdS, the only light ray that falls into the singularity is the one with zero impact parameter \cite{Cruz:2011yr}, although it is not clear whether these null curves are prompt paths between the endpoints, which would make them null-separated.\footnote{A null path is considered ``prompt'' if there is no causal path between the endpoints that intersects the interior of the past light cone of the final point; see \cite{Witten:2019qhl} for a detailed exposition.} We will see that for rotating BTZ there are indeed points on boundaries of the upper and lower level parts of the Penrose diagram that are  null-separated, and others that are actually spacelike separated.\footnote{It might seem confusing that some points that appear time-like separated on the Penrose diagram are in fact space-like separated, but it is just a consequence of the need to project the diagram to two dimensions. For a simpler example, note that a space-like geodesic that stays on a single boundary, but connects points with slightly different Killing time will appear as a vertical line on these diagrams.} 

The procedure is illustrated on Fig. \ref{fig:movingoperators}. Suppose that we wish to move an operator from the lower left boundary to a point on the upper right boundary that is time-like separated from it.\footnote{There is an analogous procedure for space-like separated points.} First we need to make sure that the operator we wish to move is to the future of all other operators in the correlator in order to avoid light-cone singularities.  This is done by choosing the appropriate Lorentzian sheet of the complexified correlator when we pass around the the branch point corresponding to the light cone. Once this is done, we can follow the orange path on Fig. \ref{fig:movingoperators} without crossing light cones in the bulk. The inner horizon is reached from the interior at a second (smaller) simple zero of $F$ of \eqref{eq:generalBH}, $F(r_-)=0$. Here we have $r_*\rightarrow +\infty$ for the tortoise coordinate. Around the inner bifurcate horizon, it is also possible to pick Kruskal coordinates in a similar way to \eqref{eq:generalkruskal}, but now the temperature of the inner horizon appears. It is then easy to track the required imaginary shifts in the coordinates as in \eqref{eq:exampleshift}. In general, there could be an ambiguity in the signs of the shifts around different bifurcate horizons, but we can fix this by requiring that at the end of the path in Fig. \ref{fig:movingoperators}, there is no net shift in the tortoise coordinate. The upshot is that the orange path in Fig. \ref{fig:movingoperators} corresponds to sending
\beq
\label{eq:onelevelup}
t\mapsto t-i \frac{\beta_{>}+\beta_{<}}{2}.
\eeq
In other words, the analogue of \eqref{eq:TFDprescription} would be
\beq
\langle V_{R,\rm top}(t,x)V_{L,\rm bottom}(0,0)\rangle = \langle V_{L,\rm bottom}(t-i \frac{\beta_{>}+\beta_{<}}{2},x)V_{L,\rm bottom}(0,0)\rangle.
\eeq

Note again that the consistency of this prescriptions requires both the inner and outer KMS conditions to be satisfied by the boundary correlation functions. In practice, this means that we will only apply it (more precisely the rotating generalization of it) to the rotating BTZ black hole.

\section{Embedding Space and the Hyperboloid}\label{sec:embedding}

To prepare for our discussion of rotating BTZ inner horizon stability, we review the embedding space quotient construction.
Our discussion closely follows the seminal work of \cite{Banados:1992gq,Banados:1992wn}.
We begin with a review of AdS$_3$ before progressing to rotating BTZ, and conclude with a brief discussion of some unintuitive aspects of the causal structure of rotating BTZ which may not be obvious from the Penrose diagram.

\subsection{AdS$_3$}
The construction of global AdS$_3$ begins by studying a hyperboloid in the embedding space $\mathbb{R}^{2,2}$, with metric
\begin{equation}
ds^2 = -dT_1^2 - dT_2^2 + dX_1^2 + dX_2^2.
\label{eq:embedding-metric}
\end{equation}
The hyperboloid embedding equation is
\begin{equation}
-T_1^2 - T_2^2 + X_1^2 + X_2^2 = -\ell_{\text{AdS}}^2 ,
\label{eq:hyperboloid}
\end{equation}
and from now on we will set $\ell_{\text{AdS}} = 1$.
The induced metric on the hyperboloid \eqref{eq:hyperboloid} is obtained by pulling back \eqref{eq:embedding-metric}.
A set of coordinates which covers all of AdS$_3$ is $(t, r, \theta)$ with $t,\theta \in [0,2\pi)$ and $r \in [0, \infty)$, with the embedding space coordinate maps given by
\begin{equation}
\begin{alignedat}{2}
T_1 & = \sqrt{r^2 + 1} \cos t, \hspace{.5cm} & T_2 & = \sqrt{r^2 + 1} \sin t , \\
X_1 & = r \cos \theta , \hspace{.5cm} & X_2 & = r \sin \theta . 
\end{alignedat}
\label{eq:global-ads-maps}
\end{equation}
These coordinates naturally satisfy \eqref{eq:hyperboloid}, and the pullback of \eqref{eq:embedding-metric} via \eqref{eq:global-ads-maps} is
\begin{equation}
ds^2  =  -(r^2+1) dt^2 + \frac{dr^2}{r^2+1} + r^2 d\theta^2 .
\label{eq:global-ads-metric}
\end{equation}
This is almost the metric we are familiar with, but the time coordinate is periodic.
This periodicity implies the existence of closed timelike curves, for example $\gamma(s) = (s,r_0, \theta_0)$ has both $\gamma(s+2\pi) = \gamma(s)$ and tangent vector $\dot{\gamma} = (1,0,0)$, which in \eqref{eq:global-ads-metric} has strictly negative constant norm $\dot{\gamma}^2 = -(r_0^2+1)$.
To avoid this issue, we usually pass to the universal covering space $\widehat{\text{AdS}}_3$ where we unwrap $t \in S^1$ to $t \in \mathbb{R}$.
In what follows, by AdS$_3$ we will always mean the universal covering space, and we will refer to \eqref{eq:hyperboloid} simply as the hyperboloid.
Before proceeding, we discuss the isometries of \eqref{eq:global-ads-metric}.
The hyperboloid \eqref{eq:hyperboloid} is invariant under the rotation and boost symmetries of \eqref{eq:embedding-metric}, so its isometry group is $SO(2,2)$.
In the embedding coordinates $X^A \equiv (T_1,T_2,X_1,X_2)^A$, the 6 Killing vectors $V_{AB}$ of \eqref{eq:global-ads-metric} are
\begin{equation}
V_{AB} = X_B \partial_A - X_A \partial_B.
\end{equation}
Note that $X_A = (-T_1,-T_2,X_1,X_2)^A$ since we must apply \eqref{eq:embedding-metric} to $X^A$.
These can be pulled back to the hyperboloid using \eqref{eq:global-ads-maps} to obtain Killing vectors of \eqref{eq:global-ads-metric}.

\subsection{Rotating BTZ}

\begin{figure}[ht]
\begin{minipage}[c]{0.45\linewidth}
\centering
\vspace{0pt}
\includegraphics[scale=1]{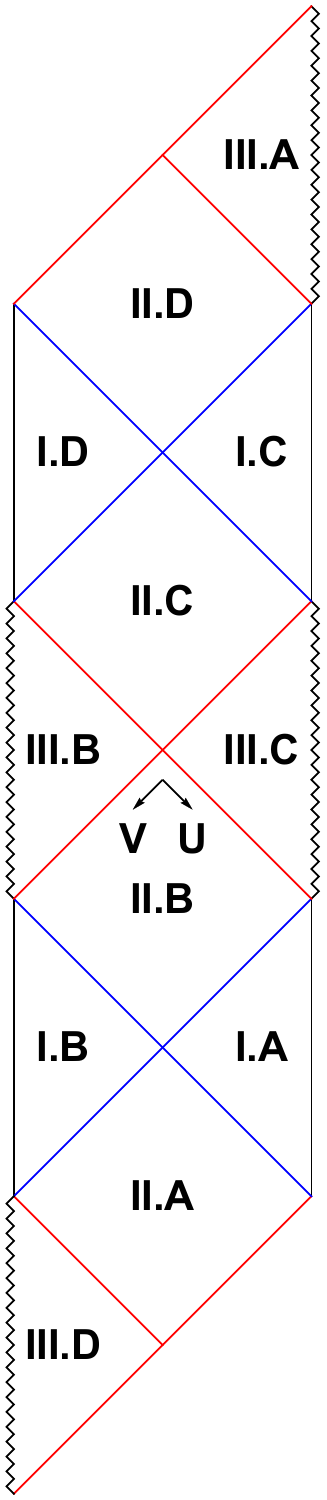}
\end{minipage}
\begin{minipage}[c]{0.45\linewidth}
\centering
\vspace{0pt}
\begin{tabular}{l*{6}{c}r}
Patch              & $X_1^+$ & $X_1^-$ & $X_2^+$ & $X_2^-$ \\
\hline
I.A  & $+$ & $+$ & $+$ & $-$ \\
II.A  & $-$ & $+$ & $+$ & $-$ \\
III.A  & $-$ & $+$ & $+$ & $+$ \\
I.B  & $-$ & $-$ & $+$ & $-$ \\
II.B  & $+$ & $-$ & $+$ & $-$ \\
III.B & $+$ & $-$  & $-$ & $-$ \\
I.C  & $+$ & $+$ & $-$ & $+$ \\
II.C & $+$ & $-$ & $-$ & $+$ \\
III.C  & $+$ & $-$ & $+$ & $+$ \\
I.D  & $-$ & $-$ & $-$ & $+$ \\
II.D  & $-$ & $+$ & $-$ & $+$ \\
III.D   & $-$ & $+$ & $-$ & $-$\\
\end{tabular}
\end{minipage}
\caption{Left: a Penrose diagram.
If the top and bottom of the diagram are identified, it represents the hyperboloid after excising the region $\xi^2 < 0$ and compactifying the angle $\phi$.
If instead the diagram is repeated infinitely in either direction, it represents the rotating BTZ black hole spacetime, the universal cover of the excised and compactified hyperboloid.
Blue edges are outer horizons and red edges are inner horizons.
Right: the signs of the lightcone coordinates in each patch.
An embedding coordinate map for any region can be constructed by putting appropriate signs on and swapping hyperbolic functions in the maps for regions I.A, II.A, or III.A which are given in \eqref{eq:region-1}-\eqref{eq:region-3}.}
\label{fig:penrose}
\end{figure}

The rotating BTZ spacetime is famously obtained from AdS$_3$ by a discrete identification of the geometry.
This identification can be formulated entirely in embedding space.
A one-parameter subgroup of isometries (associated with some Killing vector $\xi$) acts on embedding space points as $X \to e^{s \xi} X$.
A discrete subgroup is then a choice of both a Killing vector $\xi$ along with a set of equally spaced values of $s$, say $s = 2\pi n$ for $n \in \mathbb{Z}$.
Quotienting AdS$_3$ by a discrete subgroup of this form compactifies (with period $2\pi$) the integral curves of the Killing vector $\xi$, so one must ensure the resulting closed curves are neither timelike nor null.
The metric is still well-defined since flows along integral curves of a Killing vector preserve the metric by definition.
A necessary condition for preventing closed timelike curves is for $\xi$ to be everywhere spacelike, $\xi^2 > 0$.\footnote{
For the identification which gives the rotating BTZ black hole, this condition turns out to be sufficient.
It is not sufficient in general since we can take a one-parameter subgroup generated by, for example, a Killing vector which generates spacelike but not achronal curves.
One encounters a similar consideration in properly defining the notion of a Cauchy hypersurface.
}
The appropriate Killing vector to choose to pass from AdS$_3$ to the rotating BTZ black hole is, in a particular embedding frame,
\begin{equation}
\begin{split}
\xi & \equiv r_+ V_{T_2X_2} - r_- V_{T_1X_1} \\
& = - r_- X_1 \partial_{T_1} + r_+ X_2 \partial_{T_2} - r_- T_1 \partial_{X_1} + r_+ T_2\partial_{X_2} .
\end{split}
\label{eq:killing-quotient}
\end{equation}
We will not prove that the identification $X \sim e^{2\pi n \xi} X$ does what we claim, and instead refer the interested reader to \cite{Banados:1992wn}.\footnote{
The choice of Killing vector in \cite{Banados:1992wn} differs from our \eqref{eq:killing-quotient} by exchanging the pairs $(T_1,X_1) \leftrightarrow (T_2,X_2)$.
Notice that, up to this point, all expressions were symmetric under such interchange.
The choice of Killing vector breaks the symmetry between these pairs of coordinates, and this manifests in the asymmetric form of the Schwarzschild embedding patches which we give in \eqref{eq:region-1}-\eqref{eq:region-3}.
}
Notice that, despite the previous discussion, the norm of $\xi$ in \eqref{eq:embedding-metric} is
\begin{equation}
\xi^2 = r_+^2 (T_2^2 - X_2^2) + r_-^2 (T_1^2 - X_1^2) ,
\label{eq:excision-norm}
\end{equation}
so there are indeed regions of the hyperboloid where $\xi^2 \leq 0$.
In particular, if we use \eqref{eq:hyperboloid} to simplify, we find $\xi^2 \leq 0$ is equivalent to
\begin{equation}
T_2^2 - X_2^2  \leq \frac{-r_-^2}{r_+^2-r_-^2} .
\label{eq:excision}
\end{equation}
Since we have chosen to quotient by a discrete subgroup generated by a Killing vector with $\xi^2 \leq 0$ in some areas, the correct thing to do is to truncate the global AdS$_3$ spacetime at the hypersurface $\xi^2 = 0$.
We will call this hypersurface the excision surface; it is defined on the hyperboloid as well as its universal cover AdS$_3$.
We now define three types of regions in AdS$_3$:
\begin{equation}
\begin{split}
\text{I:} \hspace{1cm} & 1 < T_2^2 - X_2^2, \\
\text{II:} \hspace{1cm}& 0 < T_2^2 - X_2^2 < 1 , \\
\text{III:} \hspace{1cm} & -\frac{r_-^2}{r_+^2-r_-^2} < T_2^2 - X_2^2 < 0 .
\end{split}
\end{equation}
These regions are separated by null hypersurfaces $T_2^2 - X_2^2 = 1$ and $T_2^2 - X_2^2 = 0$ which will become the outer and inner horizons at $r = r_+$ and $r = r_-$, respectively.
Notice the norm of $\xi$ is precisely $\xi^2 = r_+^2$ and $\xi^2 = r_-^2$ at these surfaces.
Having excised the region $\xi^2 \leq 0$ from AdS$_3$, we now define three coordinate patches to cover regions of type I, II, and III.
These three types of regions can be infinitely tiled to cover the portion of AdS$_3$ which obeys $\xi^2 > 0$.
Before passing to the universal cover, the portion of the hyperboloid obeying $\xi^2 > 0$ can be covered by four patches of each type of region, for a total of twelve patches.
A set of connected patches (one of each type) is given by
\begin{enumerate}
\item[I.A]: $r_+ < r$
\begin{equation}
\begin{alignedat}{2}
T_1 & = \sqrt{\frac{r^2-r_+^2}{r_+^2-r_-^2}} \sinh \left( r_+ t - r_-\varphi \right) , \hspace{1cm}  & T_2 & = \sqrt{\frac{r^2-r_-^2}{r_+^2-r_-^2}} \cosh \left( r_+\varphi - r_- t \right)  , \\
X_1 & = \sqrt{\frac{r^2-r_+^2}{r_+^2-r_-^2}} \cosh \left( r_+ t - r_-\varphi \right) , \hspace{1cm}  & X_2 & = \sqrt{\frac{r^2-r_-^2}{r_+^2-r_-^2}} \sinh \left( r_+\varphi - r_- t \right)  .
\end{alignedat}
\label{eq:region-1}
\end{equation}

\item[II.A]: $r_- < r < r_+$
\begin{equation}
\begin{alignedat}{2}
T_1 & = -\sqrt{\frac{r_+^2-r^2}{r_+^2-r_-^2}} \cosh \left( r_+ t - r_-\varphi \right) , \hspace{1cm}  & T_2 & = \sqrt{\frac{r^2-r_-^2}{r_+^2-r_-^2}} \cosh \left( r_+\varphi - r_- t \right) , \\
X_1 & = -\sqrt{\frac{r_+^2-r^2}{r_+^2-r_-^2}} \sinh \left( r_+ t - r_-\varphi \right) , & X_2 & = \sqrt{\frac{r^2-r_-^2}{r_+^2-r_-^2}} \sinh \left( r_+\varphi - r_- t \right) .
\end{alignedat}
\label{eq:region-2}
\end{equation}

\item[III.A]: $0 < r < r_-$
\begin{equation}
\begin{alignedat}{2}
T_1 & = -\sqrt{\frac{r_+^2-r^2}{r_+^2-r_-^2}} \cosh \left( r_+ t - r_-\varphi \right) , \hspace{1cm}  & T_2 & = \sqrt{\frac{r_-^2-r^2}{r_+^2-r_-^2}} \sinh \left(r_+\varphi - r_- t \right) , \\
X_1 & = -\sqrt{\frac{r_+^2-r^2}{r_+^2-r_-^2}} \sinh \left( r_+ t - r_-\varphi \right) , & X_2 & = \sqrt{\frac{r_-^2-r^2}{r_+^2-r_-^2}}  \cosh \left(r_+\varphi - r_- t \right) .
\end{alignedat}
\label{eq:region-3}
\end{equation}
\end{enumerate}
These patches have $t,\varphi \in (-\infty, \infty)$ and are particularly convenient for imposing the discrete identification generated by \eqref{eq:killing-quotient}, because the pullback of \eqref{eq:killing-quotient} to the hyperboloid via any of \eqref{eq:region-1}-\eqref{eq:region-3} is simply
\begin{equation}
\xi = \partial_\varphi .
\end{equation}
Therefore, the effect of the identification is to compactify $\varphi \in [0,2\pi)$.
Notice that the only differences  between \eqref{eq:region-1}-\eqref{eq:region-3} are in the overall sign of a coordinate and whether the coordinate involves hyperbolic sine or cosine; the radial prefactors are fixed by the region type and the hyperbolic function arguments are fixed regardless of region.
This information can be stored in a single set of signs, the signs of the lightcone coordinates
\begin{equation}
\label{eq:embeddinglightconecoord}
X_1^\pm \equiv X_1 \pm T_1 , \hspace{1cm} X_2^\pm \equiv X_2 \pm T_2 ,
\end{equation}
which are fixed in each patch.
The induced metric in any of the three patch types is
\begin{equation}
ds^2 = \frac{-(r^2-r_+^2)(r^2-r_-^2)}{r^2} dt^2 + \frac{r^2 dr^2}{(r^2-r_+^2)(r^2-r_-^2)} + r^2 \left( d\varphi - \frac{r_+ r_-}{r^2} dt \right)^2 .
\label{eq:global-rbtz-metric}
\end{equation}
The mass $M$, angular momentum $J$, and inverse chiral temperatures $\beta_\pm$ are given by
\begin{equation}
\label{eq:charges}
    M = \frac{r_+^2+r_-^2}{8}, \hspace{.5cm} J = \frac{r_+r_-}{4}, \hspace{.5cm} \beta_\pm = \frac{2\pi}{r_+\mp r_-} ,
\end{equation}
where $\beta_{\pm}$ couples to $x^\pm = \varphi \pm t$ and we picked the direction of rotation so that $\beta_+>\beta_-$.
As emphasized previously, $\beta_\pm$ are \textit{not} the inverse horizon temperatures, but rather correspond to the chiral temperatures in the dual CFT state.
The Penrose diagram for the rotating BTZ spacetime (and the hyperboloid) is given in Fig.~\ref{fig:penrose}.
We also define the following co-rotating angle
\begin{equation}
\label{eq:btzcorotating}
\phi \equiv \varphi - \frac{r_-}{r_+} t .
\end{equation}
 For all regions of type I in the rotating BTZ spacetime, the co-rotating embedding coordinates are given by 
\begin{enumerate}
\item[I.A]:
\begin{equation}
\begin{alignedat}{2}
T_1 & = \sqrt{\frac{r^2-r_+^2}{r_+^2-r_-^2}} \sinh \left( \kappa t - r_-\phi \right) , \hspace{1cm}  & T_2 & = \sqrt{\frac{r^2-r_-^2}{r_+^2-r_-^2}} \cosh \left( r_+\phi \right)  , \\
X_1 & = \sqrt{\frac{r^2-r_+^2}{r_+^2-r_-^2}} \cosh \left(\kappa t - r_-\phi \right) , \hspace{1cm}  & X_2 & = \sqrt{\frac{r^2-r_-^2}{r_+^2-r_-^2}} \sinh \left( r_+\phi \right)  .
\end{alignedat}
\label{eq:region-1A-corotate}
\end{equation}

\item[I.B]:
\begin{equation}
\begin{alignedat}{2}
T_1 & = -\sqrt{\frac{r^2-r_+^2}{r_+^2-r_-^2}} \sinh \left( \kappa t - r_-\phi \right) , \hspace{1cm}  & T_2 & = \sqrt{\frac{r^2-r_-^2}{r_+^2-r_-^2}} \cosh \left( r_+\phi \right)  , \\
X_1 & = -\sqrt{\frac{r^2-r_+^2}{r_+^2-r_-^2}} \cosh \left(\kappa t - r_-\phi \right) , \hspace{1cm}  & X_2 & = \sqrt{\frac{r^2-r_-^2}{r_+^2-r_-^2}} \sinh \left( r_+\phi \right)  .
\end{alignedat}
\label{eq:region-1B-corotate}
\end{equation}

\item[I.C]:
\begin{equation}
\begin{alignedat}{2}
T_1 & = \sqrt{\frac{r^2-r_+^2}{r_+^2-r_-^2}} \sinh \left( \kappa t - r_-\phi \right) , \hspace{1cm}  & T_2 & = -\sqrt{\frac{r^2-r_-^2}{r_+^2-r_-^2}} \cosh \left( r_+\phi \right)  , \\
X_1 & = \sqrt{\frac{r^2-r_+^2}{r_+^2-r_-^2}} \cosh \left(\kappa t - r_-\phi \right) , \hspace{1cm}  & X_2 & = -\sqrt{\frac{r^2-r_-^2}{r_+^2-r_-^2}} \sinh \left( r_+\phi \right)  .
\end{alignedat}
\label{eq:region-1C-corotate}
\end{equation}

\item[I.D]:
\begin{equation}
\begin{alignedat}{2}
T_1 & = -\sqrt{\frac{r^2-r_+^2}{r_+^2-r_-^2}} \sinh \left( \kappa t - r_-\phi \right) , \hspace{1cm}  & T_2 & = -\sqrt{\frac{r^2-r_-^2}{r_+^2-r_-^2}} \cosh \left( r_+\phi \right)  , \\
X_1 & = -\sqrt{\frac{r^2-r_+^2}{r_+^2-r_-^2}} \cosh \left(\kappa t - r_-\phi \right) , \hspace{1cm}  & X_2 & = -\sqrt{\frac{r^2-r_-^2}{r_+^2-r_-^2}} \sinh \left( r_+\phi \right)  .
\end{alignedat}
\label{eq:region-1D-corotate}
\end{equation}
\end{enumerate}
In all of these patches we have $t \in (-\infty,\infty)$, $r \in (r_+,\infty)$, and $\phi \in [0,2\pi)$, and we have defined the outer surface gravity
\begin{equation}
\kappa \equiv \kappa_+ = \frac{r_+^2-r_-^2}{r_+} .
\end{equation}
The metric in co-rotating coordinates is
\begin{equation}
ds^2 = \frac{-(r^2-r_+^2)(r^2-r_-^2)}{r^2} dt^2 + \frac{r^2 dr^2}{(r^2-r_+^2)(r^2-r_-^2)} + r^2 \left( d\phi + \frac{r_-(r^2-r_+^2)}{r_+ r^2} dt \right)^2 .
\label{eq:global-rbtz-metric-corotate}
\end{equation}
We will also need a Kruskal patch, though not the usual one which covers two asymptotic boundary regions.
The usual ``outer" Kruskal patch is given in embedding space by
\begin{equation}
\begin{alignedat}{2}
T_1 & = \frac{V+U}{1+UV}\cosh(r_-\phi) - \frac{V-U}{1+UV}\sinh(r_-\phi) , & T_2 & = \frac{1-UV}{1+UV}\cosh(r_+\phi)  , \\
X_1 & = \frac{V-U}{1+UV}\cosh(r_-\phi) - \frac{V+U}{1+UV}\sinh(r_-\phi) , \hspace{1cm}  & X_2 & = \frac{1-UV}{1+UV}\sinh(r_+\phi)  .
\end{alignedat}
\label{eq:outer-kruskal-embedding}
\end{equation}
This patch covers four wedges of the Penrose diagram, including regions I.A and I.B.
However, in this paper we will be concerned with the inner horizon, and it is more useful to have an ``inner" Kruskal patch which includes a set of four wedges that touch the timelike singularity.
This inner Kruskal patch is given in embedding space by
\begin{equation}
\begin{alignedat}{2}
T_1 & = \frac{1-UV}{1+UV}\cosh(r_-\phi) , \hspace{1cm}  & T_2 & = \frac{V+U}{1+U V} \cosh(r_+\phi) - \frac{V-U}{1+UV}\sinh(r_+\phi)  , \\
X_1 & = \frac{UV-1}{1+UV} \sinh(r_-\phi) , \hspace{1cm}  & X_2 & = \frac{V+U}{1+UV}\sinh(r_+\phi) - \frac{V-U}{1+UV}\cosh(r_+\phi)  .
\end{alignedat}
\label{eq:inner-kruskal-embedding}
\end{equation}
Notice that the inner Kruskal patch is constructed by simply exchanging the 1 and 2 subscripts, along with some minus signs to orient the patch correctly.
The only other necessary change is switching $r_+ \leftrightarrow r_-$; the need for this will become clear shortly.
This patch covers regions II.B, II.C, III.B, and III.C, and the positive U-V axes are oriented as shown in Fig.~\ref{fig:penrose}.
The inner Kruskal patch metric is\footnote{Note that the inner co-rotating angle, which appears in the inner Kruskal metric, differs from the outer one by the same $r_+ \leftrightarrow r_-$ interchange.}
\begin{equation}
ds^2 = \frac{1}{(1+UV)^2} \bigl[ -4dU dV + 4r_+(V dU - U dV) d\phi + (4r_+^2 UV + r_-^2(1-UV)^2) d\phi^2 \bigr] .
\label{eq:inner-kruskal-metric}
\end{equation}
It is now clear why we needed to exchange $r_+ \leftrightarrow r_-$ with respect to the outer Kruskal patch; the radius of the fibered circle should be $r_-$ on the horizon at $UV = 0$ in this patch, and $r_+$ at the boundaries $UV = 1$.
The $UV$ coordinate range is also more restricted compared to the outer Kruskal patch; we have
\begin{equation}
1 > UV > 1 + \frac{2 r_+^2}{r_-^2} \left( \sqrt{1- \frac{r_-^2}{r_+^2}} - 1 \right) .
\end{equation}
At the lower end of the coordinate range for $UV$ we encounter the excision surface $\xi^2 = 0$.
This is a causal singularity of the geometry; the metric is regular, the curvature does not diverge, and there is no conical singularity either.

\subsection{Causal structure}
Before proceeding, we will demonstrate that there exist both timelike and spacelike separated points on different levels of the Penrose diagram for rotating BTZ.
This may be counterintuitive because the Penrose diagram in Fig.~\ref{fig:penrose} seems to imply that all points beyond the inner bifurcate horizon are timelike separated from the asymptotic boundaries below it.
The resolution is related to the fact that we have really been drawing projection diagrams.
To gain some intuition, consider the simpler case of a Lorentzian cylinder with flat metric $-dt^2+d\phi^2$; we can imagine this cylinder as essentially the conformal boundary which appears in AdS$_3$.
The points $(0,0)$ and $(\epsilon, \pi)$ are spacelike separated for small enough $\epsilon$, but if we project out the fibered circle then $(\epsilon,\pi)$ is projected to $(\epsilon,0)$ which is timelike separated from the origin.
A more sophisticated version of this effect occurs in the full rotating BTZ geometry, as we will now show.

We will have two warmups, involving boundaries on the same level and then boundaries on different levels but the same side of the diagram, before moving to the interesting case of different levels and opposite sides.
For all examples we consider the inner product in embedding space between points $(t_1,r_\infty,\phi_1)$ and $(t_2,r_\infty,\phi_2)$ at large $r_\infty$ and drop subleading pieces.
This amounts to computing a distance between boundary points $P$.

\begin{figure}[ht]
\begin{minipage}[c]{0.45\linewidth}
\centering
\vspace{0pt}
\includegraphics[width=0.9\textwidth]{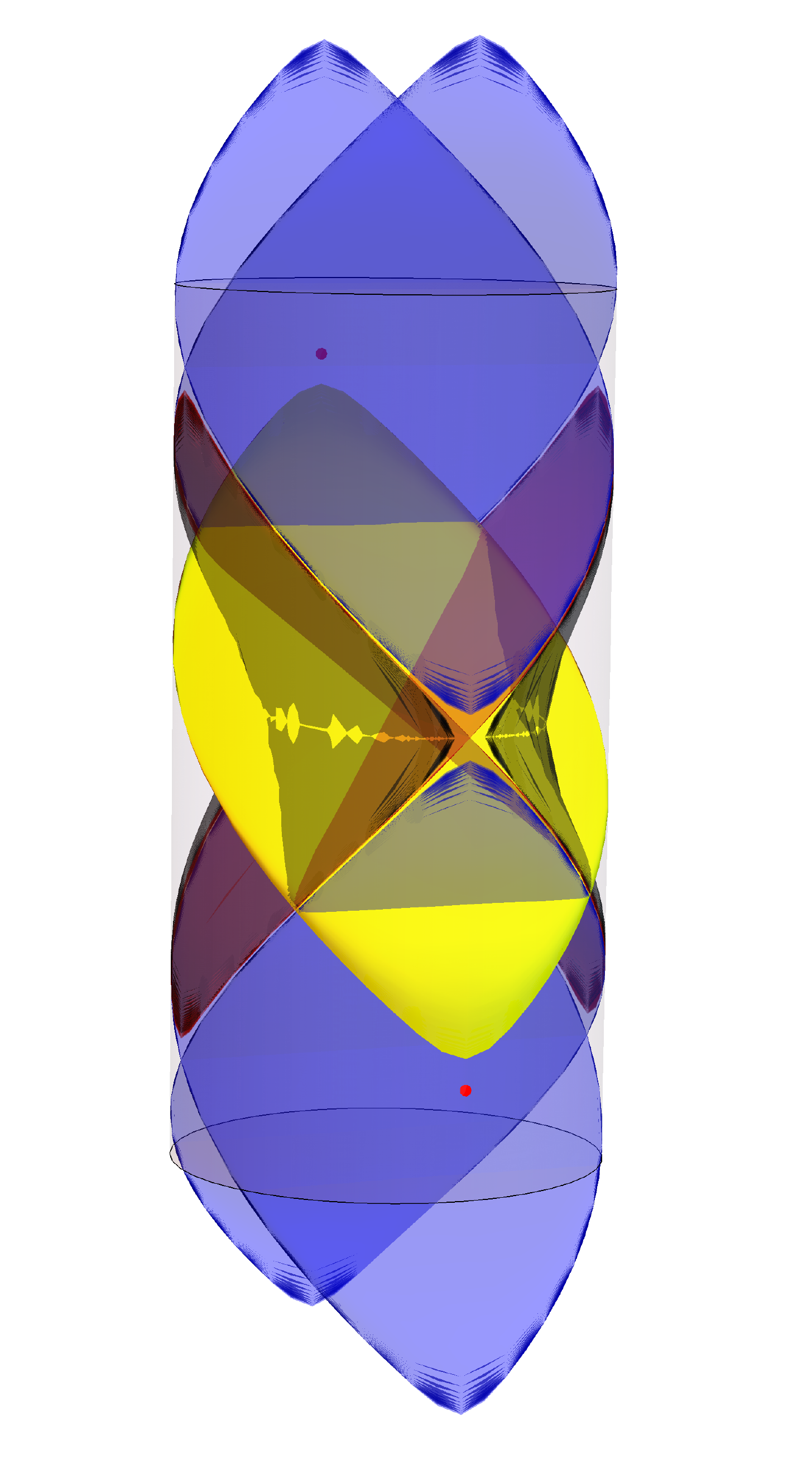}
\end{minipage}
\begin{minipage}[c]{0.45\linewidth}
\centering
\vspace{0pt}
\includegraphics[width=0.9\textwidth]{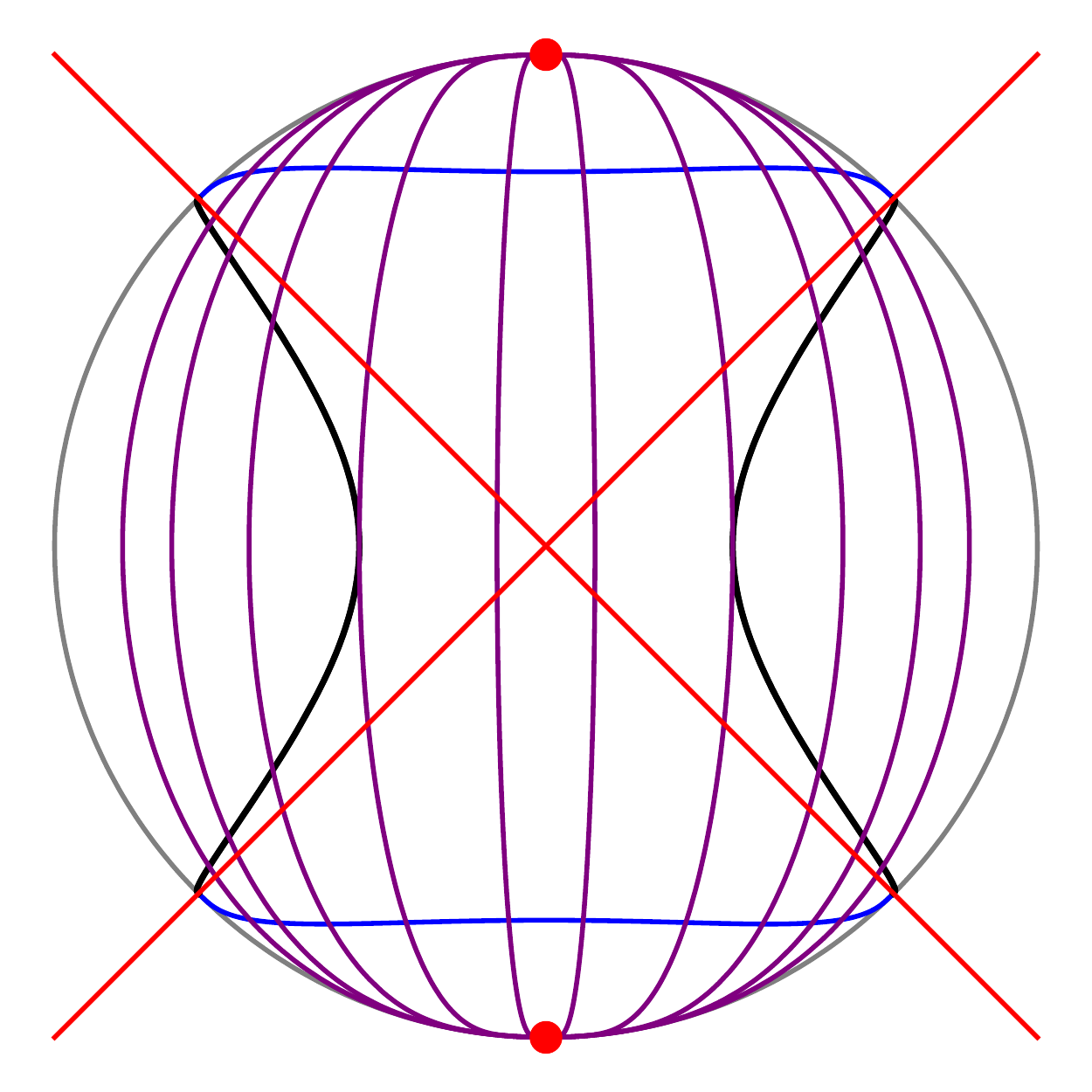}
\end{minipage}
\caption{Left: Features of the rotating BTZ geometry represented in global AdS coordinates of \eqref{eq:global-ads-maps}. A point in I.B and its ``image" on I.C are shown with red dots. The bulk lightcone connecting them is the yellow surface. The excision surface is shown in black, the outer horizons in blue, the inner horizons in red. One may recognize the Penrose diagram, but as we proceed upwards, we need to twist it. Right: Various features as they intersect the light sheet from the left figure. The purple lines are the actual null geodesics that make up the light sheet and one sees that many of them do not fall behind the excision surface.
}
\label{fig:bulklightcones}
\end{figure}

Consider regions I.A and I.B.
We do not expect any timelike separated points, as these boundaries are on the same level.
The inner product is
\begin{equation}
    -P_{\text{I.A}} P_{\text{I.B}} = \frac{r_\infty^2}{r_+^2-r_-^2} \left( \cosh(\kappa \Delta t - r_-\Delta \phi) + \cosh (r_+ \Delta \phi) \right) ,
\end{equation}
where we have defined $\Delta t \equiv t_1 - t_2$ and $\Delta \phi \equiv \phi_1 - \phi_2$.
Since region I.C is related to I.B by an overall minus sign in embedding coordinates, the result for I.A and I.C is simply negative of our previous result.
\begin{equation}
    -P_{\text{I.A}} P_{\text{I.C}} = -\frac{r_\infty^2}{r_+^2-r_-^2} \left( \cosh(\kappa \Delta t - r_-\Delta \phi) + \cosh (r_+ \Delta \phi) \right).
\end{equation}
Clearly the first result is strictly positive and greater than zero, as a sum of hyperbolic cosines.
The second result is similarly purely negative.\footnote{One might wonder whether we can trust the embedding space distance after all the modifications (excision, quotient) we have made to global AdS.  In fact, we can, as demonstrated in Fig.~\ref{fig:bulklightcones}.}
One can probably guess what the result will be for the mixed case.
It is
\begin{equation}
\label{eq:difflevnullsep0}
    -P_{\text{I.B}} P_{\text{I.C}} = \frac{r_\infty^2}{r_+^2-r_-^2} \left( \cosh(\kappa \Delta t - r_- \Delta\phi) - \cosh(r_+ \Delta\phi) \right) .
\end{equation}
This implies points on the boundaries of I.B and I.C are null separated when\footnote{We are in co-rotating coordinates, which is why this expression looks different from the usual null relation $\Delta t = \Delta \varphi$.}
\begin{equation}
\label{eq:difflevnullsep1}
    \Delta \phi = \left( 1 - \frac{r_-}{r_+} \right) \Delta t .
\end{equation}
Notice that there is only a small range of $\Delta t$ for which this is possible, since the angular coordinate is bounded.
However, it is clearly possible to obtain timelike separated points by taking (say) $\Delta t =0$ and $\Delta \phi \neq 0$.
Likewise, examples of spacelike separated points can be obtained with $\Delta \phi = 0$ and $\Delta t \neq 0$.
This is essentially the negation of how points on the \textit{same} boundary are related.
If we consider two points on the boundary of I.B which are timelike (spacelike) separated, and then send one of the points to its ``image" point on I.C (by image point we mean the point with the same coordinate values), the separation becomes spacelike (timelike).
This behavior is related to the fact that the forward lightcone of a point on the boundary of I.B becomes the backward lightcone of its image point on I.C.

We can visualize the casual structure better by plotting various features of the rotating geometry in global AdS coordinates of \eqref{eq:global-ads-maps}, see Fig. \ref{fig:bulklightcones}. On this figure, the radial coordinate is compactified with the tanh function. The yellow light sheet goes from the projective boundary coordinate $Q=(1,0,0,1)$ in I.B to its image $-Q$ in I.C. It is given by the equation $X\cdot Q=0$ where $X$ is an embedding space vector parametrized as \eqref{eq:global-ads-maps}. In terms of light cone embedding coordinates \eqref{eq:embeddinglightconecoord}, the singularity is given by $X_1^+ X_1^-=r_-^2/(r_+^2-r_-^2)$, the outer horizon is $X_2^-=0$ and the inner horizon is $X_1^-=0$. We can recognize on the left of Fig. \ref{fig:bulklightcones} the structure of the Penrose diagram from Fig. \ref{fig:penrose}, but we need to twist it as we proceed upwards.

\section{Geodesics in the Shockwave Background}\label{sec:geodesics}

As discussed in the introduction, there have been many classical analyses of inner horizon stability.
In particular, a phenomenon called mass inflation occurs, where even a mildly perturbed black hole mass function diverges at the inner horizon \cite{Poisson:1990eh}.
Here we will focus on a more coarse measure of stability to gain intuition about whether an impassable singularity forms due to highly blueshifted infalling matter.
As a precursor to our quantum gravity analysis of stability in Sec.~\ref{sec:otoc}, we wish to study whether one can send signals between levels of the Penrose diagram after a perturbation has been applied to the state.
A small perturbation applied from one boundary at late times will be blueshifted, and we model this blueshift effect as a backreacting shockwave traveling along the inner horizon.
In this section we analyze boundary-anchored null geodesics in rotating BTZ and demonstrate that there are null geodesics which travel between levels of the Penrose diagram even in the presence of such a shockwave.
Of course, these techniques will not prove or disprove singularity formation, but one signature of such physics is whether any null geodesics at all can traverse the inner horizon region.
If none can make it through, we may expect that the perturbing shockwave effectively ``closes up" the inner horizon region to all observers, and this  effect manifests as a redirecting of null geodesics into the singularity \cite{Marolf:2011dj}.
We will see, however, that this does not happen for rotating BTZ.

We can compute null geodesics which run from region I.A to region I.D through the inner Kruskal region \eqref{eq:inner-kruskal-embedding}.
Within the outer horizon, the metric can be written as a gluing of two rotating BTZ black holes with a shift in the $V$ coordinate across the shockwave, which we take to lie on the $U=0$ surface.
We imagine this shockwave is sourced by some highly boosted local object, so we include a transverse profile function $\alpha(\phi)$ which we do not specify.\footnote{
We will specify this function, along with the stress tensor for the solution, in our calculation of the eikonal phase in Sec.~\ref{sec:otoc}.
}
\begin{equation}
\begin{split}
ds^2 & = \frac{1}{[1+u (v+\alpha(\phi)\Theta(u))]^2} \biggl( -4 du (dv + \alpha'(\phi) \Theta(u)d\phi) + 4 r_+ (v + \alpha(\phi) \Theta(u)) dud\phi \\
& - 4r_+ u (dv + \alpha'(\phi) \Theta(u) d\phi) d\phi + (4r_+^2 u (v+\alpha(\phi)\Theta(u)) + r_-^2[1-u(v+\alpha(\phi)\Theta(u))]^2) d\phi^2 \biggr) ,
\end{split}
\label{eq:shockwave-discontinuous}
\end{equation}
where $\Theta$ is the Heaviside step function.
We have made the null coordinates lowercase as we will reserve the capital versions for perturbed metrics of the form $ds^2 \to ds^2 + 4 \alpha(\phi) \delta(U) dU^2$ whose coordinates are continuous across the shockwave. 
We address metrics of this type in Sec.~\ref{sec:otoc}; these are equivalent to metrics like \eqref{eq:shockwave-discontinuous} by a coordinate change $U = u$, $V = v+\alpha(\phi)\Theta(u)$.

\begin{figure}
\centering
\includegraphics[scale=.75]{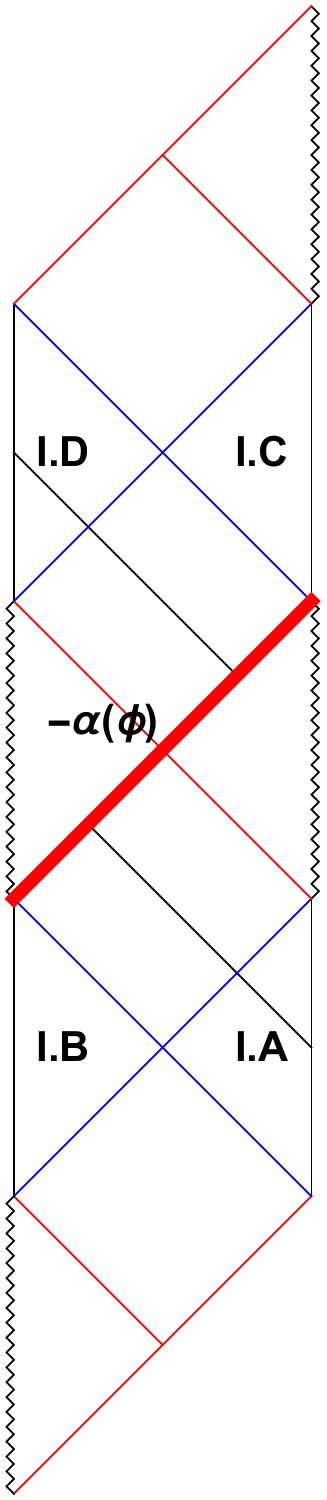}
\caption{The thick red line on $u=0$ in the inner Kruskal region is a shockwave.
The black line from region I.A is a null geodesic which encounters the shockwave at an angle $\phi$, receives a kick of size $\alpha(\phi)$, and continues through to region I.D.
If $\alpha$ is negative, the kick is directed upward on the diagram due to the orientation of the Kruskal axes.
}
\label{fig:penrose-shock}
\end{figure}

The metric \eqref{eq:shockwave-discontinuous} is simply the inner Kruskal metric \eqref{eq:inner-kruskal-metric} where we understand the coordinate $v$ jumps to $v + \alpha(\phi)$ when crossing the $u=0$ surface.
The relevant portion of the Penrose diagram is shown in Fig.~\ref{fig:penrose-shock}.
Null geodesics can be understood directly in embedding space by extremizing the constrained functional
\begin{equation}
\mathcal{L} = \dot{X}^2 + \lambda (X^2 + 1) ,
\end{equation}
where $\lambda$ is a Lagrange multiplier.
In this language, null geodesics are given by solutions to the Euler-Lagrange equations with $\lambda = 0$, which are 
\begin{equation}
X(s) = Q s + P.
\end{equation}
The coefficients are constrained to obey $Q^2 = 0$, $P^2 = -1$, and $QP = 0$.\footnote{
$Q^2 = 0$ arises because null geodesics obey $\dot{X}^2 = 0$, $P^2 = -1$ because we must have $X(0)^2 = -1$ on the hyperboloid at $s=0$, and $QP = 0$ since we must actually have $X(s)^2 = -1$ for any $s$.
}
We want to start the null geodesic in region I.A and follow it across the shock to see if it reaches region I.D or falls into the timelike singularity.
As boundary conditions, we take $s=0$ at the shock to make sure that the geodesic crosses the shock somewhere.
As usual, the conformal boundary of asymptotic AdS$_3$ (sometimes called the projective null cone) is reached by going to large radial distance and then removing the diverging factor.
In our language, the magnitude of the affine parameter $|s|$ itself is a proxy for radial distance, and indeed in type I regions we have $|s| \approx r$ at large $r \gg r_+$.
So, the past boundary point is determined by $\lim_{s\to -\infty} X(s)/|s|$ and the future boundary point is similarly $\lim_{s \to \infty} X(s)/|s|$. 
Therefore these points are $-Q$ and $Q$ respectively.
A boundary point is therefore parametrized by evaluating  \eqref{eq:region-1A-corotate} at large radius and then removing the diverging radial prefactor:
\begin{equation}
-Q(t,\phi') = \biggl( \sinh(\kappa t-r_-\phi'),\ \cosh(r_+\phi'),\ \cosh(\kappa t - r_-\phi'),\ \sinh(r_+\phi') \biggr).
\end{equation}
As required, we see $Q^2(t,\phi') = 0$.
Points on the shockwave are parametrized by the inner Kruskal embedding coordinates at $u = 0$, which gives
\begin{equation}
P(v,\phi) = \biggl( \cosh(r_-\phi),\ e^{-r_+\phi} v,\ -\sinh(r_-\phi),\ -e^{-r_+\phi} v  \biggr).
\end{equation}
Again as required, we find $P^2(v,\phi) = -1$.
Finally we must enforce $Q(t,\phi') P(v,\phi) = 0$ before the shockwave (in regions II.B and III.C).
This yields one parameter in terms of the other three, namely
\begin{equation}
v_B(t,\phi',\phi) = - e^{r_+(\phi - \phi')} \sinh ( \kappa t + r_- (\phi-\phi')) .
\end{equation}
We can plug this back into the expression for $P(v,\phi)$ and find a reduced form
\begin{equation}
P_B(t,\phi',\phi) = P(v_B(t,\phi',\phi), \phi) .
\end{equation}
Since the metric \eqref{eq:shockwave-discontinuous} is identical on both sides of $u=0$ except for a shift in $v$, the total effect of the shockwave on the geodesic is a kick, or a shift in the $v$ coordinate. The prescription in embedding space is therefore to send $P(v,\phi) \to P(v-\alpha(\phi), \phi)$.
This means that above the shockwave, we must require $P(v-\alpha(\phi),\phi) Q(t,\phi') = 0$, which yields a different relation for $v$:
\begin{equation}
v_A(t,\phi',\phi) = v_B(t,\phi',\phi) + \alpha(\phi).
\end{equation}
The reduced form for $P$ above the shock is therefore
\begin{equation}
P_A(t,\phi',\phi) = P(v_A(t,\phi',\phi), \phi) .
\end{equation}
In summary, the total piecewise null geodesic with initial conditions $t,\phi',\phi$ is given by\footnote{
The conditions are simply the initial boundary point $(t,\phi')$ and the angle $\phi$ at which the geodesic intersects the $u=0$ surface.
}
\begin{equation}
X_{t,\phi',\phi}(s) = Q(t,\phi') s + \left[ \Theta(-s) P_B(t,\phi',\phi) + \Theta(s) P_A(t,\phi',\phi) \right] ,
\end{equation}
where the affine parameter takes values $s \in \mathbb{R}$.

Now,   question we wish to  answer is whether there exists a finite $s_0 > 0$ such that $X_{t,\phi',\phi}(s_0)$ is on the excision surface.
If there does not exist such an $s_0$, the geodesic must avoid the singularity and make it through unscathed to region I.D.
The singularity is given in embedding space by the vanishing of the norm \eqref{eq:excision-norm} of $\xi$.
This vanishing norm defines the singularity hypersurface in embedding space, and we can check whether a geodesic intersects this hypersurface by computing the particular combination \eqref{eq:excision-norm} of its components.
This combination can be computed by contracting the geodesic path with a  tensor $E_{MN}$ whose entries are
\begin{equation}
E = \text{diag}(r_-^2,r_+^2,-r_-^2,-r_+^2) .
\end{equation}
Therefore, the equation that we must solve is
\begin{equation}
E_{AB} X_{t,\phi',\phi}^A(s_0) X_{t,\phi',\phi}^B(s_0) = 0, 
\end{equation}
for any $s_0 > 0$.
This reduces to
\begin{equation}
(r_+^2-r_-^2)s_0^2 + \left( 2 (r_+^2-r_-^2) \sinh(\kappa t + r_-(\phi-\phi')) - 2 r_+^2 \alpha(\phi) e^{-r_+(\phi-\phi')} \right) s_0 + r_-^2 = 0 .
\end{equation}
By understanding the roots of this quadratic equation we can determine whether or not there is a positive real solution.

In order to have at least one positive real root, we can simply require the larger root to be positive and real.
Since the quadratic and constant coefficients are both strictly positive, the only way for this to occur is
\begin{equation}
 (r_+^2 - r_-^2) \sinh (\kappa t + r_-(\phi - \phi')) - r_+^2 \alpha(\phi) e^{-r_+(\phi-\phi')} \leq - r_-\sqrt{r_+^2-r_-^2}  .
\label{eq:condition-singularity}
\end{equation}
This is the necessary and sufficient condition on the parameters $t,\phi',\phi$ which, if true, says the geodesic $X_{t,\phi',\phi}$ falls into the timelike singularity after getting kicked by the shockwave.

Notice that for late times $t$ in\eqref{eq:condition-singularity} it will be almost impossible to satisfy this condition because the sinh becomes very large on the left hand side.  Therefore, there will always be at least some null geodesics which are able to travel between different levels of the Penrose diagram.  In other words, the shock wave does not redirect all geodesics into the singularity, meaning that cosmic censorship is violated at this level.
In addition, for very slow rotation $r_- \ll r_+$, there is a wider range of $t$ which satisfies the condition since the right hand side is becoming very slightly negative and the left hand side remains dominated by $r_+^2 \sinh \kappa t$.
So, more geodesics are falling into the singularity, which is in accord with our intuitions about the singularity closing up and becoming spacelike in the static case.

Note that, the shock profile must be \textit{negative}, $\alpha(\phi) < 0$, in order for the kick to send the geodesic away from the singularity and not toward it. 
Indeed, such an effect is necessary to for the shockwave to not violate causality, and it shouldn't if we assume the matter sourcing the shockwave obeys the null energy condition.
Of course, the form of $\alpha(\phi)$ must be determined by Einstein's equation for the metric \eqref{eq:shockwave-discontinuous}.
It can be shown that such a metric implies a delta function stress energy tensor, and $\alpha(\phi)$ is then determined by a differential equation with delta function source.
In the static and outer horizon cases \cite{Shenker:2014cwa,Jahnke:2019gxr}, $\alpha(\phi)$ was found to be a positive function.
In our analysis of the eikonal phase in Sec.~\ref{sec:otoc}, we will see that the opposite is true for inner horizon shockwaves, where $\alpha(\phi)$ is in fact negative. These sign differences just come from the fact that the inner Kruskal coordinates point downwards on the Penrose diagram in Fig. \ref{fig:penrose}.

\section{Stability and Inner Horizon Shockwaves}\label{sec:otoc}

In this section we explain why the OTOC is relevant for probing stability of the inner horizon of a rotating BTZ black hole.
We then calculate a four sided correlator in the boosted black brane using  CFT methods, and our prescription from Sec.~\ref{sec:multiboundary}. This correlator turns out to be a second sheet correlator from a single sided point of view, that is the same as the one used to compute the OTOC in \cite{Roberts:2014ifa}. 
This result suggests that the inner horizon region remains stable to perturbations even when one includes the backreaction as a shockwave.\footnote{In some sense this is not surprising for the boosted black brane, since it is just empty AdS written in different coordinates.}
We repeat this calculation in the rotating BTZ bulk to study finite size effects, employing the elastic eikonal approximation of \cite{Shenker:2014cwa}, and using the method of images propagator to move particles into the scattering region. This corresponds to a particular choice of boundary condition at the singularity that obeys the inner KMS condition.\footnote{For heavy enough operators, we expect that the result is the same for any boundary condition that preserves the property that the two point function localizes to the geodesic.} We find similar results as for the boosted black brane.
Finally, we show the finite size OTOC obtained from the bulk calculation reduces to our boosted black brane OTOC upon taking a decompactification limit.

\subsection{Chaos, stability, and the boosted black brane}\label{sec:bndyotoc}

In this subsection, we will restrict to the case when both the horizon and the boundary are planar; so bulk geometry will describe the boosted black brane. This is because in this case the boundary calculation is tractable. We can think about the boosted black brane as the high temperature limit of the rotating black hole.

The rough intuition for why chaos and stability are related is as follows. Let us consider a two point function between different levels of the Penrose diagram, obtained via the procedure outlined in Fig. \ref{fig:movingoperators}. To leading order in $G_N$, this two point function just probes the background geometry. We can then ask what happens with this two point function when we slightly perturb the background.
To this end, let us consider the four-point function 
\begin{equation}
\label{eq:boostedthermal}
    \tr \left[ e^{-\beta_- L_0 - \beta_+ \bar{L}_0} W^\dagger V(x^+,x^-) V(0) W \right]  .
\end{equation}
We interpret this correlator as a two-point function of probe operators $V$ in a boosted thermal background perturbed by $W$ operators.
The positions of the perturbing $W$ operators are not so important; only the final Lorentzian operator ordering will be crucial, namely, we begin with a \textit{time ordered} correlator on a single boundary. 
In a two-sided purification $\ket{\beta_+\beta_-}$ (the boosted thermofield double), we can write the four point function \eqref{eq:boostedthermal} as
\begin{equation}
\label{eq:perturbed2pt1}
    \bra{\beta_+\beta_-} W^\dagger V(x^+,x^-) V(0) W \ket{\beta_+\beta_-} .
\end{equation}
The state $\ket{\beta_+\beta_-}$ in $\mathcal{H}_{\text{CFT}} \otimes \mathcal{H}_{\text{CFT}}$ is dual to a two-sided boosted black brane geometry in the bulk, the planar version of the two-sided rotating BTZ black hole studied in Sec.~\ref{sec:embedding} and \ref{sec:geodesics}.
Unlike the one-sided case, the two-sided boosted black brane has a relative boost between the two sides, and so has both an outer and inner temperature scale just like the finite-size geometry. We interpret \eqref{eq:perturbed2pt1} as a two point function in the perturbed state $W \ket{\beta_+\beta_-}$.

By a conformal transformation, correlators such as \eqref{eq:perturbed2pt1} map to vacuum expectation values on the plane
\begin{equation}
    z = e^{\frac{2\pi}{\beta_-}x^-}, \hspace{.5cm} \bar{z} = e^{\frac{2\pi}{\beta_+} x^+},
\end{equation}
where $x^\pm = x \pm t$. We pick imaginary parts for times $t_i\rightarrow t_i-i\epsilon_i$ such that we have the ordering in \eqref{eq:perturbed2pt1}, that is, the correlator is time ordered. Then we follow the prescription of Sec.~\ref{sec:multiboundary} and move one of the $V$ operator into the bulk, via the path illustrated on Fig. \ref{fig:movingoperators}. Now we do expect to cross a light cone in the bulk, and we wish keep track of this by just following the boundary coordinates of the operator.\footnote{We can imagine representing the bulk operator via an HKLL smearing kernel, which would correspond to ``thickening" the lines on certain parts of the boundary trajectory of the operator. This should be possible without issues for the black brane since it is just AdS in disguise.} On Fig. \ref{fig:movingoperators}, we first cross the outer horizon at $t\rightarrow -\infty$. Here, $z$ goes to infinity while $\bar z$ goes to zero. When we cross this horizon, we shift $x^\pm \rightarrow x^{\pm}\mp i \beta_{\pm}/4$, as explained before. Therefore, we make quarter turns in opposite directions in the $z$ and $\bar z$ plane. Then we go to the inner horizon at $t\rightarrow \infty$. Now $z$ goes to the origin and $\bar z$ goes to infinity, and the direction of the turns is the same for $z$ and $\bar z$ because crossing the inner horizon requires $x^\pm \rightarrow x^{\pm}- i \beta_{\pm}/4$. We go on to cross the inner horizon again, but at $t\rightarrow -\infty$, and then the outer horizon at $t\rightarrow -\infty$ again. 

\begin{figure}
\centering
\includegraphics[width=0.3\textwidth]{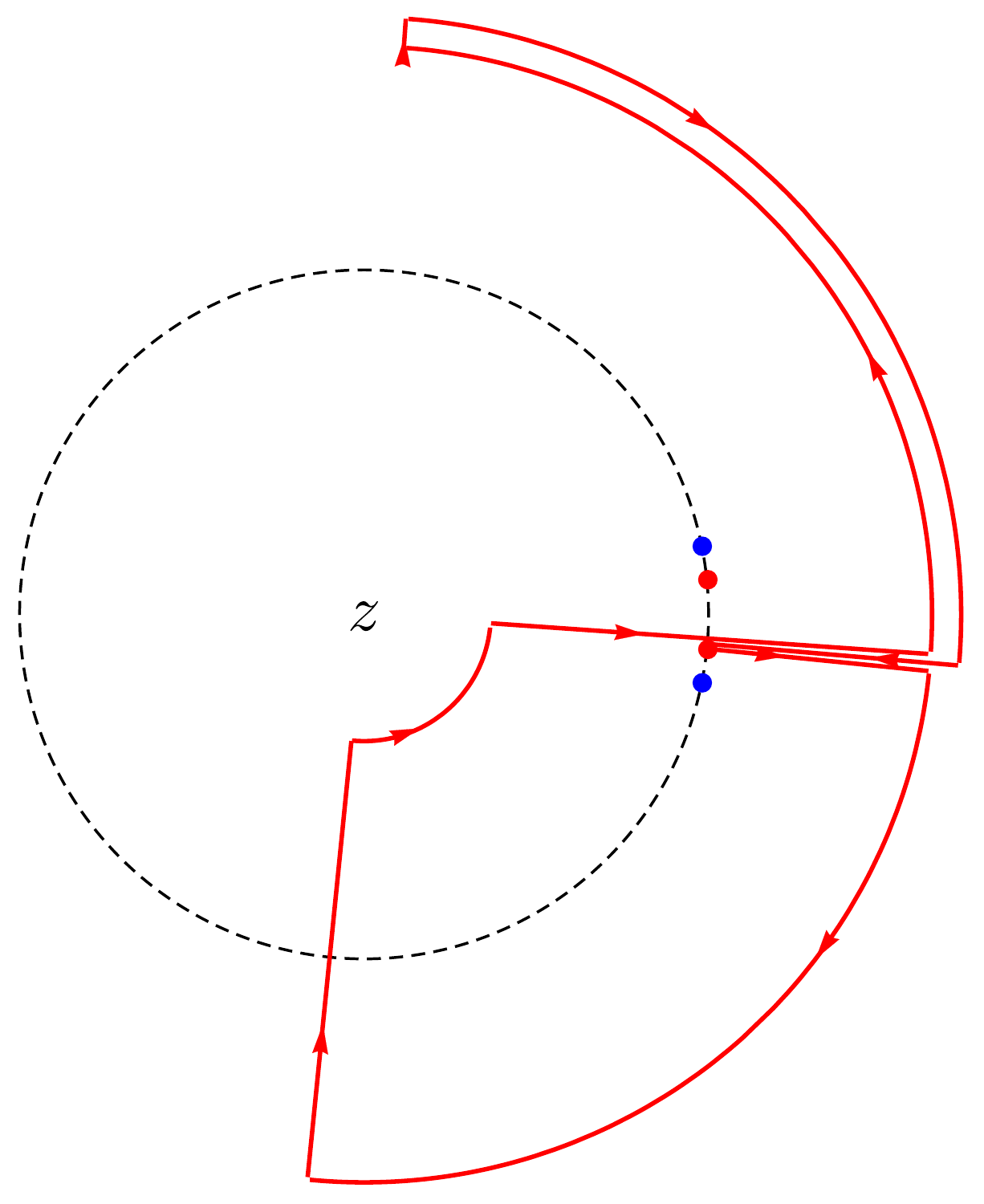}\hspace{1cm}
\includegraphics[width=0.3\textwidth]{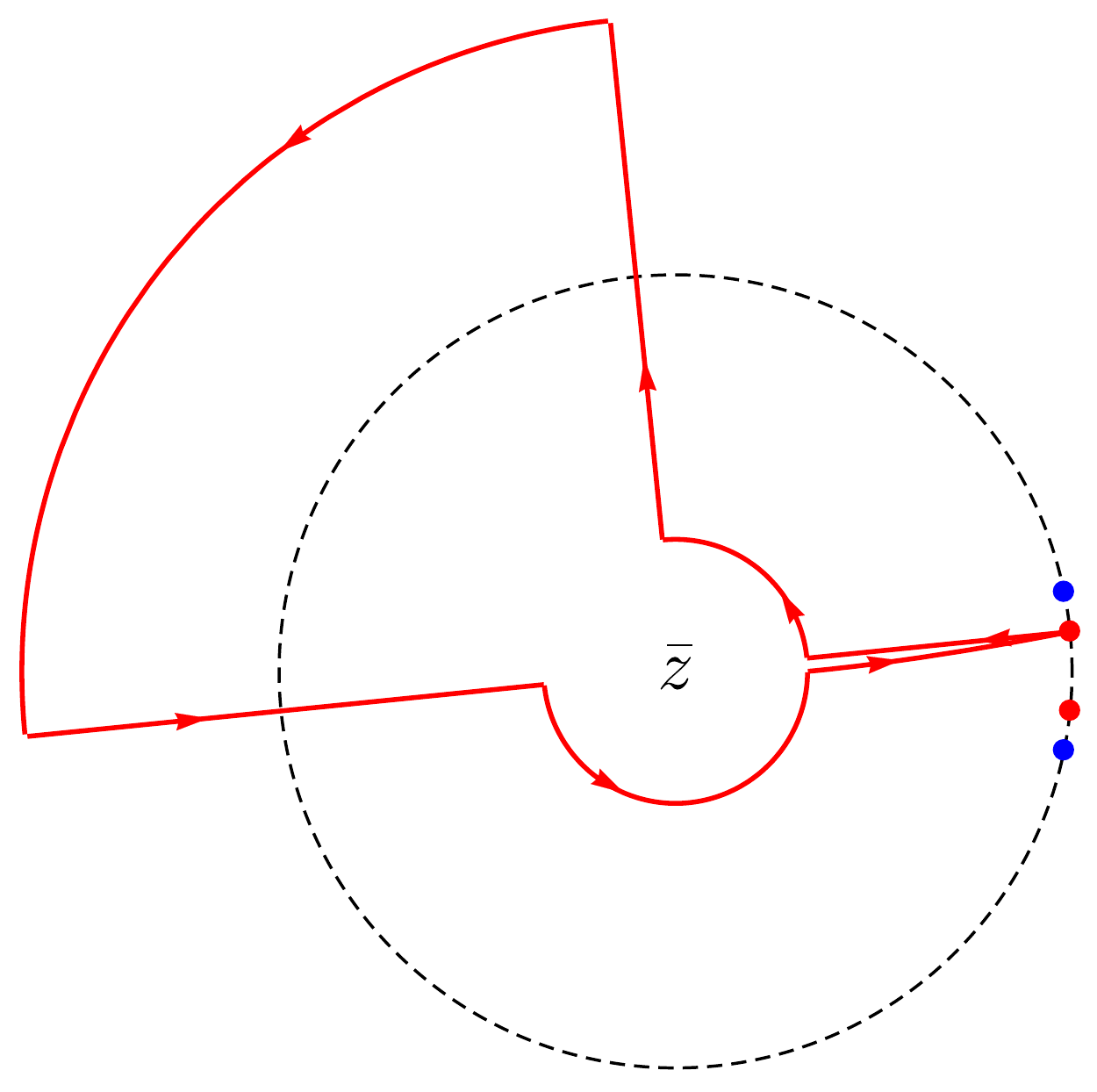}
\caption{We map the boundary coordinates to the plane and follow the trajectory of the red operator (red contour) as we send it along the orange path in Fig. \ref{fig:movingoperators}. The large arcs represent infinity, the small circle arcs represent zero, while the dashed circle is the unit circle. Left is the $z$ and right is the $\bar z$ coordinate. The former encircles a blue operator, the latter does not.
}
\label{fig:continuations}
\end{figure}

It is important during this to keep track of the vorticity of the KMS monodromies, and the fact that they flip between the two outer horizons as in Fig. \ref{fig:movingoperators}. On the particular choice of orange curve in Fig. \ref{fig:movingoperators}, we always cross the horizons such that we are going against the vorticities. The contour that the $V$ operator follows on the $z,\bar z$ plane is illustrated on Fig. \ref{fig:continuations}. The initial configuration is time ordered, with the blue dots representing $W$ and the red dots representing $V$ operators. Whenever the perturbing blue operators are inserted in a Euclidean reflection symmetric configuration (such that the interpretation of \eqref{eq:perturbed2pt1} as a two point function in a perturbed state is valid), the $V$ operator (red dot) encircles one blue operator for exactly one of the two cross ratios $z$ and $\bar z$. This amounts to crossing to the second sheet of the plane correlator, and the continuation is the same as the one required to compute the OTOC \cite{Roberts:2014ifa}. We can therefore interpret this second sheet as a correlation function between two different levels as on Fig. \ref{fig:movingoperators} in a perturbed black brane geometry.

Before discussing the possible implications of this observation for the stability of the BTZ inner horizon, let us first briefly describe how to turn \eqref{eq:perturbed2pt1} into a four sided correlator, that can be interpreted in the bulk as a high energy scattering experiment near the inner horizon. There could be different ways to continue into a four sided correlator, but in order to describe the same physics as above, we use the following precription. We first move the last insertion (i.e. $W^\dagger$ in \eqref{eq:perturbed2pt1}) from the bottom left boundary to the top right one. We can do this without crossing light cones, and we just need to follow the contours in Fig. \ref{fig:movingoperators}, but now for the blue operators. It is clear that in this procedure, none of the red operators get encircled in either $z$ or $\bar z$ so we end up with a trivial operation. After this, we do a half outer KMS shift on both $W$ operators that send them to the opposite side on the same level. This amounts to rotating the blue dots to the other half of the unit circle compared to Fig. \ref{fig:continuations}. Now we have a three sided correlator with two $V$ operators on the bottom left and no operators on the top right. We then send one $V$ operator to the top right using the contour on Fig. \ref{fig:continuations} to reach a four sided configuration similar to the one on Fig. \ref{fig:penrose-operators}. Now the $W$ insertion gets encircled in $\bar z$ instead of $z$, but the net effect is the same, we end up with the OTOC.

Now let us examine the possible implications for the stability of the inner horizon under perturbations. First note that time ordered correlators factorize to leading order in $G_N$. The effect we are looking for in a four sided correlator probing stability is a kinematical enhancement of $G_N$ corrections, suggesting that backreaction is becoming important. This is precisely what the second sheet correlator gives us; in the OTOC language, the $G_N$ corrections undergo a Lyapunov growth, and we need to start resumming them around the scrambling time. More precisely, in terms of the plane cross ratios
\beq
\chi = \frac{z_{12}z_{34}}{z_{13}z_{24}}, \quad \bar \chi = \frac{\bar z_{12}\bar z_{34}}{\bar z_{13}\bar z_{24}},
\eeq
the OTOC continuation corresponds to $1-\chi \rightarrow e^{-2\pi i}(1-\chi)$ while keeping $\bar \chi$ fixed (or vice versa). On this second sheet, $1/c$ corrections get enhanced in the OPE-like limit $\chi \rightarrow 0$, $\chi/\bar \chi=\text{fixed}$.\footnote{This is called the Regge limit of the flat space four point function.} In a vacuum block approximation, the normalized correlator in this limit looks like \cite{Roberts:2014ifa}
\beq
\label{eq:virasoroblockotoc}
\frac{\langle W(z_2,\bar z_2)V(z_1,\bar z_1)V(z_3,\bar z_3)W(z_4,\bar z_4) \rangle}{\langle W(z_2,\bar z_2)W(z_4,\bar z_4) \rangle \langle V(z_1,\bar z_1)V(z_3,\bar z_3)\rangle} \sim \left( \frac{1}{1-\frac{12\pi i \Delta_W}{c \chi}}\right)^{\Delta_V},
\eeq
and we see that large $c$ factorization fails when $\chi \sim 1/c$. For our black brane setup, this kinematic regime corresponds to either sending the perturbing operators to late times with fixed co-rotating coordinates (such that they create a shockwave localized on the inner horizon) or sending the probe $V$ operator on the upper level to the location where bulk light rays coming from the lower left $V$ operator go, i.e. when they approach the configuration of red dots in Fig. \ref{fig:bulklightcones}.\footnote{This corresponds to the pair of points $\Delta \phi=0$ and $\Delta t=0$ in \eqref{eq:difflevnullsep0}, \eqref{eq:difflevnullsep1}, or the ``image point".} Away from these limits, the geometry as seen by the probe $V$ operators seems unaffected by the perturbation in the large $c$ limit.

\subsection{Bulk eikonal scattering}

Having obtained the black brane answer, we now turn to the question of finite-size effects.
The global rotating BTZ geometry is dual to a CFT state on the circle, but direct CFT calculations are not under control there.
So, we employ the elastic eikonal bulk scattering interpretation of the OTOC developed in \cite{Shenker:2014cwa}. This calculation was generalized for the outer horizon of rotating BTZ in \cite{Jahnke:2019gxr} and here we present a version of this for the inner horizon.
The CFT correlator that we are interested in is
\begin{equation}
\frac{\avg{V_{\text{I.D}}(t_1,\phi_1) W_{\text{I.C}}(t_2,\phi_2) V_{\text{I.A}}(t_3,\phi_1) W_{\text{I.B}}(t_4,\phi_2)}_{\beta_+ \beta_-}}{\avg{V(t_1,\phi_1)V(t_3,\phi_1)}_{\beta_+\beta_-} \avg{W(t_2,\phi_2)W(t_4,\phi_2)}_{\beta_+\beta_-} }  ,
\end{equation}
where we have labeled the numerator operators with the asymptotic region on Fig.~\ref{fig:penrose} that we wish to place them on.
We will compute this quantity in the bulk by interpreting pairs of operators as creating 2-particle in and out states on a background state $\ket{\beta_+ \beta_-}$.
Recall that we defined $\ket{\beta_+ \beta_-}$ as the two-sided purification of the thermal charged ensemble $e^{-\beta_- L_0 - \beta_+ \bar{L}_0}$.
Since the particles created by the $V$ and $W$ operators are highly boosted by the time they reach the inner bifurcation surface, a convenient basis for the states is one of definite momentum in the respective null directions and definite position in the transverse directions.
Thus the 2-particle states are given in a tensor product Hilbert space, with individual basis element $\ket{p,\phi}$, with normalization
\begin{equation}
\inner{p,\phi}{q,\phi'} \propto p\ \delta(p-q)\delta(\phi-\phi') .
\label{eq:ket-norm}
\end{equation}
The in and out states are
\begin{equation}
\begin{split}
V(t_3,\phi_1)W(t_4,\phi_2) \ket{\beta_+\beta_-} = \int d\phi_3' d\phi_4' dp_3^U dp_4^V \psi_3(p_3^U,\phi_3') \psi_4(p_4^V,\phi_4') \ket{p_3^U,\phi_3'} \otimes \ket{p_4^V,\phi_4'} , \\
W(t_2,\phi_2)^\dagger V(t_1,\phi_1)^\dagger \ket{\beta_+\beta_-} = \int d\phi_1' d\phi_2' dp_1^U dp_2^V \psi_1(p_1^U,\phi_1') \psi_2(p_2^V,\phi_2') \ket{p_1^U,\phi_1'} \otimes \ket{p_2^V,\phi_2'} ,
\end{split}
\end{equation}
where we have labeled the null momenta with their direction.
The wavefunctions $\psi$ can be thought of as an LSZ reduction for the bulk scattering process.
The crucial approximation we must make is that the full matrix element in the basis \eqref{eq:ket-norm} is a simple phase
\begin{equation}
\ket{p,\phi} \otimes \ket{p',\phi'}_{\text{out}} \approx e^{i\delta(s,\phi-\phi')} \ket{p,\phi} \otimes \ket{p',\phi'}_{\text{in}} ,
\end{equation}
where $s$ is the relevant Mandelstam variable.
This is the eikonal approximation, and thus the scattering element is
\begin{equation}
\label{eq:bulkscattering}
\avg{VWVW}_{\beta_+\beta_-} = \int d\phi d\phi' dp_1^U dp_2^V e^{i\delta(s,\phi-\phi')} \left[ p_1^U \psi_1^*(p_1^U,\phi) \psi_3(p_1^U,\phi) \right] \left[ p_2^V \psi_2^*(p_2^V,\phi') \psi_4(p_2^V,\phi') \right] .
\end{equation}
The disconnected correlator $\avg{VV}\avg{WW}$ can be obtained by setting $\delta = 0$.

The calculation proceeds in three steps:
\begin{enumerate}
\item
Define bulk-to-boundary propagators using embedding coordinates and compute their Fourier transforms to obtain wavefunctions of 2-particle in and out states.
\item
Compute the eikonal phase, the second-order on-shell Einstein-Hilbert action for a two-shockwave perturbed background.
\item
Perform the integrals over momentum and angular variables which enter in the definition of the scattering matrix element.
\end{enumerate}

\subsubsection{Propagators and wavefunctions}

Let us start by commenting on the dependence of the bulk prescription \eqref{eq:bulkscattering} on the boundary conditions imposed on the singularities. The dependence can come from two places: the eikonal scattering phase and the wavefunctions that we use to propagate the particles to the scattering region. For the eikonal phase, we will assume that our particles are the only objects scattering on top of the analytically extended spacetime, that is, there is no additional shockwave coming from the excision surface. For the wave functions, we will use the method of images propagator obtained from empty AdS. 
This choice of propagator amounts to a choice of transparent boundary conditions on the singularity. 
There is no guarantee that this is the boundary condition that the CFT imposes, but it is a choice that satisfies both the inner and outer KMS conditions, which is required for the interpretation of the four sided correlator in terms of a single CFT, as discussed in Sec.~\ref{sec:multiboundary}. It is also the (unique) maximal analytic extension of the two point function between the outer and inner horizons, when written in terms of inner Kruskal coordinates. So our strategy is to assume the ``mildest possible" effect from the singularity and see if we run into trouble. These assumptions are also partially justified by the fact that we are able to recover the vacuum block result \eqref{eq:virasoroblockotoc} in the high temperature limit.

\begin{figure}
\centering
\includegraphics[scale=.75]{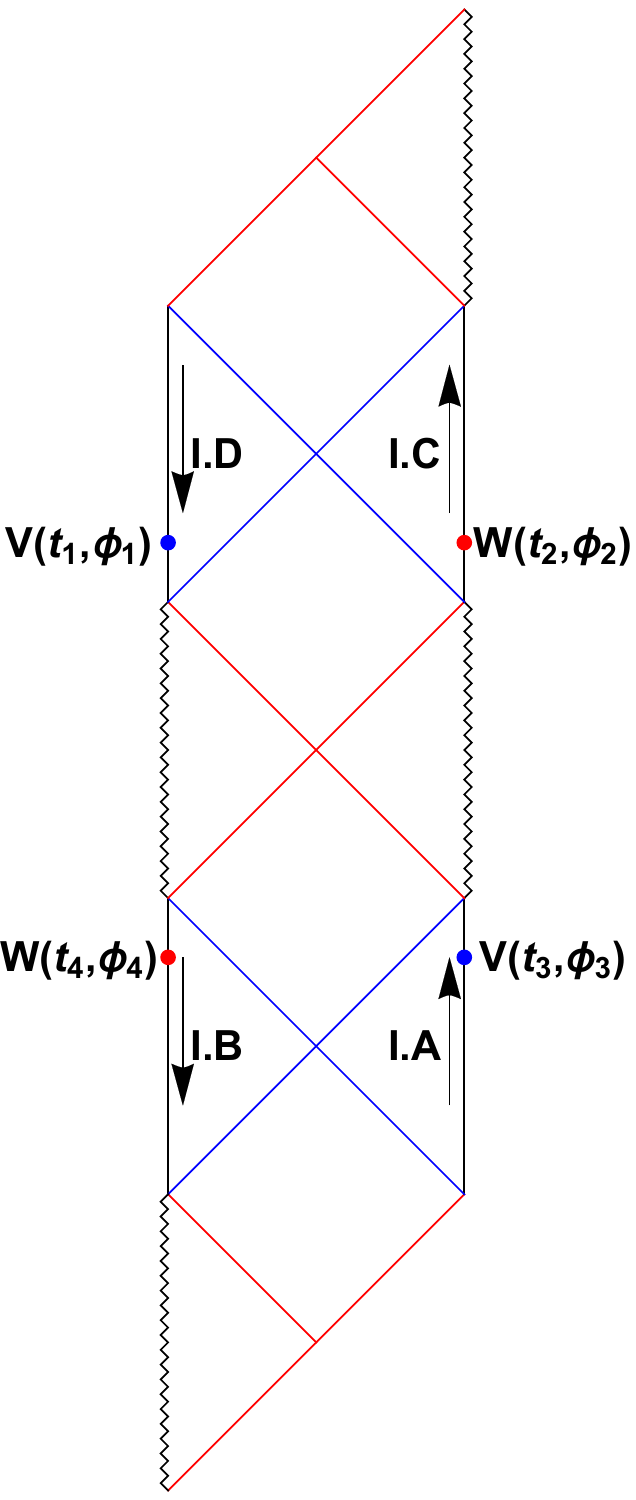}
\caption{The configuration of $V$ and $W$ operators in the rotating BTZ black hole spacetime.
The arrows represent the direction of increasing Killing time $t$.
To place the $W$ operators in the shockwave limit, we must take $t_2 = t_4 \approx -t$ for $t$ large.
}
\label{fig:penrose-operators}
\end{figure}

Bulk-to-boundary propagators can be computed by writing the embedding space distance $d$, 
\begin{equation}
\cosh(d) = -P_1 P_2, 
\end{equation}
and then the propagator $\avg{\mathcal{O}_\Delta (P_1) \Phi (P_2)}$ is given by removing the decaying mode from $(-P_1 P_2)^{-\Delta}$.
From the asymptotic regions $(t,r_\infty,\phi')$ to the inner horizon region $(U,V,\phi)$ we use \eqref{eq:region-1A-corotate}-\eqref{eq:region-1D-corotate} and \eqref{eq:inner-kruskal-embedding} to find (defining $\Delta\phi \equiv \phi - \phi'$)
\begin{equation}
\begin{split}
\cosh(d_{\text{I.A}}) & = \frac{r_\infty (r_+^2 - r_-^2)^{-1/2}}{1+UV} \left[ U e^{r_+\Delta\phi } + V e^{-r_+\Delta\phi } + (1-UV)\sinh(\kappa t + r_-\Delta\phi )  \right],\\
\cosh(d_{\text{I.B}}) & = \frac{r_\infty (r_+^2 - r_-^2)^{-1/2}}{1+UV} \left[ U e^{r_+\Delta\phi } + V e^{-r_+\Delta\phi } - (1-UV)\sinh(\kappa t + r_-\Delta\phi )  \right],\\
\cosh(d_{\text{I.C}}) & = \frac{r_\infty (r_+^2 - r_-^2)^{-1/2}}{1+UV} \left[ -U e^{r_+\Delta\phi } - V e^{-r_+\Delta\phi } + (1-UV)\sinh(\kappa t + r_-\Delta\phi )  \right],\\
\cosh(d_{\text{I.D}}) & = \frac{r_\infty (r_+^2 - r_-^2)^{-1/2}}{1+UV} \left[ -U e^{r_+\Delta\phi } - V e^{-r_+\Delta\phi } - (1-UV)\sinh(\kappa t + r_-\Delta\phi )  \right] .
\end{split}
\end{equation}
We take the $W$ particles to be in regions I.B and I.C, and the $V$ particles to be in regions I.A and I.D.
This four-sided configuration is shown in Fig.~\ref{fig:penrose-operators}.
The bulk-to-boundary propagators $\avg{\mathcal{O}_\Delta \Phi}$ are then given by $\cosh(d)^{-\Delta}$, where $\Delta$ is the conformal dimension of the boundary operator $\mathcal{O}$ and $\Phi$ is the dual bulk field sourced by $\mathcal{O}$.
We will use the method of images to enforce periodicity in the angular variable. 
The relevant propagators are evaluated on the inner horizon $UV = 0$, so their simplified form is
\begin{equation}
\begin{split}
\avg{V(t_1,\phi_1)^\dagger \Phi_V(U,V,\phi)} & = c_V \sum_{n_1=-\infty}^\infty \left[ -U e^{r_+\Delta\phi_1 } - V e^{-r_+\Delta\phi_1 } - \sinh(\kappa t_1^* + r_-\Delta\phi_1 ) \right]^{-\Delta_V} , \\
\avg{W(t_2,\phi_2)^\dagger \Phi_W(U,V,\phi')} & = c_W \sum_{n_2=-\infty}^\infty \left[ -U e^{r_+\Delta\phi'_2 } - V e^{-r_+\Delta\phi'_2 } + \sinh(\kappa t_2^* + r_-\Delta\phi'_2 ) \right]^{-\Delta_W} , \\
\avg{V(t_3,\phi_3)\Phi_V(U,V,\phi)} & = c_V \sum_{n_3=-\infty}^\infty \left[ U e^{r_+\Delta\phi_3 } + V e^{-r_+\Delta\phi_3 } + \sinh(\kappa t_3 + r_-\Delta\phi_3 )  \right]^{-\Delta_V} , \\
\avg{W(t_4,\phi_4)\Phi_W(U,V,\phi')} & = c_W \sum_{n_4=-\infty}^\infty \left[ U e^{r_+\Delta\phi'_4 } + V e^{-r_+\Delta\phi'_4 } - \sinh(\kappa t_4 + r_-\Delta\phi'_4 ) \right]^{-\Delta_W},
\end{split}
\end{equation}
where $\Delta\phi_1 \equiv \phi - \phi_1 + 2\pi n_1$, $\Delta\phi_3 \equiv \phi - \phi_3 + 2\pi n_3$, $\Delta\phi_2' \equiv \phi' - \phi_2 + 2\pi n_2$, and $\Delta\phi_4' \equiv \phi' - \phi_4 + 2\pi n_4$.
The wavefunctions are simply Fourier transforms of these propagators along the opposite lightcone coordinate axis.
\begin{equation}
\begin{split}
\psi_1(p^U,\phi) & = \int_{-\infty}^\infty dV\ e^{2i V p^U} \avg{V(t_1,\phi_1)^\dagger \Phi_V(0,V,\phi)} ,\\
\psi_2(p^V,\phi') & = \int_{-\infty}^\infty dU\ e^{2i U p^V} \avg{W(t_2,\phi_2)^\dagger \Phi_W(U,0,\phi')} ,\\
\psi_3(p^U,\phi) & = \int_{-\infty}^\infty dV\ e^{2i V p^U} \avg{V(t_3,\phi_3) \Phi_V(0,V,\phi)} ,\\
\psi_4(p^V,\phi') & = \int_{-\infty}^\infty dU\ e^{2i U p^V} \avg{W(t_4,\phi_4) \Phi_W(U,0,\phi')}  ,
\end{split}
\end{equation}
where the exponentials are $e^{-ig_{UV}V p^U}$ or $e^{-ig_{VU}U p^V}$, $g$ being the unperturbed inner Kruskal metric \eqref{eq:inner-kruskal-metric}, and we have evaluated each expression at either $U=0$ or $V=0$, depending on whether the particles travel along the $U$ or $V$ axis respectively.
The general formula for this type of Fourier transform is \cite{Shenker:2014cwa}\footnote{For $\Delta > 2 \in \mathbb{Z}$, this result can be derived by changing variables $u = x + \frac{b}{a}$ and then applying the residue theorem.
Notice the right hand side is analytic in $\Delta$.}

\begin{equation}
\int_{-\infty}^\infty dx\ e^{ipx} [ax+b]^{-\Delta} = \text{sign}\left(\text{Im}\left(\frac{b}{a}\right)\right) \Theta\left[-p\  \text{sign}\left(\text{Im}\left(\frac{b}{a}\right)\right)\right] \frac{2\pi i}{a^\Delta \Gamma(\Delta)} (ip)^{\Delta-1} e^{-ipb/a} .
\label{eq:fourier-transform}
\end{equation}
The wavefunctions are therefore
\begin{equation}
\begin{split}
\psi_1(p^U, \phi) & = \sum_{n_1=-\infty}^\infty \Theta(-p^U) \frac{-2\pi i c_V e^{r_+\Delta \phi_1}}{\Gamma(\Delta_V)} \left(- 2 i p^U e^{r_+\Delta \phi_1} \right)^{\Delta_V-1} e^{-2 i p^U e^{r_+\Delta\phi_1} \sinh (\kappa t_1^* + r_- \Delta \phi_1 )} , \\
\psi_2(p^V,\phi') & = \sum_{n_2=-\infty}^\infty \Theta(-p^V) \frac{-2\pi i c_W e^{-r_+\Delta \phi_2'}}{\Gamma(\Delta_W)} \left( - 2 i p^V e^{-r_+\Delta \phi'_2} \right)^{\Delta_W-1} e^{2 i p^V e^{-r_+\Delta\phi'_2} \sinh (\kappa t_2^* + r_- \Delta \phi'_2 )} , \\
\psi_3(p^U,\phi) & = \sum_{n_3=-\infty}^\infty \Theta(-p^U) \frac{2 \pi i c_V e^{r_+\Delta \phi_3}}{\Gamma(\Delta_V)} \left( 2 i p^U e^{r_+\Delta \phi_3} \right)^{\Delta_V-1} e^{-2 i p^U e^{r_+\Delta\phi_3} \sinh (\kappa t_3 + r_- \Delta \phi_3 )} , \\
\psi_4(p^V,\phi') & = \sum_{n_4=-\infty}^\infty \Theta(-p^V) \frac{2 \pi i c_W e^{-r_+\Delta \phi_4'}}{\Gamma(\Delta_W)} \left(2 i p^V e^{-r_+\Delta \phi'_4} \right)^{\Delta_W-1} e^{2 i p^V e^{-r_+\Delta\phi'_4} \sinh (\kappa t_4 + r_- \Delta \phi'_4 )}  .
\end{split}
\end{equation}
Notice that we want $\Theta(-p)$ for forward propagation, since the U-V axes are oriented downward on the Penrose diagram (Fig.~\ref{fig:penrose}).

Before moving on, we address a subtle point about the Fourier transform \eqref{eq:fourier-transform}.
At the end of the OTOC calculation, our plan is to set the times $t_{1,2,3,4}$ to real values plus small imaginary parameters $i\epsilon_{1,2,3,4}$.
These imaginary parameters control the operator ordering in the Lorentzian correlator, if we were to compute it by analytic continuation from Euclidean signature.
Though the overall sign of the wavefunctions will cancel from the normalized correlator, the sign inside the step function has the potential to change depending on several things like the sign of $\epsilon_{1,2,3,4}$ or the relative sign between the sinh and exponential functions.
In writing the above wavefunctions we have taken 
\begin{equation}
    \epsilon_2 > \epsilon_3 > 0 > \epsilon_1 > \epsilon_4 .
\end{equation}
This may seem confusing from the perspective of a one-sided correlator, since this choice leads to a time-ordered correlator there.
However, as we saw in Sec.~\ref{sec:bndyotoc}, we cross a single branch cut when we continue the one-sided correlator to the four-sided correlator.
Unlike in \cite{Shenker:2014cwa}, we have set large (of order $\beta_+$, $\beta_-$) imaginary values for $t_{1,2,3,4}$ already to place the operators on the appropriate boundaries, and we consider the $\epsilon$ parameters to be very small.
As such, we do not touch them during the continuation, and therefore this choice leads to a four-sided correlator that we expect to be related to inner horizon stability as discussed above.\footnote{Recall that in the two-sided case discussed in \cite{Shenker:2014cwa}, no branch cut was crossed when moving two operators to the other side of the thermal circle, so the OTO two-sided ordering coincided with the OTO one-sided ordering.
In that discussion, the branch cut was crossed during continuation to positive Lorentzian times.
In our discussion, the branch cut is crossed when continuing across the inner horizon.} 

\subsubsection{Eikonal phase}

\begin{figure}
\centering
\includegraphics[scale=.75]{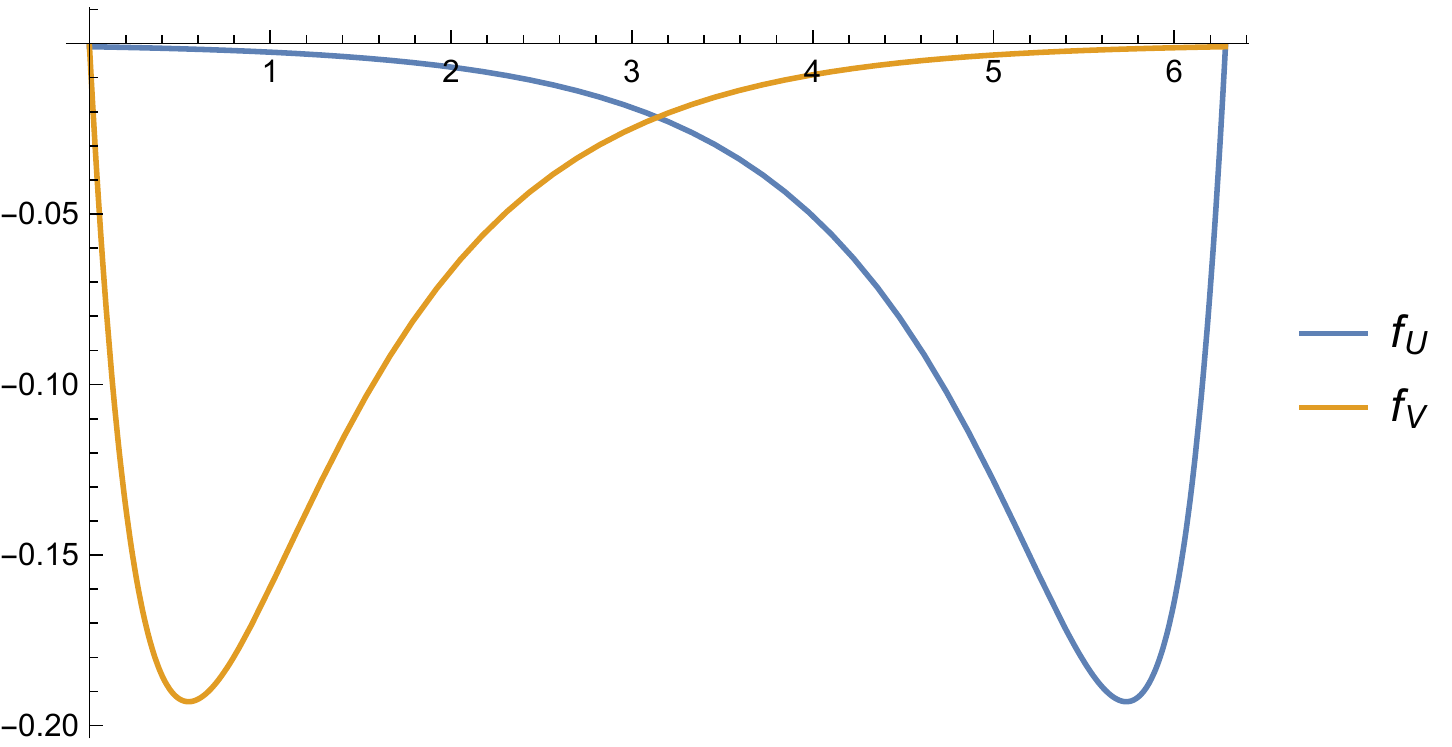}
\caption{Profiles for the $U$ and $V$ shocks for $r_+ = 2$, $r_- = 1$.
The asymmetry between the two results from enhancement due to the rotation direction.
The overall scale of either profile could be changed by including a scaling factor in the appropriate stress tensor.}
\label{fig:shock-profiles}
\end{figure}

We now turn to the eikonal phase $\delta(s,\phi-\phi')$ which is given by the classical saddle-point action $S_{\text{cl}}$ on the shockwave-plus-black-hole background.
\begin{equation}
\delta(s,\phi-\phi') = S_{\text{cl}} .
\end{equation}
We will construct a solution of the linearized Einstein equations\footnote{
In three bulk dimensions, these are actually solutions to the full nonlinear Einstein equations.
} 
that takes the form of a shockwave propagating on the inner horizon.
The perturbation ansatz for a $W$-particle with $p^V_2$ is
\begin{equation}
h_{UU} = 32\pi G_N r_- p^V_2 \delta(U) f_U(\phi-\phi'') .
\end{equation}
For the $V$-particles we have
\begin{equation}
h_{VV} = 32\pi G_N r_- p^U_1 \delta(V) f_V(\phi-\phi') .
\end{equation}
The stress tensors are
\begin{equation}
\begin{split}
T_{UU} & = \frac{2}{r_-}p_2^V \delta(U) \delta(\phi-\phi'') ,\\
T_{VV} & = \frac{2}{r_-}p_1^U \delta(V) \delta(\phi-\phi') .
\end{split}
\end{equation}
Note that the up-index stress tensors, which will appear in the on-shell action, are (for example) $T^{VV} = g^{VU} g^{VU} T_{UU}$.\footnote{We have discarded the $g^{VV}g^{VV}T_{VV}$ term since it goes like $V^2 \delta(V) = 0$.  Similarly for $T^{UU}$.}
\begin{equation}
\begin{split}
T^{VV} & = \frac{1}{2r_-} p_2^V \delta(U) \delta(\phi-\phi'') , \\
T^{UU} & = \frac{1}{2r_-} p_1^U \delta(V) \delta(\phi-\phi') .
\end{split}
\end{equation}
Adding such a perturbation to the metric\footnote{The $h_{UU}$-perturbed metric can be put into the form \eqref{eq:shockwave-discontinuous} by a coordinate transformation $u = U$, $v = V-\alpha(\phi)\Theta(U)$ where $\alpha(\phi)$ is $h_{UU}/4$ without the delta function $\delta(U)$.
}
\eqref{eq:inner-kruskal-metric} and plugging into $R_{\mu\nu} - \frac{1}{2} R g_{\mu\nu} + \Lambda g_{\mu\nu} = 8\pi G_N T_{\mu\nu}$ (with $\Lambda = -1$ since we have set $\ell_{\text{AdS}} = 1$ in AdS$_3$), we find the equation governing the $h_{UU}$ shock profile\footnote{It is necessary to use delta function identities such as $\delta'(U) = - \frac{\delta(U)}{U}$ and $U^2 \delta(U)^2 = 0$.}
\begin{equation}
-f_U''(\phi) + 2r_+ f_U'(\phi) - (r_+^2 - r_-^2) f_U(\phi) = \delta(\phi) .
\label{eq:shock-profile-u}
\end{equation}
We caution that in the case of the rotating black hole, the shock profile for $h_{VV}$ is actually different than that of $h_{UU}$ due to the cross-terms in the Kruskal metric \eqref{eq:inner-kruskal-metric}, which break the $U \leftrightarrow V$ symmetry of the non-rotating case.
This is the reason we have defined separate profile functions $f_U$ and $f_V$ in the perturbations.
The $h_{VV}$ shock profile is determined by
\begin{equation}
-f_V''(\phi) - 2r_+ f_V'(\phi) - (r_+^2 - r_-^2) f_V(\phi) = \delta(\phi) .
\end{equation}
These equations can be solved with linear combinations of exponentials, and we must remember to impose $f(0) = f(2\pi)$ which fixes the relative constant, so we have\footnote{These functions contain all angular dependence of the abstract shock profile $\alpha(\phi)$ discussed in Sec.~\ref{sec:geodesics}.
}
\begin{equation}
\begin{alignedat}{2}
f_U(\phi) & = \alpha_U \left( (e^{2\pi(r_+-r_-)}-1) e^{(r_+ + r_-)\phi} - (e^{2\pi(r_+ + r_-)}-1) e^{(r_+ - r_-)\phi} \right),\hspace{1cm} & \phi\text{ mod }2\pi , \\
f_V(\phi) & = \alpha_V \left( (1- e^{-2\pi(r_+-r_-)}) e^{-(r_++r_-)\phi} - (1- e^{-2\pi(r_++r_-)} ) e^{-(r_+-r_-)\phi} \right), \hspace{1cm} & \phi\text{ mod }2\pi .
\end{alignedat}
\label{eq:shockprofiles}
\end{equation}
Notice that the overall constant is set by the coefficient which appears in the shock stress tensor ansatz; omitting this extra parameter (as we have done) is equivalent to setting the overall constant to 1, as otherwise the Einstein equations are not obeyed.
To fix $\alpha_U$ and $\alpha_V$ then, we integrate the shock equations over an epsilon window of $\phi = 0$ which instructs us to equate $f'(2\pi) - f'(0) = 1$.
(The 1 can be modified by scaling up the shock stress tensor, but we will not do this.)
This gives
\begin{equation}
\begin{split}
\alpha_U^{-1} & = 2r_- (e^{2\pi(r_+ + r_-)}-1) (e^{2\pi(r_+-r_-)}-1) , \\
\alpha_V^{-1} & = 4r_- e^{-2\pi r_+} (\cosh (2\pi r_+) - \cosh(2\pi r_-) ) .
\end{split}
\end{equation}
Observe that these values of the overall coefficient will exactly cause $f_U(\phi) = f_V(-\phi)$ where this equation is understood with $\phi$ mod $2\pi$.
A plot of both shock profiles is shown in Fig.~\ref{fig:shock-profiles}. While this profile appears qualitatively different from the outer horizon case discussed in \cite{Jahnke:2019gxr,Mezei:2019dfv}, it can be formally obtained from that by exchanging $r_+$ and $r_-$.
Solutions of this type have a divergent stress tensor, and are perfectly consistent with results stating that perturbations on the inner horizon develop an infinite stress-energy.  However, this divergence need not indicate a gravitational catastrophe.
Indeed, all shock solutions have infinite stress-energy on the shock but do not signal the formation of a singularity of the same sort as the spacelike black hole singularity.

The eikonal phase is given by the classical action
\begin{equation}
S_{\text{cl}} = \frac{1}{2} \int d^3x \sqrt{-g} \left[  \frac{1}{16\pi G_N} h_{UU} \mathcal{D}^2 h_{VV} + h_{UU} T^{UU} + h_{VV} T^{VV} \right] .
\end{equation}
The first term is the quadratic contribution of $h$ to the Ricci scalar; there is no linear contribution since $h$ solves the linearized equations of motion.
The last two terms are in fact equal, since we have $f_U(\phi'-\phi'') = f_V(\phi''-\phi')$.
Varying the quadratic action by $h_{UU}$, we see that in this form the linearized equation is $\mathcal{D}^2 h_{VV} + T^{UU} = 0$, thus the first two terms in the action will cancel each other.
We are left with
\begin{equation}
\delta(s,\phi'-\phi'') = 16 \pi G_N r_- p_1^U p_2^V f_U(\phi'-\phi'') = 4\pi G_N r_- s f_U(\phi'-\phi''), \hspace{.5cm} s = 4p_1^U p_2^V.
\end{equation}

\subsubsection{Momentum and angular integrals}

The correlator is given by the overlap
\begin{equation}
F = \int d\phi d\phi' dp_1^U dp_2^V\ e^{i\delta(s,\phi-\phi')} \left[ p_1^U \psi_1^*(p_1^U,\phi) \psi_3(p_1^U,\phi) \right] \left[ p_2^V \psi_2^*(p_2^V,\phi') \psi_4 (p_2^V, \phi') \right] ,
\end{equation}
Throughout this subsection, we will drop various overall constants with the understanding that we are interested in $F$ divided by the factorized correlator, which arises if we set $\delta = 0$.
We also drop the momentum subscripts and set $\phi_1 = \phi_3$ and $\phi_2 = \phi_4$ to ease the calculation (though we will continue to use the notation e.g. $\Delta\phi_3$, which is now $\phi - \phi_1 + 2\pi n_3$ since the images stay separate).
After various reductions, we find
\begin{equation}
\begin{split}
F & = \sum_{n_1,\dots,n_4} \int_{-\infty}^0 dp^U dp^V \int_0^{2\pi} d\phi d\phi'\ e^{i\delta} (p^U)^{2\Delta_V-1} (p^V)^{2\Delta_W-1} \frac{ e^{\Delta_V r_+ (2\phi -2\phi_1 +2\pi(n_1+n_3))} }{ e^{\Delta_W r_+(2\phi' -\phi_2 + 2\pi(n_2+n_4))} }   \\
& \times \frac{\exp \left[ 2ip^U (e^{r_+\Delta\phi_1}\sinh(\kappa t_1  + r_-\Delta\phi_1)- e^{r_+\Delta\phi_3}\sinh(\kappa t_3 + r_-\Delta\phi_3))  \right]}{\exp \left[ 2ip^V (e^{-r_+\Delta\phi_2'}\sinh(\kappa t_2 + r_-\Delta\phi_2') - e^{-r_+\Delta\phi_4'}\sinh(\kappa t_4 + r_-\Delta\phi_4')) \right]}
\end{split}
\end{equation}
If we do not have $n_1 = n_3$ and $n_2=n_4$, the integrand oscillates wildly and the contribution of the image will be suppressed.  So we set these variables to be equal, and
then perform the following shifts in the angular variables to disentangle the time coordinates.\footnote{
We might worry that this shift is adding a small imaginary piece to our angles, since we will eventually include $i\epsilon$'s in the times.  But since we imagine these parameters to be truly small and our expression is only a function of exponentials of $\phi$ and $\phi'$, this shift should not affect the integral beyond some overall prefactors which we can drop.
}
\begin{equation}
\phi \to \theta \equiv \phi -\phi_1 + \frac{\kappa}{2r_-}(t_1+t_3), \hspace{1cm} \phi' \to \theta' \equiv \phi' -\phi_2 + \frac{\kappa}{2r_-}(t_2+t_4) .
\end{equation}
Due to the method of images sum, the integration region can remain fixed under this shift, and the Jacobian is trivial.
Dropping overall prefactors, we have
\begin{equation}
\begin{split}
F & = \sum_{n_1,n_2} \int_{-\infty}^0 dp^U dp^V \int_0^{2\pi} d\theta d\theta' \ e^{i\delta} (p^U)^{2\Delta_V-1} (p^V)^{2\Delta_W-1} \frac{ e^{\Delta_V r_+ (2\theta +4\pi n_1)} }{ e^{\Delta_W r_+(2\theta' + 4\pi n_2)} }   \\
& \times \frac{\exp \left[ -4ip^U e^{-\frac{\kappa r_+}{2r_-}(t_1+t_3)} \sinh \left( \kappa \frac{t_3-t_1}{2} \right) e^{r_+(\theta+2\pi n_1)} \cosh (r_-(\theta+ 2\pi n_1))  \right]}{\exp \left[ 4ip^V e^{\frac{\kappa r_+}{2r_-} (t_2+t_4)} \sinh \left( \kappa \frac{t_2-t_4}{2} \right) e^{-r_+(\theta'+2\pi n_2)} \cosh(r_-(\theta'+2\pi n_2)) \right]}
\end{split}
\end{equation}
We can now define two new momentum variables
\begin{equation}
p^U \to p \equiv ip^U e^{-\frac{\kappa r_+}{2r_-}(t_1+t_3)} \sinh \left( \kappa \frac{t_3-t_1}{2} \right) , \hspace{.5cm} p^V \to q \equiv ip^V e^{\frac{\kappa r_+}{2r_-}(t_2+t_4)} \sinh \left( \kappa \frac{t_2-t_4}{2} \right) .
\end{equation}
This shift leads to a change in sign of the momenta, but this change is just the right one to keep the integral convergent.
After this change of variables, our expression reduces to
\begin{equation}
\begin{split}
F & = \sum_{n_1,n_2} \int_0^\infty dp dq \int_0^{2\pi} d\theta d\theta' \ e^{i\delta} p^{2\Delta_V-1} q^{2\Delta_W-1} \frac{ e^{\Delta_V r_+ (2\theta +4\pi n_1)} }{ e^{\Delta_W r_+(2\theta' + 4\pi n_2)} }   \\
& \times \frac{\exp \left[ -4p e^{r_+(\theta+2\pi n_1)} \cosh (r_-(\theta+ 2\pi n_1))  \right]}{\exp \left[ 4q e^{-r_+(\theta'+2\pi n_2)} \cosh(r_-(\theta'+2\pi n_2)) \right]}
\end{split}
\end{equation}

We now must find a saddle point in the $(q,\theta')$ integral.
The function to understand is
\begin{equation}
A(q,\theta') = (2\Delta_W-1) \log q - 2\Delta_W r_+ \theta' - 4q e^{-r_+(\theta'+2\pi n_2)} \cosh(r_- (\theta'+2\pi n_2)) .
\label{eq:q-thetaprime-saddle}
\end{equation}
Solving $\partial_q A = \partial_{\theta'} A = 0$ gives a real saddle point at
\begin{equation}
\begin{split}
q & = \frac{2\Delta_W-1}{4} \exp \left[ \frac{r_+}{r_-} \text{arctanh} \left( \frac{r_+}{r_-(1-2\Delta_W)} \right) \right] \sqrt{1- \frac{r_+^2}{r_-^2(1-2\Delta_W)^2}} \\
& \approx \frac{\Delta_W}{2} - \frac{1}{4} \left(1 + \frac{r_+^2}{r_-^2} \right) + \mathcal{O}(1/\Delta_W), \\
\theta' & = \frac{1}{r_-} \text{arctanh} \left( \frac{r_+}{r_-(1-2\Delta_W)} \right) - 2\pi n_2 \approx -2\pi n_2 - \frac{r_+}{2r_-^2 \Delta_W} + \mathcal{O}(1/\Delta_W^2) .
\end{split}
\end{equation}
Notice that the region of integration for $\theta'$ contains the saddle only when $n_2 = -1$, so this is the only relevant piece of the $n_2$ images sum.\footnote{We do not expect a nice static limit for this calculation since it is constructed using the inner horizon geometry.  There is a signal of this effect in the $O(1)$ term of the saddle; for small enough $r_- \ll r_+$ this term competes with the leading $\Delta_W$ term.}
At large $\Delta_W$ we find a saddle point for the $q$ and $\theta'$ integrals $q = \Delta_W/2$ and $\theta' = 2\pi$ (with $n_2 = -1$ the contributing image).
\begin{equation}
\begin{split}
F & = \sum_{n_1} \int_0^\infty dp \int_0^{2\pi} d\theta \ e^{i\delta} p^{2\Delta_V-1} e^{\Delta_V r_+ (2\theta +4\pi n_1)} e^{-4p e^{r_+(\theta+2\pi n_1)} \cosh (r_-(\theta+ 2\pi n_1)) }
\end{split}
\end{equation}
The eikonal phase at this point looks quite different than its original form, due to the various changes of variables and redefinitions.
Using $f(\phi) = f(\phi+2\pi m)$ to eliminate $\theta' = 2\pi$ and dropping $p$ with foresight, we define the quantity
\begin{equation}
\label{eq:finaldelta}
\delta = -\frac{8\pi G_N r_- \Delta_W}{\sin\left(\kappa\frac{\epsilon_3-\epsilon_1}{2}\right) \sin\left(\kappa\frac{\epsilon_2-\epsilon_4}{2}\right)} e^{\frac{\kappa r_+}{2r_-}(t_1+t_3-t_2-t_4)} f_U \left( \theta + \phi_1-\phi_2+ \frac{\kappa}{2r_-}(t_2+t_4-t_1-t_3) \right) .
\end{equation}
We have now specified our times as $(t_1,t_2,t_3,t_4) = (i\epsilon_1, -t+i\epsilon_2, i\epsilon_3, -t + i\epsilon_4)$.
The $p$ integral can now be done exactly, and we drop an overall factor of 4 for normalization purposes.
\begin{equation}
F = \sum_{n_1} \int_0^{2\pi} d\theta \left[ -\frac{i\delta}{4} e^{-r_+(\theta+2\pi n_1)}  + \cosh(r_-(\theta+2\pi n_1)) \right]^{-2\Delta_V} .
\label{eq:initial-otoc}
\end{equation}
Now let us implicitly define $\gamma$ (which is linear in $G_N$) via
\begin{equation}
    -\frac{i\delta}{4} \equiv i \gamma f_U ,
\end{equation}
and also define the following quantity
\begin{equation}
    \zeta \equiv \phi_1-\phi_2 + \frac{\kappa}{2r_-}(t_2+t_4-t_1-t_3)\text{ mod } 2\pi .
\end{equation}
We can rewrite the argument of $f_U$ since it is defined to be a periodic function.
\begin{equation}
    f_U\left( \theta + \phi_1-\phi_2 + \frac{\kappa}{2r_-}(t_2+t_4-t_1-t_3) \right) = f_U(\theta + \zeta) .
\end{equation}
We will analyze the integrals in \eqref{eq:initial-otoc} in two ways: we give an exact expression and then we study it with saddle-point approximation in various limits.

We will perform the integrals exactly, keeping the image dependence.
Due to the sum over images, the  integral in $F$ is invariant under translations of the region of integration, so we can translate to the region where $f_U$ can be faithfully represented as a sum of exponentials.
\begin{equation}
    F = \sum_{n}  \int_{-\zeta}^{2\pi-\zeta} \left[ i\gamma f_U(\theta+\zeta) e^{-r_+(\theta+2\pi n)} + \cosh(r_-(\theta+2\pi n)) \right]^{-2\Delta_V} .
    \label{eq:shifted}
\end{equation}
We have changed the image label to $n$ in this expression because translation by an amount which is not a multiple of $2\pi$ (and $\zeta$, in our case, is not) causes what we mean by the image to change.
That is to say, the $n=0$ integral in \eqref{eq:shifted} is not equal to the $n_1 = 0$ integral in \eqref{eq:initial-otoc}.
Instead, the $n=0$ integral contains contributions from several $n_1$ images including $n_1 = 0$.
Since we have an expression for $f_U$ in terms of exponentials which is valid in our integration range, we substitute \eqref{eq:shockprofiles} and find the exact indefinite result
\begin{equation}
\begin{split}
 \int & d\theta \left[ a_n e^{r_-\theta} + b_n e^{-r_-\theta} \right]^{-2\Delta_V} \\
 & =  \frac{(a_n e^{r_-\theta} + b_n e^{-r_-\theta})^{-2\Delta_V}}{2r_-b_n \Delta_V} (a_n e^{2r_-\theta} + b_n) _{2}F_1\left( 1,1-\Delta_V,1+\Delta_V, \frac{-e^{2r_-\theta} a_n}{b_n} \right).
 \end{split}
\end{equation}
The constants $a_n$ and $b_n$ are
\begin{equation}
    \begin{split}
        a_n & = \frac{1}{2} e^{2\pi n r_-} + i\gamma \alpha_U (e^{2\pi(r_+-r_-)}-1) e^{(r_++r_-)\zeta} e^{-2\pi n r_+} ,\\
        b_n & = \frac{1}{2} e^{-2\pi n r_-} + i\gamma \alpha_U (1-e^{2\pi(r_++r_-)}) e^{(r_+-r_-)\zeta} e^{-2\pi n r_+} .
    \end{split}
\end{equation}
We can then construct $F$ as a difference of two evaluations of the indefinite result, as usual, and then a sum over images.
Unfortunately, this form of the correlator is not very enlightening since we cannot perform the sum over images and it is unclear how to pick a dominant one.
The ratio $a_n/b_n$ is growing with $n$, and the corresponding growth of the hypergeometric function competes with the suppression from the prefactor which is vanishing roughly like $a_n^{-2\Delta_V}$.

We can analyse the original expression \eqref{eq:initial-otoc} in some simplifying limits. We will consider the leading $G_N$ correction in the remainder of this section and consider the high temperature limit in the next subsection. When the time $t$ is much smaller than the scrambling time $\sim -\log G_N$, we can treat $\delta$ in \eqref{eq:initial-otoc} as a perturbation. In that case, the saddle point is approximately on the real axis $\theta=-2\pi n_1 + O(\delta)$, and we see that only the $n_1=0$ image has the saddle inside the range of integration, and the rest of the images are highly suppressed by the cosh factor. Evaluating the $n_1=0$ image at this saddle and expanding in $G_N$ gives
\bea
F&\approx 1 +\frac{i \Delta_V}{2}\delta + \cdots \\
&=1 - \frac{4\pi i G_N \Delta_V \Delta_W r_-}{\sin_{31}\sin_{24}}e^{\frac{\kappa r_+}{r_-}t}f_U\big( \phi_1-\phi_2-\frac{\kappa}{r_-} t\big) + \cdots.
\eea
Notice in particular that the $G_N$ corrections only become important when the perturbing operator is inserted at late times $t \sim -\log G_N$. Therefore, as far as stability of the inner horizon goes, this result realizes scenario \ref{pt:mild} from the introduction, namely while there are quantum gravity corrections to the correlator of operators placed on opposite side of the BTZ inner horizon, the corrections are only important in certain configurations.  This suggests that most probes will pass largely undisturbed through the inner horizon.

\subsubsection{Black brane limit}

In this section, we perform the decompactification limit of \eqref{eq:initial-otoc} (for the related outer horizon discussion, see \cite{Mezei:2019dfv}). This is useful, because it can be compared to the boundary calculation of Sec.~\ref{sec:bndyotoc}. Of course, since the heavy-heavy-light-light Virasoro vacuum block in general calculates geodesics lengths, we are not expecting disagreement, however, since for the bulk calculation we needed to make a choice of boundary condition on the singularity in our wave function factors, it is at least a basic consistency check that the bulk and boundary calculations agree in this limit.

To decompactify \eqref{eq:initial-otoc} we reintroduce the AdS length $\ell$ by the following replacements
\begin{equation}
t\mapsto \frac{1}{\ell}t, \quad r \mapsto \ell r.    
\end{equation}
Making these replacements in the BTZ metric, \eqref{eq:global-rbtz-metric}, we see that the boundary metric on the cutoff surface $r=r_c$ is $ds^2=r_c^2/\ell^2 (-dt^2+\ell^2 d\varphi^2)$, therefore in the conformal frame where we drop the prefactor, the proper length of the boundary circle is $2\pi \ell$. We may then perform the decompactification by holding $t$ and $x \equiv \ell \varphi$ fixed and sending $\ell\rightarrow \infty$. The co-rotating coordinate $\phi$ in \eqref{eq:btzcorotating} becomes a boosted boundary coordinate 
\beq
\label{eq:hatx}
\hat x\equiv \ell \phi=x- \frac{\beta_+-\beta_-}{\beta_-+\beta_+}t,
\eeq
where we have also used \eqref{eq:charges}. Now we can go ahead and take this limit in the shock profiles \eqref{eq:shockprofiles}
\bea
f_U(\phi) & \approx \frac{1}{2r_-}e^{-(r_++r_-)\hat x} && &\hat x>0,\\
f_U(\phi) & \approx -\frac{1}{2r_-} e^{-(r_+-r_-)\hat x} && &\hat x<0.
\eea
We are interested a setup with $|t|\gg |x|$, therefore the argument of the shock profile in \eqref{eq:initial-otoc} is negative, $\theta+\phi_{12}-\frac{\kappa}{2r_-} t<0$. The $\ell \rightarrow \infty$ limit also kills all the $n_1 \neq 0$ images in \eqref{eq:initial-otoc} and we end up with an integral formally identical to the one occuring in the planar non-rotating case \cite{Shenker:2014cwa}. We can then evaluate this by saddle point approximation. Translated to boundary variables via \eqref{eq:charges} this gives
\begin{equation}
\label{eq:planarlimit}
F = \left[ \frac{1}{ 1 + \frac{4\pi i G_N  \Delta_W}{\sin_{13} \sin_{24}}  e^{\frac{2\pi}{\beta}(t+i\epsilon_{1234})+\frac{2\pi}{\beta_+}\hat x} } \right]^{\Delta_V} ,
\end{equation}
where $\beta= (\beta_++\beta_-)/2$. As we will now explain, this agrees with the vacuum block result \eqref{eq:virasoroblockotoc}.
For this, we map to the plane by $z=e^{\frac{2\pi}{\beta_-}(x-t)}$, $\bar z=e^{\frac{2\pi}{\beta_+}(x+t)}$, we exchange $x$ to the boosted $\hat x$ via \eqref{eq:hatx} (here it is important that our bulk convention has $\beta_+>\beta_-$). The cross ratios are defined via (see \eqref{eq:virasoroblockotoc})
\beq
\chi = \frac{z_{12}z_{34}}{z_{13}z_{24}}, \quad \bar \chi = \frac{\bar z_{12}\bar z_{34}}{\bar z_{13}\bar z_{24}},
\eeq
which both become small for large $t$ and read as (picking operator positions as we did after \eqref{eq:finaldelta})
\beq
\chi=-4 \sin_{13}\sin_{24} e^{-\frac{2\pi}{\beta}(t+i\epsilon_{1234})+\frac{2\pi}{\beta_-}\hat x} , \quad \quad \bar \chi = -4 \sin_{13}\sin_{24} e^{-\frac{2\pi}{\beta}(t+i\epsilon_{1234})-\frac{2\pi}{\beta_+}\hat x},
\eeq
where $\sin_{ij}=\sin \frac{4\pi}{\beta}(\epsilon_i-\epsilon_j)$ as before.\footnote{Note that in the context of OTOCs, we only access the part of the second sheet where the correlator \eqref{eq:planarlimit} is smaller in magnitude than one. In the present context, we can freely adjust the phase of $\chi$. This is because formally our cross ratio is the same as in \cite{Roberts:2014ifa}, but the $\epsilon_i$ are now chosen to correspond to a time ordered configuration and therefore the sign of $\epsilon_{1234}$ is not fixed. See \cite{Hartman:2015lfa} for more general bounds that the correlator must satisfy on the second sheet.}

We are interested in the CFT correlator on the second sheet in this limit. Plugging either $\chi$ or $\bar \chi$ in the second sheet vacuum block \eqref{eq:virasoroblockotoc}, we get agreement with \eqref{eq:planarlimit} up to a choice of $\beta_{\pm}$ and a sign in the exponent for $\hat x$. 
The right choice can be understood the following way.
When we apply the vacuum block approximation, we need to take the vacuum block in the dominant channel, as explained e.g. in \cite{Asplund:2014coa}. This amounts to continuing the vacuum block across the branch cut either via $(1-\chi)\mapsto e^{-2\pi i}(1-\chi)$ or $(1-\bar \chi) \mapsto e^{-2\pi i}(1-\bar \chi)$ and picking the larger result.
These two operations give different results for the vacuum block, even though they have to give the same for the full correlator, simply because the full correlator has to be single valued on the Euclidean sheet $\bar \chi=\chi^*$. Then, the two different continuations correspond to the vacuum block in two different OPE channels. In the vacuum block approximation, we are supposed to take the larger one.\footnote{Another instance when this is important for the OTOC vacuum block is the discussion in \cite{Nakagawa:2018kvo}.} This translates to picking the larger cross ratio from $\chi$ and $\bar \chi$ and substituting that into the vacuum block formula \eqref{eq:virasoroblockotoc}. Since our bulk calculation assumes $\hat x<0$, we need to use $\bar \chi$ and then the result agrees with \eqref{eq:planarlimit}.

\section{Discussion}\label{sec:discussion}

The Penrose diagram of black holes with inner horizons extends indefinitely, and includes an infinite number of asymptotic regions.
However, in holography, the purified thermal charged ensemble is a state in a two-sided Hilbert space $\mathcal{H}_{\text{CFT}} \otimes \mathcal{H}_{\text{CFT}}$, so by the usual mechanisms of AdS/CFT there should only be two asymptotic boundaries.
It is clear then that the infinite set of asymptotic boundaries cannot be independent from each other.
Indeed, there are time-like separated points between the infinite copies, so they cannot be independent in the sense in which they would contribute more $\mathcal{H}_{\text{CFT}}$ factors to the total Hilbert space.
Our proposal for sending operators through the inner horizon with certain analytic continuations gives an explanation of how the additional time-like separated boundaries are related to the physical ones.

In general relativity, there is no prescribed way to deal with singularities.
Even solving the classical wave equation in the presence of an inner horizon requires one to have a boundary condition on the timelike singularity.
Since AdS/CFT is a UV-complete theory of quantum gravity, it should tell us (in the classical large $N$ limit) what boundary condition we need.
We have argued that our inner horizon monodromy condition is largely insensitive to this boundary condition and therefore its violation signals a deeper problem with the extension of spacetime beyond the Cauchy horizon than the boundary condition.
We note that our argument does not specify what actually goes wrong at the inner horizon from the bulk perspective; it would be very interesting to understand the possibilities.
For example, we cannot use our results to conclude that a true spacelike singularity forms, but we can say that quantum gravity predicts at least a non-analytic branch-point locus at the inner horizon for any field propagating on such a background.
This does not necessarily mean a breakdown of smoothness, since the solution to the wave equation could be of the form $\sqrt{U V}e^{-1/U^2} e^{-1/V^2}$ in inner Kruskal coordinates, where the inner horizon locus is smooth but branched non-analytic. However, it does imply the breakdown of predictability if any spacetime exists beyond the inner horizon.
In rotating BTZ, however, we have found that the analytic extension seems consistent from the quantum gravitational point of view.

\subsection{Charge and rotation}
Our techniques for dealing with charged black holes differed greatly from those which applied to the rotating cases.
This is a bit odd from the string theoretic perspective, since it is a basic principle of Kaluza-Klein (KK) reductions that angular momentum in higher dimensions can be exchanged for charge in one fewer dimensions.
So, we might have expected angular momentum and charge to be treated on a more equal footing.
Indeed, in famous AdS/CFT instances like $\mathcal{N}=4$ super Yang-Mills, the Kaluza-Klein perspective on charge looms large since the harmonics coming from $S^5$ reduction to AdS$_5$  actually correspond to  BPS multiplets of the gauge theory.  We will leave further investigation for future work. 

We have also not addressed the case when both rotational and gauge charges are turned on at the same time.
Rotating, charged AdS black holes certainly exist, and have a similar Penrose diagram to the ones we have encountered in this work.
We expect that arguments similar to the ones we have made will apply in those cases, but it would be good to check this explicitly.
Along the same lines, we have not considered extremal black holes.
In this case, the Penrose diagram looks quite different, and it is not clear how our techniques will fare in those cases.
The main issue is that extremality is quite unstable, in the sense that if we send a probe operator into the interior then extremality is lost unless this probe is quite special.

\subsection{Strong cosmic censorship}
We have argued that strong cosmic censorship is violated for rotating BTZ.
Are there other cases where it can be violated?
The key feature in rotating BTZ which evaded our monodromy arguments was the non-degeneracy of the inner KMS condition in the static limit.  Of course the uncharged BTZ black hole is topological in nature because it is constructed via a global identification, and this may have had something to do with the apparent  stability of its inner horizon.  Note that the 3d charged black hole does not have a topological construction since it requires a $U(1)$ gauge field to exist, and the inner KMS condition cannot be satisfied in this case consistently with boundary causality and unitarity.
We might therefore be tempted to conjecture that the inner KMS condition can be topologically ``protected".
In higher dimensions, there are  topological black holes which have exotic features like high genus horizon topology and no Hawking-Page transition.
Nevertheless, they can be studied holographically \cite{Emparan:1999gf}. It would be interesting to understand if they show the same signatures of strong cosmic censorship violation.

It would also be interesting to extend our discussion to lower dimensions. There is really only one lower dimension to go, the 2d case, but there are nontrivial black hole geometries even here, like coset black holes \cite{Mukhi:1993zb}.

\subsection{Lightcone singularities}

A peculiar feature of the rotating BTZ geometry is that there exist light rays between boundaries separated by the inner horizons that do not fall behind the singularity, see Fig. \ref{fig:bulklightcones}. These light rays go between a boundary point ($x^-,x^+)$ and its ``image" $(x^- +i\beta_-,x^+)$, and the corresponding light cone singularities are present in the boundary two point function \eqref{eq:BTZmethodofimages}. In the static limit, the time-like singularities touch,  closing up the spacetime so that there is only one null geodesic that marginally makes it through. If we identify the upper boundary which is reached by this marginal light ray with a second sheet of the lower boundary, this looks like a bouncing null ray, as discussed in \cite{Fidkowski:2003nf}.\footnote{For BTZ, this bouncing null ray cannot be approached by space-like geodesics connecting the two boundaries, which is the meat of the discussion in \cite{Fidkowski:2003nf} for the higher dimensional cases.} It is interesting to ask about the fate of this lightcone singularity if $1/c$ corrections are taken into account. While we do not have an answer to this question in holographic CFTs, it is interesting to note that for e.g. the Ising CFT, one can check\footnote{For Ising correlators see chapter 12 of \cite{DiFrancesco:1997nk}.} that this chiral copy of the lightcone singularity does survive in the analytically continued torus correlator (even though the inner KMS condition does not).

\subsection{Entanglement and KMS conditions}
Our work is closely related to the recently proposed criterion for smoothness of the inner horizon \cite{Papadodimas:2019msp} which involved essentially testing for vacuum-like entanglement across the inner horizon at small enough distance scales.
The basic reasoning was that if the inner horizon is smooth, then on small enough length scales the state looks like the vacuum (as all states do in quantum field theory), and so the entanglement structure should simply be that of the vacuum.
Our inner KMS condition is sensitive to similar physics; if the inner horizon region is analytic, our inner KMS collapse argument goes through and we find a contradiction in the CFT. One of the explicit conditions of \cite{Papadodimas:2019msp} was that Rindler-like free field modes $a_\omega, a_\omega^\dagger$ that are sufficiently localized near the horizon should have a thermal two point function
\beq
\langle a^\dagger_\omega a_\omega \rangle = \frac{1}{e^{\frac{2\pi}{\beta_{<} }\omega}-1}.
\eeq
It is easy to see that this two point function follows from the KMS condition for a near (inner) horizon Rindler Hamiltonian $\tilde H$ and that we have a free field $[\tilde H,a_\omega]=-\omega a_\omega$, $[a_\omega,a_\omega^\dagger]=1$.

\subsection*{Acknowledgments}
We are grateful to Alex Belin, Onkar Parrikar, Amit Sever, Tomonori Ugajin, and Sasha Zhiboedov 
for useful discussions, and to Onkar Parrikar for collaboration in the early stages of this work.
VB, AK, and GS were supported in part by the Simons Foundation through the It From Qubit Collaboration (Grant No. 38559) and also by the DOE grants FG02-05ER-41367 and QuantISED grant DE-SC0020360.
GS was partially supported by FWO-Vlaanderen through project G044016N, and by Vrije Universiteit Brussel through the Strategic Research Program ``High-Energy Physics”.

\bibliographystyle{JHEP}
\bibliography{inner}

\end{document}